\documentclass[11pt, a4paper]{article}
\pdfoutput=1

\usepackage{amsmath}
\usepackage{amssymb}
\usepackage{comment}
\usepackage{multirow}
\usepackage[utf8]{inputenc}
\usepackage{bm}
\usepackage{cite}
\usepackage{physics}
\usepackage{slashed}
\usepackage{braket}
\usepackage{enumitem}
\usepackage{xargs}
\usepackage{tcolorbox}
\usepackage{xfrac}
\usepackage[footnotesize]{caption}
\usepackage{fourier}
\usepackage{cancel}
\usepackage{soul}

\usepackage{ifpdf}
\ifpdf
\usepackage{graphicx, hyperref, xcolor}           
\else                                             
\usepackage[dvipdfmx]{graphicx, hyperref, xcolor} 
\fi
\usepackage{subcaption}

\usepackage{tikz}
\usepackage{tikz-feynman}
\tikzfeynmanset{compat=1.0.0}

\definecolor{rossoferrari}{HTML}{D9073D}
\definecolor{mediumblue}{HTML}{0000CD}
\definecolor{forestgreen}{HTML}{228B22}
\definecolor{desy_blue}{HTML}{009EE2}
\definecolor{desy_orange}{HTML}{FD8800}
\definecolor{light_pink}{rgb}{1,0.4,0.4}
\definecolor{light_blue}{rgb}{0.284602,0.317763,0.963947}
\hypersetup{
setpagesize=false,
bookmarksnumbered=true,%
bookmarksopen=true,%
colorlinks=true,%
linkcolor=light_blue,
urlcolor=rossoferrari,
citecolor=rossoferrari,
linktocpage=false,
}

\usepackage[height=8.85in,width=6.8in]{geometry}
\renewcommand{\baselinestretch}{1.14}

\makeatletter
\@addtoreset{equation}{section}

\makeatother

\begin{document}


\begin{titlepage}


~\vspace{-2cm}

\hfill CERN-TH-2022-162

\hfill RESCEU-17/22 

\hfill KEK-TH-2455 

\hfill MS-TP-22-37 

\hfill TU-1170 

\begin{center}

\vskip 0.5in

{\huge \bf Wash-in leptogenesis after axion inflation}

\vskip .8in

{\Large
Valerie Domcke$^{1}$, Kohei Kamada$^{2}$, Kyohei Mukaida$^{3,\,4}$,\\[.5em]
Kai Schmitz$^{1,\,5}$, Masaki Yamada$^{6,\,7}$
}

\vskip .6in

{\small
\begin{tabular}{ll}
$1$&\!\!\!\!\!\! \emph{Theoretical Physics Department, CERN, 1211 Geneva 23, Switzerland}\\[.3em]
$2$&\!\!\!\!\!\! \emph{Research Center for the Early Universe (RESCEU), Graduate School of Science,}\\[.3em]
   &\!\!\!\!\!\! \emph{The University of Tokyo, Hongo 7-3-1 Bunkyo-ku, Tokyo 113-0033, Japan}\\[.3em]
$3$&\!\!\!\!\!\! \emph{KEK Theory Center, Tsukuba 305-0801, Japan}\\[.3em]
$4$&\!\!\!\!\!\! \emph{Graduate University for Advanced Studies (Sokendai), Tsukuba 305-0801, Japan}\\[.3em]
$5$&\!\!\!\!\!\! \emph{University of M\"unster, Institute for Theoretical Physics, 48149 M\"unster, Germany}\\[.3em]
$6$&\!\!\!\!\!\! \emph{Department of Physics, Tohoku University, Sendai, Miyagi 980-8578, Japan}\\[.3em]
$7$&\!\!\!\!\!\! \emph{FRIS, Tohoku University, Sendai, Miyagi 980-8578, Japan}
\end{tabular}
}
\end{center}


\vskip .6in

\begin{abstract}
$CP$ violation and the violation of baryon-minus-lepton number $B\!-\!L$ do not necessarily have to occur simultaneously in order to accomplish successful leptogenesis. Instead, it suffices if new $CP$-violating interactions at high energies result in primordial charge asymmetries, which are then reprocessed into a nonvanishing $B\!-\!L$ asymmetry by right-handed neutrinos (RHNs) at lower energies. In this paper, we study this novel mechanism known as \textit{wash-in leptogenesis}, utilizing axion inflation as the source of high-scale $CP$ violation. We specifically consider axion inflation coupled to the Standard Model hypercharge sector, which results in the dual production of hypermagnetic helicity and fermionic charge asymmetries. Although the survival of these charges is endangered by sphaleron processes, magnetic diffusion, and the chiral plasma instability, we find a large range of viable scenarios. We consistently account for RHN flavor effects and coherence among the Standard Model lepton flavors across a wide range of RHN masses. We find a lower bound of $10^{5\cdots9}\,\textrm{GeV}$ on the mass of the lightest RHN involved in wash-in leptogenesis, depending on the onset of turbulence in the chiral plasma and the Hubble scale of inflation. Our model is representative of a broader class of new leptogenesis scenarios and suggests interesting observational signatures with regard to intergalactic magnetic fields, primordial black holes, and gravitational waves.
\end{abstract}


\end{titlepage}


\renewcommand{\thepage}{\arabic{page}}
\setcounter{page}{1}


{\hypersetup{linkcolor=black}\renewcommand{\baselinestretch}{1}\tableofcontents}

\renewcommand{\thepage}{\arabic{page}}
\renewcommand{\thefootnote}{$\natural$\arabic{footnote}}
\setcounter{footnote}{0}


\section{Introduction}
\label{sec:introduction}


The observed \textit{baryon asymmetry of the Universe} (BAU), typically quantified in terms of the baryon-to-photon ratio $\eta_B^{\rm obs} = n_{\rm b}/n_\gamma = \left(6.12 \pm 0.04\right) \times 10^{-10}$~\cite{Aghanim:2018eyx,Zyla:2020zbs}, cannot be created within the \textit{Standard Model} (SM) and hence provides clear evidence for new physics.
One of the most attractive possibilities to explain the origin of the BAU consists in baryogenesis via leptogenesis~\cite{Fukugita:1986hr}, which naturally occurs in the type-I seesaw extension of the Standard Model~\cite{Minkowski:1977sc,Yanagida:1979as,Yanagida:1980xy,GellMann:1980vs,Mohapatra:1979ia,Schechter:1980gr,Schechter:1981cv} and thus establishes a close connection between early-Universe cosmology and neutrino physics~\cite{Chun:2017spz,Bodeker:2020ghk,Dasgupta:2021ies}.
The main idea behind leptogenesis is to employ the $CP$-violating couplings of \textit{right-hand neutrinos} (RHNs) in order to generate a primordial lepton asymmetry\,---\,either via RHN decays~\cite{Fukugita:1986hr} or oscillations~\cite{Akhmedov:1998qx}\,---\,which is then partly reprocessed into a baryon asymmetry in consequence of the chemical transport in the SM plasma.
Here, a key role is played by the weak sphaleron processes~\cite{Kuzmin:1985mm}, which violate baryon-plus-lepton number $B\!+\!L$ and hence allow for the generation of nonzero baryon number $B$.
Meanwhile, the generation of a primordial lepton asymmetry during leptogenesis can also be regarded as a violation of baryon-minus-lepton number $B\!-\!L$, \textit{i.e.}, the linear combination of global charges that is orthogonal to $B\!+\!L$.
Unlike other baryogenesis scenarios, such as, \textit{e.g.}, GUT baryogenesis in the standard $SU(5)$ \textit{grand unified theory} (GUT)~\cite{Yoshimura:1978ex,Dimopoulos:1978kv,Toussaint:1978br,Weinberg:1979bt,Barr:1979ye}, leptogenesis therefore does not suffer from disastrous sphaleron wash-out.
Instead, the initial $B\!-\!L$ asymmetry remains conserved throughout and sets the scale for the final $B$ and $L$ asymmetries at the time of sphaleron freeze-out during the \textit{electroweak phase transition} (EWPT). 


Standard leptogenesis assumes that $CP$ violation and the violation of $B\!-\!L$ occur simultaneously, namely, whenever the RHN interactions are active in the thermal bath. 
In Ref.~\cite{Domcke:2020quw}, we, however, recently pointed out that this is, in fact, not a necessary condition in scenarios \textit{beyond the Standard Model} (BSM) that include heavy Majorana neutrinos.
In BSM models that build upon the type-I seesaw extension of the Standard Model rather than the Standard Model itself, there may, instead, exist a large hierarchy between the temperature scales of $CP$ and $B\!-\!L$ violation, which significantly relaxes Sakharov's conditions for successful baryogenesis~\cite{Sakharov:1967dj} and thus opens up a new window for model building.
The key observation in Ref.~\cite{Domcke:2020quw} was that new $CP$-violating interactions at high energies can readily lead to the generation of primordial charge asymmetries, which are then converted by RHN interactions at lower energies to a new chemical equilibrium that features nonzero $B\!-\!L$.%
\footnote{See also Refs.~\cite{Campbell:1992jd,Cline:1993vv,Cline:1993bd,Fukugita:2002hu} for related earlier work as well as Refs.~\cite{Fong:2015vna,Fong:2021xmi} for some related discussions in the more recent literature.}
This mechanism, which we dubbed \textit{wash-in leptogenesis} no longer relies on any $CP$ violation in the RHN sector, but merely utilizes RHN interactions in or close to thermal equilibrium in order to modify the chemical transport in the SM plasma.
Wash-in leptogenesis is thus based on the assumption that, just like the theoretical discovery of the weak sphaleron in the 1980s necessitated a first revision of the SM chemical transport, the experimental discovery of neutrino oscillations now calls for a second revision:
RHN interactions should be treated on the same footing as weak sphaleron processes; just like weak sphalerons can \textit{wash in} a baryon asymmetry in the standard leptogenesis scenarios, RHNs can \textit{wash in} a $B\!-\!L$ asymmetry. 


Going back in time in the early Universe, the number of conserved global charges in the SM grows 
as a function of temperature, as less and less SM interactions are still able to keep up with the Hubble expansion.
This offers a wealth of possibilities to set the stage for wash-in leptogenesis at lower temperatures. 
High-scale $CP$ violation only needs to create one or a few among the large number of available global charges; the existence of RHNs at a lower mass scale will then always automatically ensure the generation of $B\!-\!L \neq 0$. 


In this paper, we shall demonstrate the efficiency of this mechanism for a concrete and well-motivated source of high-scale $CP$ violation: axion inflation coupled to the SM hypercharge sector~\cite{Anber:2015yca,Adshead:2016iae,Jimenez:2017cdr,Domcke:2019mnd}, which spontaneously breaks $CP$ invariance by means of the nonzero and time-dependent value of the axion inflaton field during inflation. 
Axion inflation coupled to gauge fields has been extensively studied in the literature and gives rise to a rich phenomenology. 
In the version of the model that we are interested in, the violation of $CP$ invariance is communicated to the Standard Model via the axion--vector coupling $\phi Y_{\mu\nu}\widetilde{Y}^{\mu\nu}$, where $\phi$ is the axion inflaton field, $Y_{\mu\nu}$ is the hypercharge field strength tensor, and $\widetilde{Y}^{\mu\nu}$ denotes its dual.
This coupling results in the generation of helical hypermagnetic fields (\textit{i.e.}, primordial hypermagnetogenesis)~\cite{Turner:1987bw,Garretson:1992vt, Anber:2006xt}, which in turn leads to the nonperturbative production of SM fermions via the Schwinger effect~\cite{Domcke:2018eki,Gorbar:2021rlt,Gorbar:2021zlr}. 
The fermionic charge asymmetries generated during inflation are dictated by the SM chiral anomaly~\cite{Adler:1969gk,Bell:1969ts} and set the initial conditions for the chemical transport after inflation.
At the same time, the helicity stored in the hypermagnetic field is approximately conserved, as long as processes such as magnetic diffusion~\cite{Pouquet:1976zz,Kahniashvili:2012uj,Banerjee:2004df} and the \textit{chiral plasma instability} (CPI)~\cite{Joyce:1997uy,Boyarsky:2011uy,Akamatsu:2013pjd,Hirono:2015rla,Yamamoto:2016xtu,Rogachevskii:2017uyc,Kamada:2018tcs} can be neglected.
If it survives all the way down to the EWPT, its decay around the time of sphaleron freeze-out yields another relevant contribution to the BAU~\cite{Fujita:2016igl,Kamada:2016eeb,Kamada:2016cnb,Kamada:2018tcs}, in addition to the leptogenesis contribution generated at higher temperatures.
Following up on Ref.~\cite{Domcke:2020quw}, the aim of the present paper is to provide a unified description of these different mechanisms.


The remainder of this paper is organized as follows.
In Sec.~\ref{sec:ic}, we will first review the generation of hypermagnetic helicity and fermionic charge asymmetries during axion inflation and compute the initial conditions at the end of inflation.
Then, in Sec.~\ref{sec:helicity}, we will formulate the conditions under which hypermagnetic helicity has a chance to survive all the way down to the EWPT, before turning to the different possibilities of violating baryon and lepton number after axion inflation in Sec.~\ref{sec:bau}.
In this section, we will specifically discuss the chemical transport in the SM plasma (see Sec.~\ref{subsec:transport}) as well as the mechanisms of wash-in leptogenesis and baryogenesis from helicity decay (see Secs.~\ref{subsec:washin} and \ref{subsec:decay}, respectively). 
In this analysis, we will mostly stick to an explicit benchmark scenario and focus on the lowest possible RHN mass scale, \textit{i.e.}, a lightest RHN mass of around 100 TeV.
In Sec.~\ref{sec:scenarios}, we will then generalize our analysis and discuss the whole range of viable scenarios.
To this end, we will first consider the entire allowed range of RHN masses in Sec.~\ref{subsec:regimes}, collecting our main results in a compact format in Tab.~\ref{tab:coefficients}.
Similarly, we will discuss the general implications of our analysis that go beyond the simple case of axion inflation coupled to the SM hypercharge sector in Sec.~\ref{subsec:independent}, before studying the viable parameter space of this specific model in more detail in Sec.~\ref{subsec:estimates}.
Sec.~\ref{sec:conclusions} contains our conclusions. 


\section[\texorpdfstring{\boldmath{$CP$}}{CP}-violating initial conditions]{\boldmath{$CP$}-violating initial conditions}
\label{sec:ic}


Let us first review the generation of hypermagnetic helicity and chiral fermions during axion inflation. In doing so, we will also illustrate how our results can be generalized to alternatives to axion inflation that are as well capable of setting the initial conditions for wash-in leptogenesis. A slightly more extended discussion of such alternative scenarios will be provided in the context of our model-independent analysis in Sec.~\ref{subsec:independent}.


\subsection{Gauge-field production}


We are interested in axion inflation in a generic scalar potential $V\left(\phi\right)$, which we do not need to specify for our purposes, and in the presence of an axion--vector coupling to the SM hypercharge gauge field of the form 
\begin{equation}
\label{eq:phiYY}
\mathcal{L} \supset \frac{\alpha_Y}{4\pi}\frac{\phi}{f_\phi}\,Y_{\mu\nu}\widetilde{Y}^{\mu\nu} \,.
\end{equation}
Here, $\phi$ is the pseudoscalar axion field that drives inflation, while $Y_{\mu\nu}$ and $\widetilde{Y}^{\mu\nu} = \epsilon^{\mu\nu\rho\sigma}Y_{\rho\sigma}/2$ (with totally antisymmetric Levi-Civita symbol $\epsilon^{0123} = +1$) denote the hypercharge field strength tensor and its dual; $\alpha_Y = g_Y^2/\left(4\pi\right)$ is the hypercharge fine structure constant, with running hypercharge gauge coupling constant $g_Y$; and $f_\phi$ represents the axion decay constant.
We stress that the operator $Y_{\mu\nu}\widetilde{Y}^{\mu\nu}$ is topological, which means that its definition is not affected by the choice of coordinate system.
For concreteness, however, we will mostly work in the conformal frame in this section, such that indices are raised and lowered by the Minkowski metric.
Unless explicitly stated otherwise, all quantities carrying Lorentz or spatial indices are therefore understood to denote comoving quantities in dependence of conformal time $\tau$ and comoving spatial coordinates $\bm{x}$.
Physical quantities will enter our discussion whenever we perform spatial averages in the \textit{Friedmann--Lema\^itre-Robertson-Walker} (FLRW) spacetime. 
Specifically, this means that the quantities $Y_{\mu\nu}$, $A^\mu = \left(A^0,\bm{A}\right)$, $J^\mu = \left(J^0,\bm{J}\right)$, $\bm{E}$, $\bm{B}$ (see below) are by default comoving, while the densities $h_Y$, $q_i$, $q_{\rm CS}$ (see below) as well as all other quantities in our discussion such as $H$, $T$, etc.\ (see below) are by default physical.


The coupling in Eq.~\eqref{eq:phiYY} results in the explosive production of helical hypermagnetic fields during inflation. 
To see this, consider the equation of motion for the comoving hypercharge vector field $A^\mu$ as a function of $\tau$ and $\bm{x}$. 
In radiation gauge, $A^0 = \bm{\nabla}\cdot\bm{A} = 0$, this equation of motion obtains the following form,
\begin{equation}
\label{eq:AEOM}
\bm{A}''\left(\tau,\bm{x}\right) - \bm{\nabla}^2 \bm{A}\left(\tau,\bm{x}\right) = - \frac{\alpha_Y}{\pi}\frac{\phi'\left(\tau\right)}{f_\phi}\,\bm{\nabla} \times \bm{A}\left(\tau,\bm{x}\right) + g_Y\bm{J}\left(\tau,\bm{x}\right) \,,
\end{equation}
where a prime indicates the derivative with respect to $\tau$, and $\bm{J}$ is the comoving hyperelectric current that is induced by fermion production, which we will discuss in more detail further below.
The first term on the right-hand side of Eq.~\eqref{eq:AEOM} corresponds to a source term that originates from the axion--vector coupling in Eq.~\eqref{eq:phiYY} and which is responsible for the exponential amplification of the gauge field .
To study the axion-induced source term in more detail, let us first neglect the effect of nonperturbative fermion production during axion inflation for a moment and set the induced fermion current to zero, $\bm{J} = 0$.
In this case, we are able to perform a Fourier transformation and analyse the equations of motion for the gauge-field modes $A_\lambda\left(\tau,\bm{k}\right)$,
\begin{equation}
\label{eq:Amodes}
A_\lambda''\left(\tau,\bm{k}\right) + k^2\left(1 - 2\lambda\,\xi\left(\tau\right)\,\frac{a\left(\tau\right)H\left(\tau\right)}{k}\right) A_\lambda\left(\tau,\bm{k}\right) = 0 \,.
\end{equation}
Here, $\lambda = \pm$ and $\bm{k}$ are the helicity and comoving momentum eigenvalues of the mode function $A_\lambda\left(\tau,\bm{k}\right)$, respectively; the absolute value of $\bm{k}$ is denoted by $k$; $a$ is the FLRW scale factor; and $H$ denotes the physical Hubble rate, $H = a'/a^2$. 
We moreover introduced the gauge-field production or instability parameter
\begin{equation}
\xi\left(\tau\right) = \frac{1}{a\left(\tau\right)H\left(\tau\right)} \frac{\alpha_Y}{2\pi} \frac{\left|\phi'\left(\tau\right)\right|}{f_\phi} \,,
\end{equation}
which typically obtains values of $\mathcal{O}\left(1\cdots10\right)$ in standard scenarios of axion inflation.
We note that, in writing down Eq.~\eqref{eq:Amodes}, we assumed a negative inflaton velocity, $\phi' < 0$, which, as we will see, is going to ensure that the final baryon asymmetry will have the correct, positive sign.%
\footnote{Alternatively, we could have also chosen a positive inflaton velocity at the cost of flipping the sign in Eq.~\eqref{eq:phiYY}.}
The important message of Eq.~\eqref{eq:Amodes} is that, given our sign conventions, the positive-helicity modes $A_+\left(\tau,\bm{k}\right)$ will become tachyonically unstable as soon as their physical momenta, $k/a$, have been redshifted to the critical value $2\xi H$.
That is, a few $e$-folds before the positive-helicity modes exit the Hubble horizon at $k/a = H$, they will begin to grow exponentially.
The negative-helicity modes $A_-\left(\tau,\bm{k}\right)$, on the other hand, will remain in the vacuum state, such that the hypermagnetic field generated during axion inflation ends up being maximally helical.


\subsection{Hypermagnetic helicity}


The helicity stored in the hypermagnetic field quantifies the spontaneous breaking of $CP$ invariance during axion inflation and therefore plays a central role in setting the size of the BAU.
We define the physical hypermagnetic helicity density $h_Y$ in terms of an average over spatial hypersurfaces in the FLRW spacetime,
\begin{equation}
\label{eq:hY}
h_Y = \frac{1}{\mathbb{V}} \int d^3\bm{x}\:\left<\epsilon^{0ijk} A_i \partial_j A_k\right>  = \frac{1}{\mathbb{V}} \int d^3\bm{x}\:\left<\bm{A}\cdot\bm{\nabla}\times\bm{A}\right> \,.
\end{equation}
Here, $\mathbb{V}$ denotes the volume of spatial hypersurfaces in FLRW coordinates, $\mathbb{V} = a^3 \int d^3\bm{x}$, and the brackets indicate the quantum vacuum expectation value during inflation.
An important property of $h_Y$ is that its derivative with respect to conformal time $\tau$ is given by the Chern--Pontryagin density $Y_{\mu\nu}\widetilde{Y}^{\mu\nu} = -4 \,\bm{E}\cdot\bm{B}$, 
\begin{equation}
\label{eq:dhdtau}
\frac{\partial}{\partial\tau}\left(\mathbb{V} h_Y\right) = \frac{1}{2}\int d^3\bm{x}\: \left<Y_{\mu\nu}\widetilde{Y}^{\mu\nu}\right> = -2 \int d^3\bm{x}\: \left<\bm{E}\cdot\bm{B}\right> \,,
\end{equation}
where $\bm{E}$ and $\bm{B}$ denote the comoving hyperelectric and hypermagnetic fields, respectively,
\begin{equation}
\bm{E}\left(\tau,\bm{x}\right) = - \frac{\partial}{\partial \tau} \bm{A}\left(\tau,\bm{x}\right) \,, \qquad \bm{B}\left(\tau,\bm{x}\right) =  \bm{\nabla}\times \bm{A}\left(\tau,\bm{x}\right) \,.
\end{equation}


For the purposes of baryogenesis, we are interested in the value of the hypermagnetic helicity at the end of primordial hypermagnetogenesis, which coincides with the end of inflation in our scenario,
\begin{equation}
\label{eq:hYint}
h_Y^{\rm end}  = -\frac{2}{\mathbb{V}} \int_{-\infty}^{\tau_{\rm end}} d\tau \int d^3\bm{x}\:\left<\bm{E}\cdot\bm{B}\right> \,.
\end{equation}
Note that $h_Y^{\rm end}$ is by construction gauge-independent.
Here and in the following, the label ``end'' will refer to the end of hypermagnetogenesis.
In order to compute the time integral in Eq.~\eqref{eq:hYint}, we make use of the fact that $\left<\bm{E}\cdot\bm{B}\right>$ is expected to reach an attractor solution, if the gauge-field production parameter $\xi$ and the Hubble rate $H$ do not vary too fast during inflation.%
\footnote{For strong axion--vector coupling, the backreaction of the gauge fields onto the axion dynamics can lead to oscillations in the axion velocity and consequently in the instability parameter $\xi$~\cite{Domcke:2020zez,Gorbar:2021rlt,Peloso:2022ovc}.
These oscillations are, however, damped once fermion production is included, which limits the efficiency of gauge-field production and hence reduces the effect on the axion motion~\cite{Gorbar:2021rlt}.}
In fact, for constant values of $\xi$ and $H$, one can show that the time dependence of $\left<\bm{E}\cdot\bm{B}\right>$ becomes trivial, $\left<\bm{E}\cdot\bm{B}\right> \propto a^4$ (see, \textit{e.g.}, Ref.~\cite{Anber:2009ua} for an explicit derivation).
This indicates that, as expected for a stationary attractor solution, the product of the electric and the magnetic field remains constant in the physical frame.
The attractor solution for $\left<\bm{E}\cdot\bm{B}\right>$ is moreover homogeneous and isotropic across the volume $\mathbb{V} \gg H^{-3}$, which allows us to drop the spatial integral in Eq.~\eqref{eq:hYint},
\begin{equation}
\label{eq:hYint2}
h_Y^{\rm end} = -\frac{2}{a_{\rm end}^3} \int_{-\infty}^{\tau_{\rm end}} d\tau\:\left<\bm{E}\cdot\bm{B}\right> \,.
\end{equation}
Because of the aforementioned scaling of the integrand, $\left<\bm{E}\cdot\bm{B}\right> \propto a^4$, we expect the time integral to be dominated by the contributions at late times close to the end of inflation.
This allows us to roughly estimate
\begin{equation}
h_Y^{\rm end} \sim -\left.\frac{2\left<\bm{E}\cdot\bm{B}\right>}{a^7}\right|_{\rm end} \times \int_{-\infty}^{\tau_{\rm end}} d\tau \: a^4\left(\tau\right) \,.
\end{equation}
Up to slow-roll corrections, the scale factor is given by the de Sitter expression $a\left(\tau\right) = - 1 / \left(\tau H\right)$, such that
\begin{equation}
\label{eq:hYresult}
h_Y^{\rm end} \sim - \left.\frac{2\left<\bm{E}\cdot\bm{B}\right>}{3a^4H} \right|_{\rm end} \,.
\end{equation}
This quantity will be an essential input for our discussion of the BAU in the remainder of this paper. 
As we will see, it specifically sets the scale for the fermionic charge asymmetries generated during inflation.
However, before we turn to these asymmetries, let us take a step back and discuss how the result in Eq.~\eqref{eq:hYresult} may be generalized to other cosmological scenarios that result in the generation of hypermagnetic helicity.


The outcome of primordial hypermagnetogenesis during axion inflation can be quantified in terms of four parameters and a sign: (i) the amplitude of the comoving hyperelectric field, $E_{\rm end}$; (ii) the amplitude of the comoving hypermagnetic field, $B_{\rm end}$; (iii) the sign of $\left<\bm{E}\cdot\bm{B}\right>$, \textit{i.e.}, the relative orientation of the comoving vectors $\bm{E}$ and $\bm{B}$; (iv) the comoving correlation length of the hypercharge gauge field, $\lambda_{\rm end}$; and (v) the physical Hubble rate, $H_{\rm end}$. 
Here, all quantities are understood to be evaluated at the end of hypermagnetogenesis.
The comoving correlation length is typically determined by the Hubble length, such that $a_{\rm end} \lambda_{\rm end} = c_\lambda H_{\rm end}^{-1}$.
In the case of axion inflation, the numerical coefficient $c_\lambda$ turns out to be of $\mathcal{O}\left(1\cdots10\right)$, indicating that the hypercharge gauge field is correlated over superhorizon distances.%
\footnote{On even larger scales, it is, however, statistically homogeneous and isotropic, which justifies the step from Eq.~\eqref{eq:hYint} to Eq.~\eqref{eq:hYint2}.}
In hypermagnetogenesis scenarios after inflation, on the other hand, $c_\lambda$ is constrained by causality to be at most $c_\lambda \sim 1$. 
Based on Eq.~\eqref{eq:hY} and making use of the coefficient $c_\lambda$, we thus expect the following rough relation to hold in general models,
\begin{equation}
\label{eq:clambda}
h_Y^{\rm end} \sim - \left.\textrm{sgn}\left(\left<\bm{E}\cdot\bm{B}\right>\right) \frac{\lambda B^2}{a^3}\right|_{\rm end} = - \left.\textrm{sgn}\left(\left<\bm{E}\cdot\bm{B}\right>\right) \frac{c_\lambda B^2}{a^4 H}\right|_{\rm end} \,,
\end{equation}
where the first relation simply follows from the general expectation that $\bm{A}\cdot\bm{B} \sim AB$ with $A \sim \lambda B$.
Alternatively, we can describe the final gauge-field configuration at the end of hypermagnetogenesis in terms of a typical length scale $c_\lambda H_{\rm end}^{-1}$ and a typical time scale $c_\tau H_{\rm end}^{-1}$.
For a typical amplitude of the vector potential, $A_{\rm end}$, we then expect $E_{\rm end} \sim A_{\rm end} a_{\rm end}H_{\rm end}/c_\tau$ and $B_{\rm end} \sim A_{\rm end} a_{\rm end}H_{\rm end}/c_\lambda$, such that $c_\tau E_{\rm end} \sim c_\lambda B_{\rm end}$ and
\begin{equation}
\label{eq:ctau}
h_Y^{\rm end} \sim - \left.\textrm{sgn}\left(\left<\bm{E}\cdot\bm{B}\right>\right) \frac{c_\tau E B}{a^4 H}\right|_{\rm end} \,.
\end{equation}
If we further assume that the gauge field is generated in a maximally helical state, such that $\bm{E}$ and $\bm{B}$ are either parallel (maximal negative helicity) or antiparallel (maximal positive helicity) to each other, we obtain
\begin{equation}
h_Y^{\rm end} \sim - \left.\frac{c_\tau\left<\bm{E}\cdot\bm{B}\right>}{a^4H} \right|_{\rm end} \,.
\end{equation}
The form of this expression matches the form of our result in Eq.~\eqref{eq:hYresult}.
We are therefore able to identity $c_\tau \sim 2/3$ in the case of axion inflation. 
Other models may be characterized by a different value of $c_\tau$.
For hypermagnetogenesis scenarios operating after inflation, we expect in general that $c_\lambda \lesssim c_\tau \lesssim 1$.


\subsection{Fermionic charge asymmetries}


The strong gauge-field background during axion inflation leads to the nonperturbative production of SM fermions, which receives $CP$-symmetric as well as $CP$-asymmetric contributions~\cite{Domcke:2018eki}.
While the former corresponds to the ordinary Schwinger effect in an inflationary background (\textit{i.e.}, Schwinger pair production), the later results in charge asymmetries whose magnitude is dictated by the SM chiral anomaly,
\begin{equation}
\label{eq:anomaly}
\partial_\mu J_i^\mu = - \varepsilon_i g_i Y_i^2\,\frac{\alpha_Y}{4\pi}\,Y_{\mu\nu}\widetilde{Y}^{\mu\nu} + \cdots \,,
\end{equation}
with $J_i^\mu$ denoting the comoving current of the $i$th SM chiral fermion species.
The prefactors $\varepsilon_i$, $g_i$, and $Y_i$ are explained and listed in Tab.~\ref{tab:factors}.
The ellipsis represents all other SM Yukawa and sphaleron processes that can impact the evolution of the fermion currents.
During inflation, these processes are inefficient, which allows us to neglect them in the computation of the fermion charges generated by the hypercharge gauge field.%
\footnote{A possible exception to this is the top-quark Yukawa coupling, which may not be fully negligible during inflation.
If efficient, this interaction could (i) reshuffle the fermionic chemical potentials, (ii) modify the top-quark contribution to the induced current, and (iii) contribute to the Higgs potential~\cite{Hook:2019vcn}.
For the purposes of this work, the first point is irrelevant, as the interactions in the thermal plasma after inflation will in any case lead to a $CP$-conserving reshuffling of the chemical potentials.
The second point will induce at the very most an error of about $16\,\%$; an estimate that we obtain if we completely drop all contributions of the third quark generation from the induced current.
For simplicity, we will also discard the third point, if necessary assuming an additional Hubble-induced contribution to the Higgs potential that stabilizes the vacuum expectation value of the Higgs field at the origin.}


The charge densities that we are interested in describe the differences of fermion and antifermion number densities, $q_i = n_i - \bar{n}_i$, and are closely related to the chemical potentials $\mu_i$ of the corresponding fermion species in the thermal plasma after inflation, $q_i = g_i \mu_i T^2/6$.
Here, $T$ denotes the temperature of the SM plasma; and the quantities $q_i$, $n_i$, $\bar{n}_i$, $\mu_i$, and $T$ are all physical.
We moreover assume that the chemical potentials always remain small at all times after inflation, $\mu_i \ll T$.
Formally, the physical charge densities $q_i$ are defined in terms of the spatial average of the temporal components of the corresponding currents,
\begin{equation}
\label{eq:qi}
q_i = \frac{1}{\mathbb{V}} \int d^3\bm{x}\:\left<J_i^0\right> \,.
\end{equation}
In order to compute these charge densities, we first note that the right-hand side of Eq.~\eqref{eq:anomaly} is proportional to the divergence of the comoving \textit{Chern--Simons} (CS) current of the hypercharge gauge field,
\begin{equation}
J_{\rm CS}^\mu = \frac{\alpha_Y}{\pi}\,\epsilon^{\mu\nu\rho\sigma}A_\nu\,\partial_\rho A_\sigma \,, \qquad \partial_\mu J_{\rm CS}^\mu = \frac{\alpha_Y}{2\pi}\,Y_{\mu\nu}\widetilde{Y}^{\mu\nu} \,.
\end{equation}
In the homogeneous and isotropic background during inflation, the average of $\bm{J}_{\rm CS}$ over the volume $\mathbb{V}$ vanishes, while the average of the temporal component is nonzero and related to the physical CS charge density,
\begin{equation}
q_{\rm CS} = \frac{1}{\mathbb{V}} \int d^3\bm{x} \: \left<J_{\rm CS}^0\right> = \frac{\alpha_Y}{\pi} \frac{1}{\mathbb{V}}\int d^3\bm{x}\:\left<\epsilon^{0ijk} A_i \partial_j A_k\right> \,.
\end{equation}
This relation allows us to immediately identify $q_{\rm CS}$ with the hypermagnetic helicity density in Eq.~\eqref{eq:hY},
\begin{equation}
\label{eq:qcshy}
q_{\rm CS} = \frac{\alpha_Y}{\pi}\,h_Y = \frac{g_Y^2}{4\pi^2}\,h_Y \,.
\end{equation}


\begin{table}
\caption{Numerical factors appearing in Eq.~\eqref{eq:anomaly}: The index $i$ labels the 15 SM fermion representations; $\varepsilon_i$ distinguishes between left- and right-handed fermions; $g_i$ counts internal gauge degrees of freedom; and $Y_i$ stands for the SM hypercharges.}
\label{tab:factors}
\centering
\begin{tabular}{|c||rrrrrrrrrrrrrrr|}
\hline
$i$ & $e$ & $\mu$ & $\tau$ & $\ell_e$ & $\ell_\mu$ & $\ell_\tau$ & $u$ & $c$ & $t$ & $d$ & $s$ & $b$ & $Q_1$ & $Q_2$ & $Q_3$ \\
\hline\hline
$\varepsilon_i$ & $+1$ & $+1$ & $+1$ & $-1$ & $-1$ & $-1$ & $+1$ & $+1$ & $+1$ & $+1$ & $+1$ & $+1$ & $-1$ & $-1$ & $-1$ \\
$g_i$ & $1$ & $1$ & $1$ & $2$ & $2$ & $2$ & $3$ & $3$ & $3$ & $3$ & $3$ & $3$ & $6$ & $6$ & $6$ \\
$Y_i$ & $-1$ & $-1$ & $-1$ & $-\sfrac{1}{2}$ & $-\sfrac{1}{2}$ & $-\sfrac{1}{2}$ & $+\sfrac{2}{3}$ & $+\sfrac{2}{3}$ & $+\sfrac{2}{3}$ & $-\sfrac{1}{3}$ & $-\sfrac{1}{3}$ & $-\sfrac{1}{3}$ & $+\sfrac{1}{6}$ & $+\sfrac{1}{6}$ & $+\sfrac{1}{6}$ \\
\hline
\end{tabular}
\end{table}


Next, we consider the quantum expectation value of Eq.~\eqref{eq:anomaly} and integrate over $\bm{x}$ on both sides.
The integral over the spatial divergence of the current vanishes because of homogeneity and isotropy, such that
\begin{equation}
\partial_\tau \int d^3\bm{x}\:\left<J_i^0\right> = - \varepsilon_i g_i Y_i^2\,\frac{\alpha_Y}{4\pi}\int d^3\bm{x}\:\left<Y_{\mu\nu}\widetilde{Y}^{\mu\nu}\right> \,.
\end{equation}
Making use of Eqs.~\eqref{eq:dhdtau}, \eqref{eq:qi}, and \eqref{eq:qcshy}, this relation results in the following conservation law,
\begin{equation}
\label{eq:qihlaw}
\partial_\tau\left(\mathbb{V}q_i\right) = - \varepsilon_i g_i Y_i^2\,\frac{\alpha_Y}{2\pi}\,\partial_\tau\left(\mathbb{V}h_Y\right) = - \varepsilon_i g_i Y_i^2\,\frac{1}{2}\,\partial_\tau\left(\mathbb{V}q_{\rm CS}\right) \,.
\end{equation}
For $CP$-symmetric conditions, such that $q_i = q_{\rm CS} = 0$ at early times during inflation, we therefore obtain
\begin{equation}
\label{eq:qiend}
q_i^{\rm end} = - \varepsilon_i g_i Y_i^2\,\frac{\alpha_Y}{2\pi}\,h_Y^{\rm end} = - \varepsilon_i g_i Y_i^2\,\frac{1}{2}\,q_{\rm CS}^{\rm end} \,,
\end{equation}
which explicitly illustrates how $h_Y^{\rm end} = \pi/\alpha_Y\,q_{\rm CS}^{\rm end}$ controls the fermion charges at the end of inflation.
We stress that Eq.~\eqref{eq:qiend} is a direct consequence of the anomaly equation in Eq.~\eqref{eq:anomaly}.
Alternative scenarios of hypermagnetogenesis, not necessarily related to axion inflation, will lead to similar relations.%
\footnote{The precise outcome will then depend on the set of equilibrated SM interactions at the time of hypermagnetogenesis.
For instance, if all SM interactions are equilibrated during hypermagnetogenesis, these will strongly suppress fermion charge generation.}


In this paper, we will work in the limit of instantaneous reheating. 
This facilitates our analysis; a more rigorous treatment would require a lattice simulation~\cite{Cuissa:2018oiw,Adshead:2019lbr,Adshead:2019igv} (see also Ref.~\cite{Caravano:2021bfn,Caravano:2022epk}).
It is moreover justified by the fact that reheating proceeds very fast after axion inflation, if a large amount of energy is already transferred to the hypercharge sector towards the end of inflation, which then quickly thermalizes in consequence of the SM gauge interactions.
In this case, we are able to employ the fermion symmetries in Eq.~\eqref{eq:qiend} as initial conditions for our description of the radiation-dominated era.
Similarly, the temperature of the SM plasma shortly after the end of inflation coincides with the reheating temperature in the limit of instantaneous reheating,
\begin{equation}
\label{eq:Treh}
T_{\rm end} \simeq T_{\rm reh} \simeq \left(\frac{90}{\pi^2 g_*}\right)^{1/4} \sqrt{H_{\rm end} M_{\rm Pl}} \,.
\end{equation}
Here, $g_* = 427/4$ denotes the effective number of relativistic degrees of freedom contributing to the energy density of the SM thermal bath and $M_{\rm Pl} \simeq 2.435 \times 10^{18}\,\textrm{GeV}$ is the reduced Planck mass.
We thus have
\begin{equation}
q_i^{\rm end} = \frac{g_i}{6}T_{\rm end}^3 \times \left.\frac{\mu_i}{T}\right|_{\rm end}  \,, \qquad  \left.\frac{\mu_i}{T}\right|_{\rm end} = - 6\,\varepsilon_i Y_i^2 \chi \,,
\end{equation}
where the dimensionless quantity $\chi$ sets the scale for the ratios $\mu_i/T$ at the onset of the radiation era,
\begin{equation}
\label{eq:chi}
\chi = \left.\frac{q_{\rm CS}}{2T^3}\right|_{\rm end} \,.
\end{equation}
The yield variable $\chi$ is directly proportional to the hypermagnetic helicity density $h_Y^{\rm end}$ [see Eq.~\eqref{eq:qcshy}] and hence the $CP$-violating expectation value $\left<\bm{E}\cdot\bm{B}\right>$ at the end of inflation [see Eq.~\eqref{eq:hYresult}].
Together with $H_{\rm end}$ and the coefficients $c_\lambda$ and $c_{\tau}$ [see Eqs.~\eqref{eq:clambda} and \eqref{eq:ctau}], the magnitude and sign of $\chi$ fully characterize the output of hypermagnetogenesis as far as the generation of the BAU at lower temperatures is concerned.
Our discussion of baryogenesis in Secs.~\ref{subsec:washin} and \ref{subsec:decay} is therefore also going to apply to alternative models of hypermagnetogenesis that allow one to calculate the expected values of these parameters at high temperatures.


\subsection{Efficiency of hypermagnetogenesis}
\label{subsec:efficiency}


Up to this point, our discussion has led to the result that both the hypermagnetic helicity and the fermion charges generated during inflation are controlled by the quantity $\left<\bm{E}\cdot\bm{B}\right>$ [see Eqs.~\eqref{eq:hYresult} and \eqref{eq:qiend}],
\begin{equation}
\label{eq:hYqi}
h_Y \sim - \left.\frac{2\left<\bm{E}\cdot\bm{B}\right>}{3a^4H} \right|_{\rm end} \,, \qquad q_i^{\rm end} \sim \left.\varepsilon_i g_i Y_i^2\,\frac{\alpha_Y}{3\pi}\frac{\left<\bm{E}\cdot\bm{B}\right>}{a^4H} \right|_{\rm end} \,.
\end{equation}
Therefore, in order to make quantitative progress, it is necessary to track the evolution of $\left<\bm{E}\cdot\bm{B}\right>$ during axion inflation all the way to its end.
This, however, represents a technical challenge because of the highly nonlinear interplay of fermion and gauge-field production.
In the following, we will therefore discuss several different estimates for the efficiency of gauge-field production during axion inflation that have recently been put forward in the literature.
A more detailed discussion of these estimates, including semianalytical fit functions that describe the exact numerical results with excellent accuracy, can be found in Ref.~\cite{Gorbar:2021zlr}.


Recall that in our discussion of the mode equations for the hypercharge gauge field, we set the induced hyperelectric current to zero [see Eq.~\eqref{eq:Amodes}]. 
In order to properly account for the nonlinear backreaction of the fermion current on the efficiency of gauge-field production, we now need to revert this step and work with the explicit expression for the current $\bm{J}$.
In the case of our interest, where the hypercharge gauge field is strongly amplified, $\left<\bm{E}^2\right> \sim \left<\bm{B}^2\right> \gg H^4$, the typical length scale of fermion production, $\left<\bm{E}^2\right>^{-1/4}$, is much shorter than the correlation length of the hypercharge gauge field, $H^{-1}$. 
Hence, one can approximate the hyperelectric and hypermagnetic fields as homogeneous and (anti) parallel, whose representative amplitudes are approximately given by $E = \left<\bm{E}^2\right>^{1/2}$ and $B = \left<\bm{B}^2\right>^{1/2}$, respectively.
Furthermore, we expect the hypercharge gauge field to reach a stationary configuration, where the tachyonic instability and the induced current balance each other, which implies that the physical amplitudes $E/a^2$ and $B/a^2$ become almost time-independent.
Under these approximations, the comoving fermion current was derived in Ref.~\cite{Domcke:2018eki} as
\begin{equation}
\label{eq:JEB}
\bm{J} = \frac{J}{E} \,\bm{E} \,,\qquad g_Y J = \frac{41g_Y^3}{72\pi^2} \frac{EB}{a H} \coth\left(\pi\frac{B}{E}\right) \,.
\end{equation}


If we now want to use Eq.~\eqref{eq:JEB} in the equation of motion for the vector field, we can choose between two possible strategies.
In the first case, which we will call the \textit{magnetic picture}, we identity the overall $B$ factor in Eq.~\eqref{eq:JEB} as the relevant dynamical quantity and treat all other factors of $E$, $B$, and $H$ in Eq.~\eqref{eq:JEB} as background quantities.
That is, we treat the current as a vector that is primarily controlled by the magnetic field, up to a proportionality factor that only depends on background quantities, $\bm{J} = - J/B\,\bm{B}$.
Given that the electric and magnetic fields generated during inflation are antiparallel to each other in our convention, $\bm{E}/E = - \bm{B}/B$ such that $EB = -\left<\bm{E}\cdot\bm{B}\right> $, this is a viable possibility.
As a consequence, Eq.~\eqref{eq:AEOM} turns into
\begin{equation}
\label{eq:AEOMm}
\bm{A}'' - \bm{\nabla}^2\bm{A} = 2aH\left[\xi - \frac{41g_Y^3}{144\pi^2}\frac{E}{a^2H^2} \coth\left(\pi\frac{B}{E}\right)\right]\,\bm{\nabla} \times \bm{A} \,.
\end{equation}
In the magnetic picture, the gauge-field production parameter $\xi$ thus receives a correction that depends on the strength of the induced fermion current.
This observation is the starting point for our first estimate of the efficiency of gauge-field production, which has been proposed for the first time in Ref.~\cite{Domcke:2018eki}.
Besides, there is also the \textit{electric picture}, developed in Refs.~\cite{Gorbar:2021rlt,Gorbar:2021zlr} and discussed in more detail below, in which the fermion current is considered to be primarily controlled by the electric field, $\bm{J} = J/E \bm{E}$.
Consequently, instead of a correction to the instability parameter $\xi$, the fermion current induces a finite generalized conductivity.

\begin{figure}[t]
\begin{center}
\includegraphics[width=0.67\textwidth]{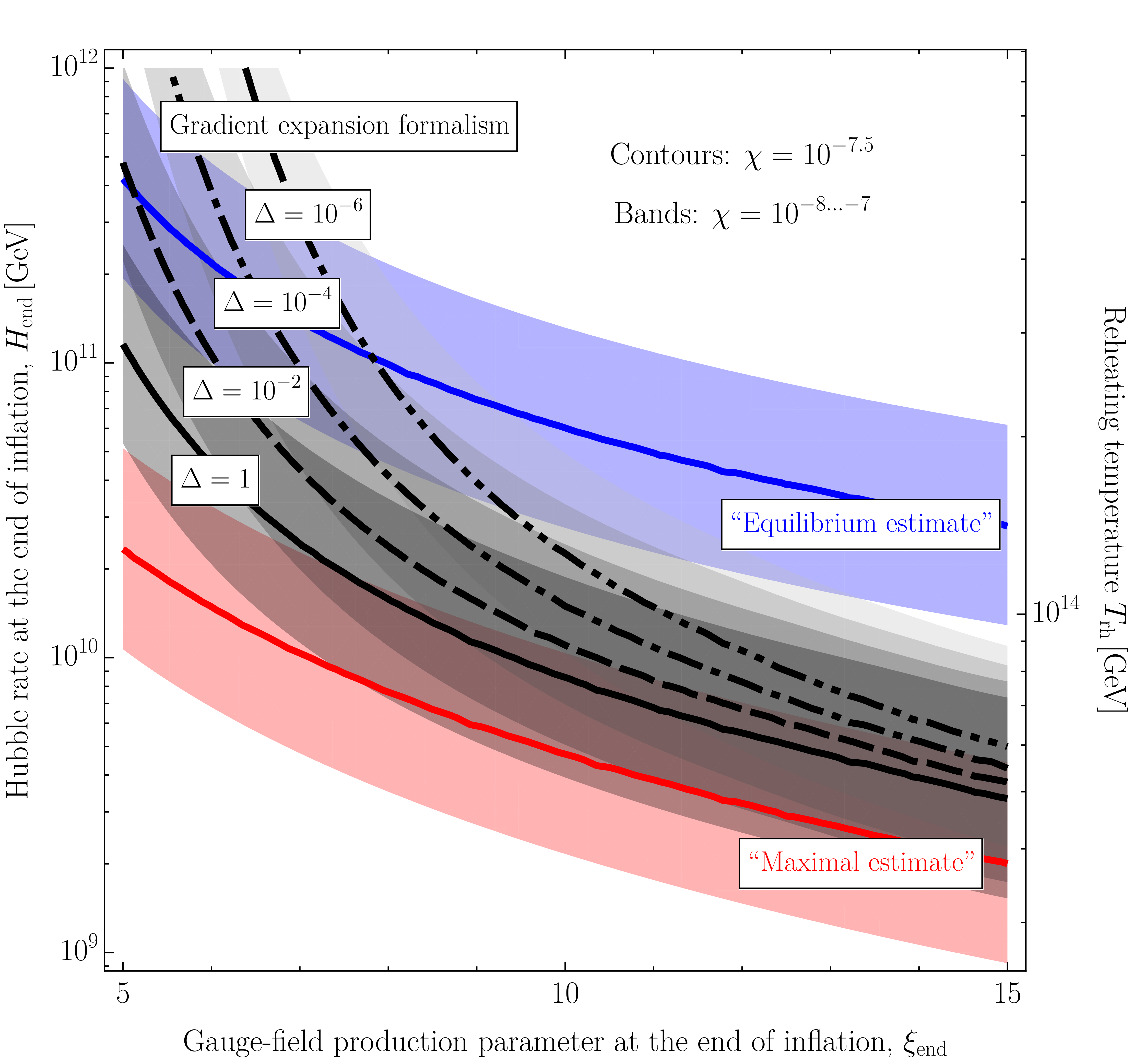}
\caption{Estimates of the parameter region where the yield parameter $\chi$ [see Eqs.~\eqref{eq:chi} and \eqref{eq:chisim}] is of the right order of magnitude for successful baryogenesis (see Secs.~\ref{subsec:washin} and \ref{subsec:decay}).
The contour lines indicate where in parameter space we expect $\chi = 10^{-7.5}$, while the shaded bands cover the corresponding range from $\chi = 10^{-8}$ to $\chi = 10^{-7}$.
Here, larger values of $H_{\rm end}$ correspond to larger values of $\chi$, according to the scaling law $\chi \propto H_{\rm end}^{3/2}$ for fixed $\xi$ [see Eqs.~\eqref{eq:Treh} and \eqref{eq:chisim}].
We compare the equilibrium estimate (blue) and maximal estimate (red) in the magnetic picture~\cite{Domcke:2018eki} to the GEF estimate (black) in the electric picture~\cite{Gorbar:2021zlr} for different values of the damping factor $\Delta$ [see Eq.~\eqref{eq:delta}].
The reheating temperature $T_{\rm rh}$ follows from the Hubble rate $H_{\rm end}$ according to Eq.~\eqref{eq:Treh}.}
\label{fig:chi}
\end{center} 
\end{figure}

Staying in the magnetic picture for the moment, we define
\begin{equation}
\label{eq:xieff}
\xi_{\rm eff} = \xi - \frac{41g_Y^3}{144\pi^2}\frac{E}{a^2H^2} \coth\left(\pi\frac{B}{E}\right) 
\end{equation}
and construct an approximate solution for $\left<\bm{E}\cdot\bm{B}\right>$ based on this effective gauge-field production parameter.
The main idea behind this construction, which we will refer to as the \textit{equilibrium estimate}, is that, after a sufficiently long time, the system reaches an equilibrium attractor, in which the electric and magnetic fields remain constant in the physical frame, $E/a^2 = \textrm{const}$, $B/a^2 = \textrm{const}$.
In this case, also $\xi_{\rm eff}$ remains constant, such that the mode equations in Eq.~\eqref{eq:Amodes} obtain the same form as in the absence of fermion production, the only difference being that $\xi$ is replaced by $\xi_{\rm eff}$.
The equilibrium estimate therefore assumes that the dependence of $\left<\bm{E}^2\right>$, $\left<\bm{B}^2\right>$, and $\left<\bm{E}\cdot\bm{B}\right>$ on $\xi_{\rm eff}$ is the same as the dependence on $\xi$ in the absence of fermions~\cite{Jimenez:2017cdr},
\begin{equation}
\label{eq:EBeq}
\frac{\left<\bm{E}^2\right>}{a^4} \simeq 2.6 \times 10^{-4}\,\frac{e^{2\pi\xi_{\rm eff}}}{\xi_{\rm eff}^3} \,, \qquad \frac{\left<\bm{B}^2\right>}{a^4} \simeq 3.0 \times 10^{-4}\,\frac{e^{2\pi\xi_{\rm eff}}}{\xi_{\rm eff}^5} \,, \qquad \frac{\left<\bm{E}\cdot\bm{B}\right>}{a^4} \simeq -2.8 \times 10^{-4}\,\frac{e^{2\pi\xi_{\rm eff}}}{\xi_{\rm eff}^4} \,. 
\end{equation}
These relations, together with the expression for $\xi_{\rm eff}$ in Eq.~\eqref{eq:xieff}, provide an implicit definition of the amplitudes of the electric and magnetic fields according to the equilibrium estimate.
For given values of $\xi$ and $H$, this set of equations can be solved numerically in order to obtain an estimate for $\left<\bm{E}\cdot\bm{B}\right>$.
The outcome of this exercise is shown in Fig.~\ref{fig:chi}, which illustrates the dependence of the equilibrium estimate for the dimensionless yield parameter $\chi$ on the gauge-field production parameter $\xi$ and Hubble rate $H$, where we used that
\begin{equation}
\label{eq:chisim}
\chi \sim - \left.\frac{\alpha_Y}{3\pi}\frac{\left<\bm{E}\cdot\bm{B}\right>}{a^4HT^3} \right|_{\rm end} \,.
\end{equation}
Our numerical results shown in Fig.~\ref{fig:chi} also take into account the running of the SM hypercharge gauge coupling constant $g_Y$ in a self-consistent way, such that $g_Y$ in Eq.~\eqref{eq:xieff} is always evaluated at the appropriate renormalization scale characterizing the energy content of the gauge field, $\mu = \left(E^2/2 + B^2/2\right)^{1/4}/a$.


In addition, we include in Fig.~\ref{fig:chi} a second estimate that can be derived in the magnetic picture and which we will refer to as the \textit{maximal estimate}~\cite{Domcke:2018eki}.
This estimate is based on the evolution equation for the energy density of the electromagnetic field, which contains an additional source term in the presence of the axion,
\begin{equation}
\left(\frac{1}{a}\partial_\tau + 4H\right)\frac{E^2 + B^2}{2a^4} = 2H\,\frac{\xi_{\rm eff}\,EB}{a^4} \,.
\end{equation}
Again assuming that the system will settle in a stationary attractor solution, we are able to drop the time derivative in this equation, which results in an algebraic relation for the field amplitudes $E$ and $B$,
\begin{equation}
E^2 + B^2 = \xi_{\rm eff}\,EB \,.
\end{equation}
While this relation must be obeyed by \textit{every} attractor solution in the magnetic picture, the maximal estimate for $\left<\bm{E}\cdot\bm{B}\right>$ corresponds to the $E$ and $B$ values that satisfy this relation while maximizing the product $EB$.
In this sense, the maximal estimate should not be regarded as a realistic proposal for an explicit solution; it rather presents an upper bound on all possible solutions for $\left<\bm{E}\cdot\bm{B}\right>$ in the magnetic picture.
We therefore show the maximal estimate in Fig.~\ref{fig:chi} only as a reference that is supposed to be compared to the equilibrium estimate.
The discrepancy between these two estimates roughly characterizes the size of the theoretical uncertainty when one attempts to estimate $\left<\bm{E}\cdot\bm{B}\right>$ solely based on the assumption of a stationary attractor solution.


A third estimate for $\left<\bm{E}\cdot\bm{B}\right>$, in this case based on the electric picture, has recently been presented in Ref.~\cite{Gorbar:2021zlr}.
This estimate is based on the \textit{gradient expansion formalism} (GEF) developed in Ref.~\cite{Gorbar:2021rlt}, which describes the evolution of all relevant background quantities during axion inflation, $\left<\bm{E}^2\right>$, $\left<\bm{B}^2\right>$, $\left<\bm{E}\cdot\bm{B}\right>$, etc., in terms of bilinear scalar functions that are constructed from the vector fields $\bm{E}$ and $\bm{B}$ in position space (see also Ref.~\cite{Sobol:2019xls} for related earlier work).
If one explicitly specifies the scalar potential $V$ and initial conditions, this approach allows one to explicitly track the dynamics of axion inflation, even in the presence of nonlinear backreaction and nonperturbative fermion production, with unprecedented accuracy.
However, for the purposes of the present paper, we are rather interested in more general, model-independent predictions. 
We will therefore not use the results of Ref.~\cite{Gorbar:2021rlt} and implement the gradient expansion formalism for a specific model.
Instead, we will resort to the model-independent results derived in Ref.~\cite{Gorbar:2021zlr}, which are based on the assumption that $\xi$ and $H$ only vary slowly during inflation.
Towards the end of inflation, which is the point in time we are most interested in, this assumption becomes violated.
Still, the analysis in Ref.~\cite{Gorbar:2021zlr} was able to show that the model-independent results continue to represent a good approximation, typically within less than one order of magnitude of the exact results and in better agreement with the exact results than the equilibrium and maximal estimates.


The GEF estimate builds upon the numerical solution of a dynamical system of equations, while the equilibrium and maximal estimates follow from simple algebraic arguments. 
We therefore expect that the GEF estimate is also suitable for more dynamical situations, while the equilibrium and maximal estimates only become relevant after a sufficiently long equilibration time, such that all quantities of interest have reached their time-independent attractor values.
Since the results in Refs.~\cite{Gorbar:2021rlt,Gorbar:2021zlr} were derived  in the electric picture, the fermion current in Eq.~\eqref{eq:AEOM} is treated  
as a vector that is primarily controlled by the electric field, $\bm{J} = J/E\,\bm{E}$.
Here, the ratio $g_YJ/E$ plays the role of a generalized comoving conductivity for the fermion gas generated during axion inflation, 
\begin{equation}
\sigma = \frac{g_YJ}{E} = \frac{41g_Y^3}{72\pi^2} \frac{B}{a H} \coth\left(\pi\frac{B}{E}\right) \,.
\end{equation}
In the electric picture, the equation of motion for the hypercharge vector field thus reads
\begin{equation}
\label{eq:AEOMe}
\bm{A}'' - \bm{\nabla}^2\bm{A} = 2aH\xi\,\bm{\nabla} \times \bm{A} - \sigma \bm{A}' \,,
\end{equation}
where the new $\sigma \bm{A}'$ term now describes the damping of the vector field in the conducting medium. 


An important observation in Refs.~\cite{Gorbar:2021rlt,Gorbar:2021zlr} was that the damping term in Eq.~\eqref{eq:AEOMe} renders the description of gauge-field production during axion inflation inherently nonlocal in time.
Gauge-field modes inside the Hubble horizon experience an exponential damping on their approach to horizon crossing because they no longer evolve in an empty de Sitter vacuum but in a conducting medium.
The amount of exponential damping, however, depends on the conductivity of the fermion gas, which in turn is sensitive to the efficiency of gauge-field and fermion production at earlier times. 
This effect can be captured by a new damping factor%
\footnote{The lower integration boundary in Eq.~\eqref{eq:delta} implies that gauge-field modes on arbitrarily small scales inside the horizon will experience damping because of the nonvanishing conductivity.
This corresponds to a technical simplification that does not necessarily reflect the actual physical situation: 
Gauge-field modes on scales smaller than the typical momenta in the fermion gas are expected to experience less damping up to no damping at all.
The integral in Eq.~\eqref{eq:delta} is, however, typically dominated by the contributions at late times, \textit{i.e.}, contributions close to the upper integration boundary. 
Our simplified treatment of the lower integration boundary is therefore expected to have no or only little phenomenological implications.
See also Ref.~\cite{Gorbar:2021zlr} for a more comprehensive discussion.}
\begin{equation}
\label{eq:delta}
\Delta\left(\tau\right) = \exp\left[- \int_{-\infty}^\tau d\tau' \sigma\left(\tau'\right)\right] \,,
\end{equation}
which describes the amount by which gauge-field modes inside the horizon get damped up to some time $\tau$.
If one performs a model-specific analysis such as the one in Ref.~\cite{Gorbar:2021rlt}, the evolution of $\Delta$ is self-consistently accounted for by the gradient expansion formalism.
In this case, $\Delta$ does not correspond to new independent parameter.
However, if one is interested in model-independent results, the parameter $\Delta$ can be used as an effective parameter that captures the unknown prehistory leading up to a certain moment in time, \textit{e.g.}, the end of inflation.
This interpretation directly applies to our analysis.
In Fig.~\ref{fig:chi}, we therefore present the GEF estimate for $\chi$ not only as a function of $\xi$ and $H$, but also for four values of the damping factor $\Delta$.
In any given model, these parameters are of course correlated; in particular, small $\xi$ typically entails $\Delta \sim 1$.
Consequently, the region in the top left of Fig.~\ref{fig:chi}, where the GEF estimate crosses beyond the equilibrium estimate for small values of $\xi$ and $\Delta$, corresponds to situations that are likely hard to achieve dynamically in realistic models.
\footnote{The efficiency of gauge-field production during axion inflation and its dependence on $\Delta$ has recently also been investigated in Ref.~\cite{Cado:2022pxk}. 
The analysis in this paper is based on the electric picture, operates in Fourier space, and confirms that small $\Delta$ values at the end of inflation are correlated with a suppression of the final helicity density, especially, if the axion decay constant $f_\phi$ is large.}


The spread in our results for the GEF estimate illustrates the range of possible outcomes that one may expect for different models that do lead to the same values of $\xi$ and $H$ at the end of inflation, but which differ in the way in which they reach the end of inflation, specifically, in the way in which the conductivity $\sigma$ evolves prior to the end of inflation.
In addition to this model-related uncertainty, the GEF estimate features a systematic uncertainty stemming from the fact that it is not capable of accounting for the time dependence of the parameters $\xi$, $H$, $\Delta$. 
(In a full, time-resolved GEF run for a specific model, this is of course not a problem.)
This uncertainty is comparable to the spread in our results for different values of $\Delta$. 
To first approximation, one may therefore also completely neglect the $\Delta$ dependence of the GEF estimate and simply work with a fixed $\Delta$ value.
This is precisely what we will do in the remainder of this paper, in which we will focus on the GEF estimate for $\Delta = 1$.
As shown in Ref.~\cite{Gorbar:2021zlr}, this estimate is still capable of approximating the outcome of specific models at a level that is comparable to the full $\Delta$-dependent GEF estimate.


Finally, we caution that all estimates presented in this section need to be taken with a grain of salt.
The \textit{ad hoc} identification of the hyperelectric current with either the electric field times a proportionality factor, $\bm{J} = J/E\,\bm{E}$, or the magnetic field times a proportionality factor, $\bm{J} = - J/B\,\bm{B}$, does not stand on firm theoretical ground. 
These two approaches merely serve the purpose to make progress by deriving simple and rough estimates for the efficiency of gauge-field production during axion inflation.
In fact, it has recently been pointed out in Ref.~\cite{Fujita:2022fwc} that a self-consistent mean-field approximation automatically gives rise to both electric and magnetic conductivities.
In the two approaches that we just explained, it is assumed that a particular form of the induced current holds even at the level of perturbations, namely, $\delta \bm{J} = \sigma_B\,\delta \bm{B}$ [Eq.~\eqref{eq:AEOMm}] or $\delta \bm{J} = \sigma_E\,\delta \bm{E}$ [Eq.~\eqref{eq:AEOMe}].
However, strictly speaking, this is not fully accurate, since the directions of the perturbed electric and magnetic fields are not necessarily (anti) parallel at the onset of the tachyonic growth around $k/a \sim 2 \xi H$.
As a consequence, both conductivities, the electric conductivity $\sigma_E$ and the magnetic conductivity $\sigma_B$, appear in the perturbed equation of motion.
In Ref.~\cite{Fujita:2022fwc}, following the approach in Ref.~\cite{Domcke:2018eki}, a dynamical and self-consistent equilibrium solution for the strength of the electric and magnetic background fields is constructed that utilizes both $\sigma_E$ and $\sigma_B$. 
Similar to the GEF estimate, this solution for the electric and magnetic fields lies well above the equilibrium estimate, and relatively close to but below the maximal estimate in Ref.~\cite{Domcke:2018eki} (see Fig.~2 of Ref.~\cite{Fujita:2022fwc}).
This is one step forward to the complete picture.
Still, the analysis is limited to the stationary equilibrium and needs to be extended to cope with the time-dependent dynamics.


In future work, it will be necessary to validate (or revise) the estimates presented in this section making use of more sophisticated methods.
This means that, in the short term, it will be important to combine the mean-field approximation with the gradient expansion formalism, which should enable one to study the time-dependent dynamical evolution of the system in a fully self-consistent way.
In the long term, more complicated first-principles calculations based on nonequilibrium quantum field theory on curved spacetime are required.
As we will see below, baryogenesis typically requires a $\chi$ value of the order of $\chi \sim \textrm{few} \times 10^{-8}$.
The key lesson from the discussion in this subsection therefore is that axion inflation is capable of generating this value; a more accurate description of the underlying parameter dependence is left for future work.


\section{Survival of the primordial helicity}
\label{sec:helicity}


In the previous section, we derived the initial conditions for the radiation-dominated era, specifically, the initial values of the hypermagnetic helicity density and fermionic charge asymmetries [see Eq.~\eqref{eq:hYqi} and Fig.~\ref{fig:chi}]. 
As outlined in the Introduction, these primordial charges set the stage for baryogenesis:
The asymmetries stored in the chiral fermions lead to wash-in leptogenesis, if they are not erased before RHN interactions become efficient; while the helicity stored in the hypermagnetic field can act as a source of baryon number, if it survives until the EWPT.
The survival of the primordial charges in the SM plasma is, however, endangered by several effects.
In addition to weak sphaleron processes, which seek to wash out any primordial baryon-plus-lepton number, we must pay attention to magnetic diffusion and the chiral plasma instability.
We will now discuss these two effects in the \textit{magnetohydrodynamics} (MHD) approximation~\cite{Banerjee:2004df,Durrer:2013pga}, closely following the more detailed discussion in Ref.~\cite{Domcke:2019mnd}.
This will provide us with lower and upper bounds on the strength of the primordial hypermagnetic field from magnetic diffusion and the chiral plasma instability, respectively.
Remarkably, the range between these two bounds remains viable and can give rise to the correct BAU.


\subsection{Magnetic diffusion}
\label{subsec:diffusion}


Due to the conductivity of the SM thermal bath, diffusion processes tend to wash out the primordial hypermagnetic fields permeating the early Universe.
These processes are particularly efficient at small scales, but reach the correlation length of the hypercharge gauge field at a temperature scale of around~\cite{Domcke:2019mnd}
\begin{equation}
\label{eq:Tdiff}
T_{\rm diff} \sim \frac{1}{c_\lambda^2} \frac{T}{\sigma_{\rm th}} H_{\rm end} \,,
\end{equation}
if the evolution is adiabatic.
Here, $H_{\rm end}$ is the physical Hubble rate at the end of hypermagnetogenesis, $c_\lambda$ parametrizes the physical correlation length at the end of hypermagnetogenesis, $c_\lambda H_{\rm end}^{-1}$ [see Eq.~\eqref{eq:clambda}], and $\sigma_{\rm th} \simeq 10^2\,T$ denotes the physical hyperelectric conductivity of the thermal plasma.%
\footnote{From now on, all quantities are going to be physical; we will no longer introduce new comoving quantities in the conformal frame.}
Note that the ratio $\sigma_\mathrm{th}/T$ is temperature-independent, up to effects related to the running of the hypercharge gauge coupling constant.
Since the diffusion processes preserve the conservation law in Eq.~\eqref{eq:qiend}, they threaten to erase both the $CP$ violation stored in the hypermagnetic field as well as in the fermion asymmetries, which means that they endanger both wash-in leptogenesis and baryogenesis from helicity decay.


Diffusion may be avoided if the hypermagnetic field is strong enough to trigger a turbulent evolution of the thermal plasma before diffusion processes reach the relevant scales~\cite{Pouquet:1976zz, Kahniashvili:2012uj,Banerjee:2004df}.
This requires an initial magnetic Reynolds number much larger than unity,%
\footnote{In the cascade regime, the magnetic Reynolds number increases monotonically.
A large initial value is therefore sufficient to ensure a turbulent regime throughout the subsequent evolution~\cite{Domcke:2019mnd}.}
such that magnetic induction dominates over diffusion,
\begin{equation}
\textrm{Re}_{\rm mag}^{\rm ini} = \left.\frac{\sigma_{\rm th} v c_\lambda}{H}\right|_{\rm end} \,.
\end{equation}
with $v$ denoting the typical magnitude of the plasma velocity field.
In order to turn the requirement $\textrm{Re}_{\rm mag}^{\rm ini} \gg 1$ into a constraint on the parameter space of our model, we need to estimate the magnitude of the velocity field at the end of hypermagnetogenesis, $v_{\rm end}$.
To this end, one may naively assume that an equipartition among kinetic energy and magnetic energy is reached in the plasma, which is typically found in MHD simulations in the regime of large kinetic Reynolds numbers $\textrm{Re}_{\rm kin}$.
However, in the region of parameter space that we are interested in, the initial kinetic Reynolds number
\begin{equation}
\textrm{Re}_{\rm kin}^{\rm ini} = \left.\frac{v c_\lambda}{\nu H}\right|_{\rm end} \,, 
\end{equation}
with $\nu \simeq 10/T$ denoting the kinetic viscosity of the plasma, is not always larger than unity.
We can thus no longer trust the MHD simulation results; in particular, it is not clear whether the energies in the velocity and magnetic fields are indeed equally partitioned, which makes it challenging to accurately determine $v_{\rm end}$.


In the following, we will therefore proceed by estimating the magnetic Reynolds number based on two alternative assumptions, which we expect the cover the range of physically conceivable scenarios,
\begin{align}
\label{eq:Rm_max}
\rho v^2 \sim \rho_B  \quad & \rightarrow \quad \textrm{Re}_{\rm mag}^{\rm ini} \sim \textrm{Re}_{\rm mag}^{\rm ini, max} \sim \left[\frac{\sigma_{\rm th}c_\lambda}{T} \left(\frac{M_*}{T}\right) \left(\frac{\rho_B}{\rho_\text{tot}}\right)^{1/2}\right]_{\rm end} \,, \\
\label{eq:Rm_visc}
\rho v^2 \sim \textrm{Re}_{\rm kin}\, \rho_B \quad &\rightarrow \quad \textrm{Re}_{\rm mag}^{\rm ini} \sim \textrm{Re}_{\rm mag}^{\rm ini, visc} \sim \left[\frac{\sigma_{\rm th}c_\lambda^2}{\nu T^2} \left(\frac{M_*}{T}\right)^2 \left(\frac{\rho_B}{\rho_\text{tot}}\right)\right]_{\rm end} \,,
\end{align}
with $M_* = \left[90/\left(\pi^2 g_*\right)\right]^{1/2} M_{\rm Pl}$, $\rho_{\rm tot} = 3H^2M_{\rm Pl}^2$, and where $\rho$ and $\rho_B = B^2/\left(2a^4\right)$ denote the energy densities of the plasma and hypermagnetic field, respectively.
In the first case, we assumed that the energy in the velocity field saturates the initial energy in the hypermagnetic field, which acts as a source term for the velocity field.
This may be regarded as an optimistic estimate that minimizes the parameter region where $\textrm{Re}_{\rm mag}^{\rm ini} \not\gg 1$.
Meanwhile, in the second case, we assumed that a viscous regime is reached, where the kinetic viscosity is balanced by the source term from the hypermagnetic field.
For more details on these analytical estimates, see Ref.~\cite{Domcke:2019mnd}.
Besides, we stress that a more refined analysis of the onset of turbulence after axion inflation requires a dedicated MHD analysis, which has thus far not been performed in the regime of interest, $\textrm{Re}_{\rm mag} \gg \textrm{Re}_{\rm kin}$.


\subsection{Chiral plasma instability}


The condition $\textrm{Re}_{\rm mag}^{\rm ini} \gg 1$ in combination with Eqs.~\eqref{eq:Rm_max} and \eqref{eq:Rm_visc} bounds the hypermagnetic energy density at the end of inflation from below.
A second, upper bound follows from the chiral plasma instability~\cite{Joyce:1997uy, Boyarsky:2011uy, Akamatsu:2013pjd,Hirono:2015rla, Yamamoto:2016xtu, Rogachevskii:2017uyc, Kamada:2018tcs}, which is based on the chiral magnetic effect~\cite{Vilenkin:1980fu, Alekseev:1998ds, Son:2004tq, Fukushima:2008xe}.
For our purposes, the essence of this effect is that large charge asymmetries in the chiral SM fermion species will trigger an instability in gauge-field modes that carry opposite helicity compared to the total helicity of the background field.
This will wash out the helicity stored in the hypermagnetic field as well as the charge asymmetries stored in the chiral fermions.


The strength of the chiral plasma instability is controlled by the chiral chemical potential
\begin{equation}
\label{eq:mu5def}
\bar{\mu}_5 = \sum_i \varepsilon_i g_i Y_i^2 \mu_i \,,
\end{equation}
whose initial value is directly proportional to the initial hypermagnetic helicity density in our model,
\begin{equation}
\label{eq:mu5Tini}
\left.\frac{\bar{\mu}_5}{T}\right|_{\rm end} = - \frac{95}{3}\chi = \left.- \frac{95}{6}\frac{\alpha_Y}{\pi}\frac{h_Y}{T^3}\right|_{\rm end} \,.
\end{equation}
Strong hypermagnetic fields therefore threaten to trigger an efficient chiral plasma instability.
In the course of the radiation-dominated era, the ratio $\bar{\mu}_5/T$ (slowly) changes by {\cal O}(1) factors
as a function of temperature, in dependence of the parity-violating SM interactions that successively enter thermal equilibrium as the Universe cools down.
The temperature scale of the chiral plasma instability then follows from the relation~\cite{Kamada:2018tcs} 
\begin{equation}
\label{eq:TCPI}
T_{\rm CPI} \sim 10^5\,\textrm{GeV}\left[\left(\frac{10^2}{g_*}\right)^{1/2} \left(\frac{\alpha_Y^{\vphantom{}}}{10^{-2}} \right)^2\left(\frac{10^2}{\sigma_{\rm th}/T}\right) \left( \frac{\bar{\mu}_5/T}{2\times10^{-3}} \right)^2\right]_{T_{\rm CPI}} \,.
\end{equation}
At this temperature, gauge-field modes with a typical wavenumber $k_{\rm CPI} = \alpha_Y \bar{\mu}_5/\pi$ begin to become unstable.


As the Universe expands more and more slowly, the chemical transport in the SM plasma becomes gradually more efficient. 
In the language of Eq.~\eqref{eq:anomaly}, this means that a steadily increasing number of terms on the right-hand side becomes active, which results in the reshuffling and erasure of the asymmetries stored in the different chiral fermion species. 
The last SM interaction to reach equilibrium is the electron Yukawa interaction, which thermalizes at a temperature of around $T_{y_e} \simeq 10^5$~GeV~\cite{Bodeker:2019ajh}.
At lower temperatures, all asymmetries are erased, $\mu_i \rightarrow 0$ for all fermion species $i$, such that no chiral chemical potential $\bar{\mu}_5$ remains. 
Therefore, if the estimate in Eq.~\eqref{eq:TCPI} returns a temperature below $T_{y_e}$ for a given input value of $\bar{\mu}_5/T$, no chiral plasma instability occurs. 
Thus, the CPI does not endanger the survival of the hypermagnetic fields until the EWPT if
\begin{equation}
\label{eq:CPI_bound}
T_{\rm CPI} <  T_{y_e} \sim 10^5\,\textrm{GeV} \,,
\end{equation}
which places an upper bound on the initial values of $\bar{\mu}_5/T$ and $h_Y$ [see Eq.~\eqref{eq:mu5Tini}].
If this bound on $T_{\rm CPI}$ as well as the bound on $\textrm{Re}_{\rm mag}^{\rm ini}$ discussed in Sec.~\ref{subsec:diffusion} [see Eqs.~\eqref{eq:Rm_max} and \eqref{eq:Rm_visc}] are both satisfied, the helicity stored in the hypermagnetic field survives until the EWPT, setting the stage for baryogenesis from helicity decay.


As we will see in Sec.~\ref{subsec:regimes}, in the parameter regime that is capable of reproducing the observed BAU, the condition~\eqref{eq:CPI_bound} is always easily fulfilled, since the large values of $\bar \mu_5$ required to trigger the CPI before the electron Yukawa interaction equilibrates would imply an overproduction of the BAU through wash-in leptogenesis for right-handed neutrino masses above $10^5$~GeV. 
In the following, we will thus focus on this situation.


\section{Baryon and lepton number violation}
\label{sec:bau}


Axion inflation does not generate any $B\!-\!L$ asymmetry.%
\footnote{This immediately follows from the fact that axion inflation does not involve any $B\!-\!L$-violating interactions.
In addition, one may explicitly convince oneself that $B-L = 0$ at the end of axion inflation by appropriately combining the charge densities in Eq.~\eqref{eq:qiend}.}
Therefore, if the radiation-dominated era after inflation is solely described by SM physics, the global $B\!-\!L$ charge always remains zero and leptogenesis plays no role in the generation of the BAU.
This situation changes as soon as RHNs are added to the picture.
As pointed out in Ref.~\cite{Domcke:2020quw}, efficient RHN interactions alter the chemical transport in the plasma such that primordial charge asymmetries can be reprocessed into a new $B\!-\!L$-violating equilibrium, even if $B\!-\!L = 0$ initially.
This mechanism was dubbed wash-in leptogenesis in Ref.~\cite{Domcke:2020quw}.
In addition, RHNs can of course lead to the generation of a $B\!-\!L$ asymmetry by means of standard thermal leptogenesis.
However, as shown in Ref.~\cite{Domcke:2020quw}, the thermal contribution to the final BAU is independent of the wash-in contribution, up to numerically negligible corrections.
It therefore suffices to simply add it to the outcome of wash-in leptogenesis.

\begin{figure}
\begin{center}
\includegraphics[width=\textwidth]{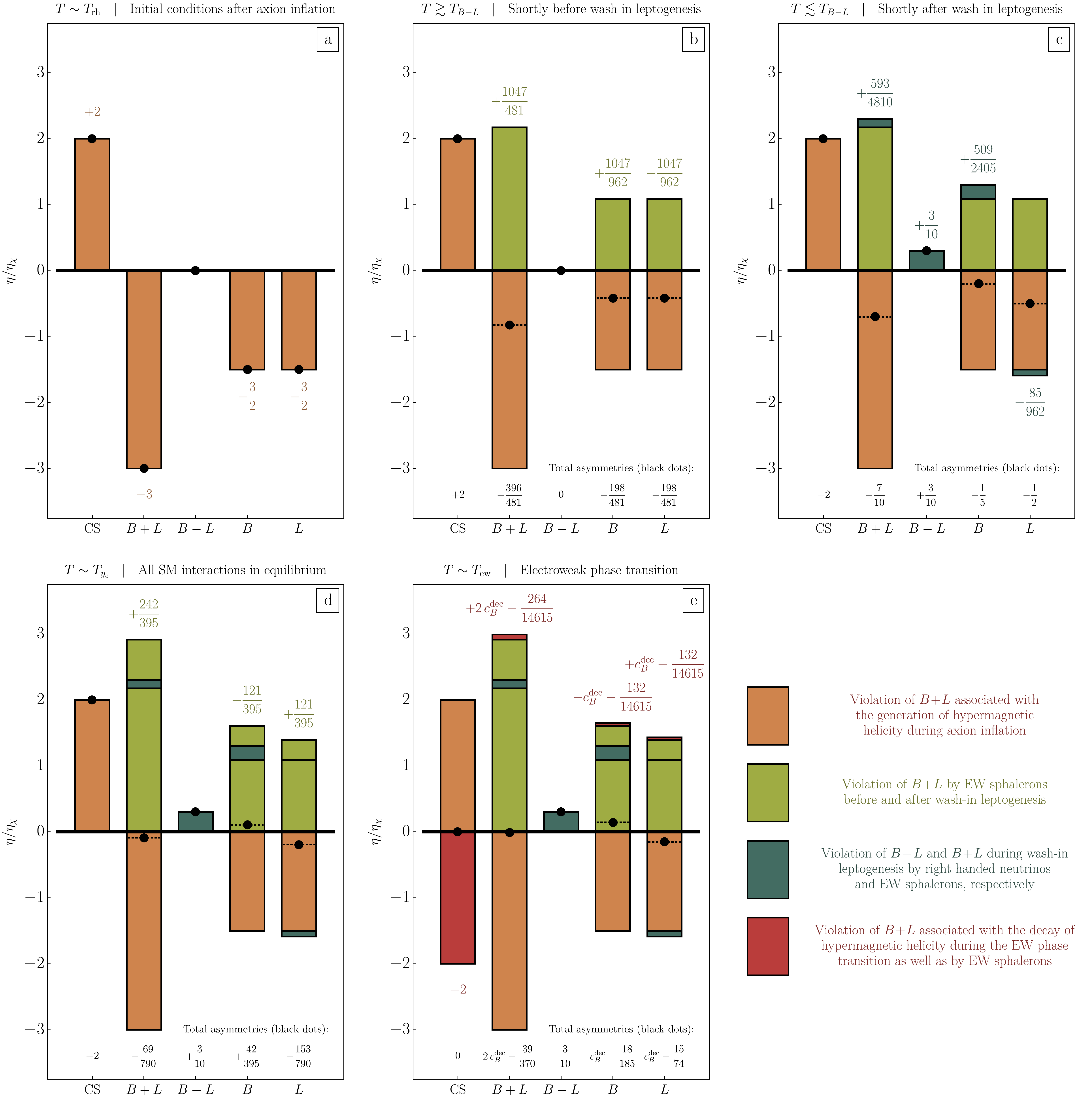}
\caption{Evolution of the global CS, $B\!+\!L$, $B\!-\!L$, $B$, and $L$ charges in the benchmark scenario discussed in Sec.~\ref{sec:bau} all the way from the end of axion inflation to the EWPT.
Panels (a) and (b) describe the initial conditions after axion inflation and the chemical equilibrium shortly before wash-in leptogenesis (see Sec.~\ref{subsec:transport}); panels (c), (d), and (e) illustrate the situation shorty after wash-in leptogenesis, below the equilibration temperature of the electron Yukawa interaction, and at the time of sphaleron freeze-out towards the end of the EWPT (see Sec.~\ref{subsec:washin}).
The red bars in panel (e) indicate the outcome of baryogenesis from helicity decay (see Sec.~\ref{subsec:decay}).
Every panel contains the bars shown in all previous panels plus new bars that indicate the relevant new contributions to the respective charges.
The total charges at the respective stages of the evolution are indicated by the black dots and horizontal black dashed lines.}
\label{fig:bars}
\end{center} 
\end{figure}

Let us now consider a particularly simple benchmark scenario: wash-in leptogenesis at temperatures shortly above the equilibration temperature of the electron Yukawa interaction, $T_{y_e}$, a regime in which standard thermal leptogenesis does not yield any appreciable contribution to the final BAU.
That is, we assume the RHNs $N_i$ ($i=1,2,3$) to have masses $M_i$ of a few 100 TeV, say, $M_1 = 200\,\textrm{TeV}$, $M_2 = 400\,\textrm{TeV}$, and $M_3 = 800\,\textrm{TeV}$, such that they all decay into SM lepton--Higgs pairs in the temperature window $T \sim 10^5\cdots10^6\,\textrm{GeV}$.
Wash-in leptogenesis in other temperature regimes as well as heavy-neutrino flavor effects will be discussed in Sec.~\ref{subsec:regimes}.
In Fig.~\ref{fig:bars}, we schematically illustrate the evolution of the global CS, $B\!+\!L$, $B\!-\!L$, $B$, and $L$ charges in our benchmark scenario all the way from the end of inflation to the EWPT in five steps, which we will discuss one by one in Secs.~\ref{subsec:washin} and \ref{subsec:decay}.
In doing so, we will express all charges in terms of charge-to-photon ratios and work in units of $\eta_\chi$, a characteristic charge-to-photon ratio whose size is controlled by the yield parameter $\chi$,
\begin{equation}
\label{eq:etachi}
\eta_C = \frac{q_C}{n_\gamma} \,, \qquad \eta_\chi = \frac{\chi T^3}{n_\gamma} \,, \qquad n_\gamma = \frac{\zeta\left(3\right)}{\pi^2}g_\gamma T^3 \,.
\end{equation}


\subsection{Chemical transport in the SM plasma}
\label{subsec:transport}


Right after axion inflation [see panel (a) in Fig.~\ref{fig:bars}], the initial CS charge is given by Eq.~\eqref{eq:chi}, which translates into a charge-to-photon ratio $\eta_{\rm CS} = 2 \,\eta_\chi$.
The corresponding $B+L$ charge can be either calculated based on Eq.~\eqref{eq:qiend} or directly deduced from the SM anomaly equations for the global $B$ and $L$ currents,
\begin{equation}
\label{eq:JBL}
\partial_\mu J_B^\mu = \partial_\mu J_L^\mu = N_g \left(\frac{g_L^2}{32\pi^2}W_{\mu\nu}^a\widetilde{W}^{a\,\mu\nu} - \frac{g_Y^2}{32\pi^2}Y_{\mu\nu}\widetilde{Y}^{\mu\nu}\right) \,.
\end{equation}
Here, $N_g = 3$ counts the number of SM fermion generations, $g_L$ is the $SU(2)_L$ isospin gauge coupling constant, and $W_{\mu\nu}^a$ ($a=1,2,3$) is the corresponding field strength tensor, with its dual denoted by $\widetilde{W}^{a\,\mu\nu}$.
Integrating the anomaly equation in Eq.~\eqref{eq:JBL} in a homogeneous background over time results in the relation
\begin{equation}
\label{eq:DeltaBL}
\Delta q_B = \Delta q_L = 3 \left(\Delta q_{\rm CS}^L - \frac{1}{4} \Delta q_{\rm CS}^Y\right) \,,
\end{equation}
with $q_{\rm CS}^L$ and $q_{\rm CS}^Y \equiv q_{\rm CS}$ denoting the $SU(2)_L$ and $U(1)_Y$ CS charge densities, respectively.
During axion inflation, long-range $SU(2)_L$ gauge-field fluctuations are not amplified and weak sphaleron processes are inefficient, such that $\Delta q_{\rm CS}^L = 0$.
This implies $\Delta q_{B+L} = -3/2\,\Delta q_{\rm CS}^Y$ at the end of axion inflation, which translates to $\eta_{B+L} = -3\,\eta_\chi$ in panel (a) in Fig.~\ref{fig:bars}.
Meanwhile, the global $B\!-\!L$ charge remains conserved during axion inflation, which means that $\eta_{B-L} = 0$ and $\eta_B = \eta_L = -3/2\, \eta_\chi$ initially.


In addition, axion inflation results in the generation of further global charges that only become violated during the radiation-dominated era as more and more SM interactions reach thermal equilibrium.
Among these charges, we will notably require the values of the following five charges in our analysis, for reasons that will become clear shortly: the charge asymmetries stored in right-handed electrons, muons, taus, and up quarks as well as the difference between the charge asymmetries of right-handed up quarks and right-handed down quarks.
At the end of axion inflation, the charge-to-photon ratios for these five charges read
\begin{equation}
\label{eq:etaC}
\eta_e = \eta_\mu = \eta_\tau = \eta_{u-d} = -\eta_\chi \,, \qquad \eta_u = -\frac{4}{3} \eta_\chi \,. 
\end{equation}
Here, the relation among the charge asymmetries of the three charged-lepton flavors, $\eta_e = \eta_\mu = \eta_\tau$, reflects the fact that the generation of fermionic charge asymmetries during axion inflation is a flavor-blind process.


In the approximation of instantaneous reheating, we expect a reheating temperature $T_{\rm rh} \sim 10^{14}\,\textrm{GeV}$ in the relevant part of parameter space (see Fig.~\ref{fig:chi}).
Let us now consider the evolution of the primordial charges generated during axion inflation from this high temperature scale all the way down to the temperature scale of wash-in leptogenesis, $T_{B-L}$, which we assume to be of the order of a few 100 TeV in our benchmark scenario.
In doing so, it will be important to keep track of the subset of SM interactions that have already reached thermal equilibrium in dependence of the decreasing temperature of the thermal bath.
In order to facilitate the discussion, we will split the temperature range from $T_{\rm rh}$ to $T_{B-L}$ into five different regimes, each of which is characterized by a certain subset of SM interactions in equilibrium.
For more details on the equilibration temperatures of the individual SM Yukawa and sphaleron interactions, see Ref.~\cite{Domcke:2020kcp}.
A second simplification consists in the fact that we will neglect the chemical potentials of the three RHNs, $\mu_{N_i}$, in our discussion.
At $T \gg M_i$, these chemical potentials may temporarily obtain nonzero values.
Whether and when this happens is, however, a model-dependent question and depends on the rate of $\Delta L = 1$ scattering processes such as, \textit{e.g.}, $N_i Q_3 \leftrightarrow \ell_\alpha t$, and hence on the RHN Yukawa couplings, as well as on the thermal history of the RHN population.
Moreover, all effects caused by $\mu_{N_i} \neq 0$ at high temperatures will be reverted at lower temperatures anyway when the RHNs turn nonrelativistic and the $\mu_{N_i}$ are driven to zero by the RHN Majorana masses. 
At the time of wash-in leptogenesis, when $T \sim M_i$, we can therefore safely work with $\mu_{N_i} = 0$. 
The model-dependent time evolution of the three $\mu_{N_i}$ will not affect our conclusions and is thus irrelevant for our purposes.


\paragraph{(i) \boldmath{$T \in \left(10^{13}, 10^{15}\right)\,\textrm{GeV}$}:}
Immediately after reheating, most SM interactions are still too slow to compete with the Hubble expansion.
In addition to the SM gauge interactions, the only process in thermal equilibrium is the top-quark Yukawa interaction.
As a consequence, a large number of global charges remains conserved in the thermal bath.
Indeed, given that the SM particle content can be described by 16 chemical potentials (15 fermion representations, see Tab.~\ref{tab:factors}, plus the SM Higgs doublet $\Phi$), one Yukawa interaction in equilibrium means that one is able to define 15 linearly independent global charges. 
One possible choice is, \textit{e.g.},~\cite{Domcke:2020quw}
\begin{equation}
\label{eq:charges}
\left\{\eta_u,\: \eta_B,\: \eta_{d-b},\: \eta_\tau,\: \eta_{u-c},\: \eta_\mu,\: \eta_{B_1-B_2},\: \eta_{d-s},\: \eta_{u-d},\: \eta_{2B_1-B_2-B_3},\: \eta_e,\:\eta_{\Delta_e},\:\eta_{\Delta_\mu},\:\eta_{\Delta_\tau},\:\eta_Y\right\} \,,
\end{equation}
where $\Delta_\alpha = B/3 - L_\alpha$ ($\alpha = e,\mu,\tau$).
These 15 charges span an orthonormal basis of a 15-dimensional vector space.
Any other orthonormal basis of this vector space also corresponds to a set of conserved charges.
The basis in Eq.~\eqref{eq:charges} is, however, a particularly convenient choice for the following reason:
As the temperature of thermal bath drops, new interactions will enter thermal equilibrium and begin to violate the charges listed in Eq.~\eqref{eq:charges} from left to right.
For instance, as the strong sphaleron processes reach thermal equilibrium, $\eta_u$ will become violated; as the weak sphaleron processes reach thermal equilibrium, $\eta_B$ will become violated; and so on and so forth, until at temperatures below $T_{y_e}$, all SM interactions are equilibrated, such that all charges in Eq.~\eqref{eq:charges} are violated, except for the three lepton flavor asymmetries $\Delta_\alpha$ and the global hypercharge $Y$.


Moreover, note that the list in Eq.~\eqref{eq:charges} contains the five charges that we introduced in Eq.~\eqref{eq:etaC}.
In view of Eq.~\eqref{eq:charges}, we can now rephrase more precisely which global charges of interest are in fact generated during axion inflation:
In the 15-dimensional vector space spanned by the charges listed in Eq.~\eqref{eq:charges}, axion inflation populates the six-dimensional subspace spanned by $\eta_u$, $\eta_B$, $\eta_\tau$, $\eta_\mu$, $\eta_{u-d}$, and $\eta_e$.
Among the nine charges in the orthogonal co-space, the six charges $\eta_{d-b}$, $\eta_{u-c}$, $\eta_{B_1-B_2}$, $\eta_{d-s}$, $\eta_{2B_1-B_2-B_3}$, and $\eta_Y$ remain zero at all times throughout the cosmological evolution, while the three lepton flavor asymmetries $\eta_{\Delta_\alpha}$ can only be generated by RHN interactions. 
In the following, we will therefore need to work with in total nine charges that can in principle obtain nonzero values: $\eta_u$, $\eta_B$, $\eta_\tau$, $\eta_\mu$, $\eta_{u-d}$, $\eta_e$, $\eta_{\Delta_e}$, $\eta_{\Delta_\mu}$, and $\eta_{\Delta_\tau}$.
The charge asymmetries of the 16 SM fermion and Higgs fields live precisely in the vector space spanned by these nine charges. 


In other words, each of the 16 SM chemical potentials, $\mu_i$ ($i = e,\mu\,\tau,\ell_e,\ell_\mu,\ell_\tau,u,c,t,d,s,b,Q_1,Q_2,Q_3,\Phi$), can be expressed in terms of a linear combination of the nine chemical potentials $\bar{\mu}_C$ ($C = u,B,\tau,\mu,u-d,e,\Delta_e,\Delta_\mu,\Delta_\tau$).%
\footnote{Here, the bar over the chemical potentials associated with global charges, $\bar{\mu}_C$, indicates that all internal gauge degrees of freedom have been summed over, which is not the case for the ordinary chemical potentials $\mu_i$.
As a consequence, we have, \textit{e.g.}, $\mu_u = \sfrac{1}{3}\,\bar{\mu}_u$.
This distinction is not necessary for quantities like the charge densities $q_i = g_i \mu_i T^2/6$ and $q_C = \bar{\mu}_C T^2/6$ or the charge-to-photon ratios $\eta_i = q_i/n_\gamma$ and $\eta_C = q_C/n_\gamma$, all of which include the same $g_i$ factors.
For more details on our conventions, see Ref.~\cite{Domcke:2020quw}.}
In order to work out these linear combinations, we employ the linear-algebra formalism developed in the appendix of Ref.~\cite{Domcke:2020quw}.
The essence of this formalism is to write down a system of 16 linear equations\,---\,one constraint equation for each conserved charge and one equilibrium condition for each interaction in chemical equilibrium\,---\,which can be solved for the 16 SM chemical potentials.
Working with 15 constraint equations and one equilibrium condition for the top-quark Yukawa interaction, we thus obtain
\begin{equation}
\begin{pmatrix}
\mu_e \\ \mu_\mu \\ \mu_\tau \\ \mu_{\ell_e} \\ \mu_{\ell_\mu} \\ \mu_{\ell_\tau} \\ \mu_u \\ \mu_c \\ \mu_t \\ \mu_d \\ \mu_s \\ \mu_b \\ \mu_{Q_1} \\ \mu_{Q_2} \\ \mu_{Q_3} \\ \mu_\Phi
\end{pmatrix}
=
\begin{pmatrix}
0 & 0 & 0 & 0 & 0 & 1 & 0 & 0 & 0 \\
0 & 0 & 0 & 1 & 0 & 0 & 0 & 0 & 0 \\
0 & 0 & 1 & 0 & 0 & 0 & 0 & 0 & 0 \\
0 & \sfrac{1}{6} & 0 & 0 & 0 & -\sfrac{1}{2} & -\sfrac{1}{2} & 0 & 0 \\
0 & \sfrac{1}{6} & 0 & -\sfrac{1}{2} & 0 & 0 & 0 & -\sfrac{1}{2} & 0 \\
0 & \sfrac{1}{6} & -\sfrac{1}{2} & 0 & 0 & 0 & 0 & 0 & -\sfrac{1}{2} \\
\sfrac{1}{3} & 0 & 0 & 0 & 0 & 0 & 0 & 0 & 0 \\
\sfrac{1}{3} & 0 & 0 & 0 & 0 & 0 & 0 & 0 & 0 \\
\sfrac{1}{27} & \sfrac{2}{27} & \sfrac{1}{9} & \sfrac{1}{9} & -\sfrac{7}{27} & \sfrac{1}{9} & -\sfrac{1}{9} & -\sfrac{1}{9} & -\sfrac{1}{9} \\
\sfrac{1}{3} & 0 & 0 & 0 & -\sfrac{1}{3} & 0 & 0 & 0 & 0 \\
\sfrac{1}{3} & 0 & 0 & 0 & -\sfrac{1}{3} & 0 & 0 & 0 & 0 \\
\sfrac{1}{3} & 0 & 0 & 0 & -\sfrac{1}{3} & 0 & 0 & 0 & 0 \\
-\sfrac{1}{3} & \sfrac{1}{6} & 0 & 0 & \sfrac{1}{6} & 0 & 0 & 0 & 0 \\
-\sfrac{1}{3} & \sfrac{1}{6} & 0 & 0 & \sfrac{1}{6} & 0 & 0 & 0 & 0 \\
-\sfrac{5}{27} & \sfrac{7}{54} & -\sfrac{1}{18} & -\sfrac{1}{18} & \sfrac{8}{27} & -\sfrac{1}{18} & \sfrac{1}{18} & \sfrac{1}{18} & \sfrac{1}{18} \\
\sfrac{2}{9} & -\sfrac{1}{18} & \sfrac{1}{6} & \sfrac{1}{6} & -\sfrac{5}{9} & \sfrac{1}{6} & -\sfrac{1}{6} & -\sfrac{1}{6} & -\sfrac{1}{6}
\end{pmatrix}
\begin{pmatrix}
\bar{\mu}_u \\ \bar{\mu}_B \\ \bar{\mu}_\tau \\ \bar{\mu}_\mu \\ \bar{\mu}_{u-d} \\ \bar{\mu}_e \\ \bar{\mu}_{\Delta_e} \\ \bar{\mu}_{\Delta_\mu} \\\bar{\mu}_{\Delta_\tau}
\end{pmatrix} \,,
\label{eq:eqI}
\end{equation}
where the chemical potentials associated with the global charges on the right-hand side are initially given by
\begin{equation}
\label{eq:muini}
\frac{\bar{\mu}_u}{T} = -8\chi \,, \quad \frac{\bar{\mu}_B}{T} = -9\chi \,, \quad \frac{\bar{\mu}_\tau}{T} = \frac{\bar{\mu}_\mu}{T} = \frac{\bar{\mu}_{u-d}}{T} = \frac{\bar{\mu}_e}{T} = -6\chi \,, \quad \frac{\bar{\mu}_{\Delta_e}}{T} = \frac{\bar{\mu}_{\Delta_\mu}}{T} = \frac{\bar{\mu}_{\Delta_\tau}}{T} = 0 \,.
\end{equation}


In passing, we also mention that, if we were to include the chemical potentials of the three RHNs in our discussion, we would have to work with 19 chemical potentials and 19 linear equations.
As explained above, this would, however, require us to fix the size of the Yukawa couplings in the RHN sector and specify the thermal history of the RHN population, while no net effect would survive down to low temperatures anyway.


\paragraph{(ii) \boldmath{$T \in \left(10^{11\cdots12}, 10^{13}\right)\,\textrm{GeV}$}:}
Around $T \sim 10^{13}\,\textrm{GeV}$, strong sphaleron processes, \textit{i.e.}, thermal fluctuations in the topological charge in the SM $SU(3)_C$ sector, reach thermal equilibrium.
As a consequence, the charge asymmetry stored in right-handed up quarks no longer represents a conserved global charge, which means that our linear system of equations now consists of 14 constraint equations and two equilibrium conditions.
Solving this system for the 16 SM chemical potentials results in the following new equilibrium solution,
\begin{equation}
\begin{pmatrix}
\mu_e \\ \mu_\mu \\ \mu_\tau \\ \mu_{\ell_e} \\ \mu_{\ell_\mu} \\ \mu_{\ell_\tau} \\ \mu_u \\ \mu_c \\ \mu_t \\ \mu_d \\ \mu_s \\ \mu_b \\ \mu_{Q_1} \\ \mu_{Q_2} \\ \mu_{Q_3} \\ \mu_\Phi
\end{pmatrix}
=
\begin{pmatrix}
0 & 0 & 0 & 0 & 1 & 0 & 0 & 0 \\
0 & 0 & 1 & 0 & 0 & 0 & 0 & 0 \\
0 & 1 & 0 & 0 & 0 & 0 & 0 & 0 \\
\sfrac{1}{6} & 0 & 0 & 0 & -\sfrac{1}{2} & -\sfrac{1}{2} & 0 & 0 \\
\sfrac{1}{6} & 0 & -\sfrac{1}{2} & 0 & 0 & 0 & -\sfrac{1}{2} & 0 \\
\sfrac{1}{6} & -\sfrac{1}{2} & 0 & 0 & 0 & 0 & 0 & -\sfrac{1}{2} \\
\sfrac{1}{12} & -\sfrac{1}{46} & -\sfrac{1}{46} & \sfrac{17}{69} & -\sfrac{1}{46} & \sfrac{1}{46} & \sfrac{1}{46} & \sfrac{1}{46} \\
\sfrac{1}{12} & -\sfrac{1}{46} & -\sfrac{1}{46} & \sfrac{17}{69} & -\sfrac{1}{46} & \sfrac{1}{46} & \sfrac{1}{46} & \sfrac{1}{46} \\
\sfrac{1}{12} & \sfrac{5}{46} & \sfrac{5}{46} & -\sfrac{16}{69} & \sfrac{5}{46} & -\sfrac{5}{46} & -\sfrac{5}{46} & -\sfrac{5}{46} \\
\sfrac{1}{12} & -\sfrac{1}{46} & -\sfrac{1}{46} & -\sfrac{2}{23} & -\sfrac{1}{46} & \sfrac{1}{46} & \sfrac{1}{46} & \sfrac{1}{46} \\
\sfrac{1}{12} & -\sfrac{1}{46} & -\sfrac{1}{46} & -\sfrac{2}{23} & -\sfrac{1}{46} & \sfrac{1}{46} & \sfrac{1}{46} & \sfrac{1}{46} \\
\sfrac{1}{12} & -\sfrac{1}{46} & -\sfrac{1}{46} & -\sfrac{2}{23} & -\sfrac{1}{46} & \sfrac{1}{46} & \sfrac{1}{46} & \sfrac{1}{46} \\
\sfrac{1}{12} & \sfrac{1}{46} & \sfrac{1}{46} & -\sfrac{11}{138} & \sfrac{1}{46} & -\sfrac{1}{46} & -\sfrac{1}{46} & -\sfrac{1}{46} \\
\sfrac{1}{12} & \sfrac{1}{46} & \sfrac{1}{46} & -\sfrac{11}{138} & \sfrac{1}{46} & -\sfrac{1}{46} & -\sfrac{1}{46} & -\sfrac{1}{46} \\
\sfrac{1}{12} & -\sfrac{1}{23} & -\sfrac{1}{23} & \sfrac{11}{69} & -\sfrac{1}{23} & \sfrac{1}{23} & \sfrac{1}{23} & \sfrac{1}{23} \\
0 & \sfrac{7}{46} & \sfrac{7}{46} & -\sfrac{9}{23} & \sfrac{7}{46} & -\sfrac{7}{46} & -\sfrac{7}{46} & -\sfrac{7}{46}
\end{pmatrix}
\begin{pmatrix}
\bar{\mu}_B \\ \bar{\mu}_\tau \\ \bar{\mu}_\mu \\ \bar{\mu}_{u-d} \\ \bar{\mu}_e \\ \bar{\mu}_{\Delta_e} \\ \bar{\mu}_{\Delta_\mu} \\\bar{\mu}_{\Delta_\tau}
\end{pmatrix} \,.
\label{eq:eqII}
\end{equation}
%


\paragraph{(iii) \boldmath{$T \in \left(10^9, 10^{11\cdots12}\right)\,\textrm{GeV}$}:}
In the temperature interval $T \sim 10^{11\cdots12}\,\textrm{GeV}$, four linearly independent SM interactions reach thermal equilibrium, which leads to the violation of two global charges that are relevant for our discussion.
More precisely, the interactions reaching equilibrium consist of the weak sphaleron processes, \textit{i.e.}, thermal fluctuations in the topological charge in the SM $SU(2)_L$ sector, as well as of the bottom-quark, charm-quark, and tau Yukawa interactions. 
Here, the weak sphalerons violate global baryon number, $\bar{\mu}_B$, while the tau Yukawa interaction violates the global charge stored in right-handed tau leptons, $\bar{\mu}_\tau$, which reduces the number of relevant global charges from eight to six.
In addition, we now observe a reshuffling of all chemical potentials across the quark and lepton sectors, apart from the chemical potentials of right-handed electrons and muons, which are not yet in contact with the rest of the thermal bath.
Our system of linear equations for the 16 SM chemical potentials now consists of ten constraint equations (for the last ten charges in Eq.~\eqref{eq:charges}) and six equilibrium conditions (for the six SM processes alluded to thus far), which yields 
\begin{equation}
\begin{pmatrix}
\mu_e \\ \mu_\mu \\ \mu_\tau \\ \mu_{\ell_e} \\ \mu_{\ell_\mu} \\ \mu_{\ell_\tau} \\ \mu_u \\ \mu_c \\ \mu_t \\ \mu_d \\ \mu_s \\ \mu_b \\ \mu_{Q_1} \\ \mu_{Q_2} \\ \mu_{Q_3} \\ \mu_\Phi
\end{pmatrix}
=
\begin{pmatrix}
0 & 0 & 1 & 0 & 0 & 0 \\
1 & 0 & 0 & 0 & 0 & 0 \\
-\sfrac{4}{589} & \sfrac{45}{589} & -\sfrac{4}{589} & \sfrac{56}{589} & \sfrac{56}{589} & -\sfrac{139}{589} \\
\sfrac{39}{589} & \sfrac{3}{589} & -\sfrac{511}{1178} & -\sfrac{503}{1178} & \sfrac{43}{589} & \sfrac{30}{589} \\
-\sfrac{511}{1178} & \sfrac{3}{589} & \sfrac{39}{589} & \sfrac{43}{589} & -\sfrac{503}{1178} & \sfrac{30}{589} \\
\sfrac{41}{589} & -\sfrac{39}{1178} & \sfrac{41}{589} & \sfrac{15}{589} & \sfrac{15}{589} & -\sfrac{195}{589} \\
\sfrac{33}{2356} & \sfrac{421}{1767} & \sfrac{33}{2356} & \sfrac{127}{2356} & \sfrac{127}{2356} & \sfrac{29}{589} \\
\sfrac{213}{2356} & -\sfrac{67}{1767} & \sfrac{213}{2356} & -\sfrac{37}{2356} & -\sfrac{37}{2356} & -\sfrac{27}{589} \\
\sfrac{129}{1178} & -\sfrac{63}{589} & \sfrac{129}{1178} & -\sfrac{39}{1178} & -\sfrac{39}{1178} & -\sfrac{41}{589} \\
\sfrac{33}{2356} & -\sfrac{56}{589} & \sfrac{33}{2356} & \sfrac{127}{2356} & \sfrac{127}{2356} & \sfrac{29}{589} \\
\sfrac{33}{2356} & -\sfrac{56}{589} & \sfrac{33}{2356} & \sfrac{127}{2356} & \sfrac{127}{2356} & \sfrac{29}{589} \\
-\sfrac{51}{1178} & \sfrac{66}{589} & -\sfrac{51}{1178} & \sfrac{125}{1178} & \sfrac{125}{1178} & \sfrac{71}{589} \\
\sfrac{123}{2356} & -\sfrac{235}{3534} & \sfrac{123}{2356} & \sfrac{45}{2356} & \sfrac{45}{2356} & \sfrac{1}{589} \\
\sfrac{33}{2356} & \sfrac{253}{3534} & \sfrac{33}{2356} & \sfrac{127}{2356} & \sfrac{127}{2356} & \sfrac{29}{589} \\
\sfrac{39}{1178} & \sfrac{3}{1178} & \sfrac{39}{1178} & \sfrac{43}{1178} & \sfrac{43}{1178} & \sfrac{15}{589} \\
\sfrac{45}{589} & -\sfrac{129}{1178} & \sfrac{45}{589} & -\sfrac{41}{589} & -\sfrac{41}{589} & -\sfrac{56}{589}
\end{pmatrix}
\begin{pmatrix}
\bar{\mu}_\mu \\ \bar{\mu}_{u-d} \\ \bar{\mu}_e \\ \bar{\mu}_{\Delta_e} \\ \bar{\mu}_{\Delta_\mu} \\\bar{\mu}_{\Delta_\tau}
\end{pmatrix} \,.
\label{eq:eqIII}
\end{equation}


\paragraph{(iv) \boldmath{$T \in \left(10^6, 10^9\right)\,\textrm{GeV}$}:}
At $T \sim 10^9\,\textrm{GeV}$, the muon Yukawa interaction and the remaining Yukawa interactions of the second and third quark generations equilibrate.
The latter include the strange-quark Yukawa interaction, but also off-diagonal Yukawa interactions such as the interaction of right-handed strange quarks with left-handed bottom quarks.
The muon Yukawa interaction violates the global charge stored in right-handed muons, which leaves us with five conserved global charges that are relevant for our discussion.
The constraint equations for the last seven charges in Eq.~\eqref{eq:charges} and nine equilibrium conditions then lead to

\begin{equation}
\begin{pmatrix}
\mu_e \\ \mu_\mu \\ \mu_\tau \\ \mu_{\ell_e} \\ \mu_{\ell_\mu} \\ \mu_{\ell_\tau} \\ \mu_u \\ \mu_c \\ \mu_t \\ \mu_d \\ \mu_s \\ \mu_b \\ \mu_{Q_1} \\ \mu_{Q_2} \\ \mu_{Q_3} \\ \mu_\Phi
\end{pmatrix}
=
\begin{pmatrix}
0 & 1 & 0 & 0 & 0 \\
\sfrac{15}{358} & \sfrac{1}{358} & \sfrac{31}{358} & -\sfrac{133}{537} & \sfrac{46}{537} \\
\sfrac{15}{358} & \sfrac{1}{358} & \sfrac{31}{358} & \sfrac{46}{537} & -\sfrac{133}{537} \\
\sfrac{1}{179} & -\sfrac{155}{358} & -\sfrac{151}{358} & \sfrac{10}{179} & \sfrac{10}{179} \\
-\sfrac{11}{716} & \sfrac{47}{716} & \sfrac{25}{716} & -\sfrac{172}{537} & \sfrac{7}{537} \\
-\sfrac{11}{716} & \sfrac{47}{716} & \sfrac{25}{716} & \sfrac{7}{537} & -\sfrac{172}{537} \\
\sfrac{91}{537} & \sfrac{6}{179} & \sfrac{7}{179} & \sfrac{5}{179} & \sfrac{5}{179} \\
- \sfrac{39}{716} & \sfrac{69}{716} & -\sfrac{9}{716} & -\sfrac{8}{179} & -\sfrac{8}{179} \\
- \sfrac{39}{716} & \sfrac{69}{716} & -\sfrac{9}{716} & -\sfrac{8}{179} & -\sfrac{8}{179} \\
- \sfrac{88}{537} & \sfrac{6}{179} & \sfrac{7}{179} & \sfrac{5}{179} & \sfrac{5}{179} \\
\sfrac{43}{716} & -\sfrac{21}{716} & \sfrac{65}{716} & \sfrac{18}{179} & \sfrac{18}{179} \\
\sfrac{43}{716} & -\sfrac{21}{716} & \sfrac{65}{716} & \sfrac{18}{179} & \sfrac{18}{179} \\
\sfrac{1}{358} & \sfrac{6}{179} & \sfrac{7}{179} & \sfrac{5}{179} & \sfrac{5}{179} \\
\sfrac{1}{358} & \sfrac{6}{179} & \sfrac{7}{179} & \sfrac{5}{179} & \sfrac{5}{179} \\
\sfrac{1}{358} & \sfrac{6}{179} & \sfrac{7}{179} & \sfrac{5}{179} & \sfrac{5}{179} \\
-\sfrac{41}{716} & \sfrac{45}{716} & -\sfrac{37}{716} & -\sfrac{13}{179} & -\sfrac{13}{179}

\end{pmatrix}
\begin{pmatrix}
\bar{\mu}_{u-d} \\ \bar{\mu}_e \\ \bar{\mu}_{\Delta_e} \\ \bar{\mu}_{\Delta_\mu} \\\bar{\mu}_{\Delta_\tau}
\end{pmatrix} \,.
\label{eq:eqIV}
\end{equation}


\paragraph{(v) \boldmath{$T \in \left(10^5, 10^6\right)\,\textrm{GeV}$}:}
Finally, at $T \sim 10^6\,\textrm{GeV}$, the Yukawa interactions of the first quark generation equilibrate, including the down-quark Yukawa interaction, but also the Yukawa interaction of right-handed down quarks with left-handed strange quarks.
The former violates the global charge accounted for by $\bar{\mu}_{u-d}$, which means that the SM chemical potentials now live in a vector space spanned by a four-dimensional basis.
Working with constraint equations for the last five charges in Eq.~\eqref{eq:charges} and eleven equilibrium conditions, we find
\begin{equation}
\begin{pmatrix}
\mu_e \\ \mu_\mu \\ \mu_\tau \\ \mu_{\ell_e} \\ \mu_{\ell_\mu} \\ \mu_{\ell_\tau} \\ \mu_u \\ \mu_c \\ \mu_t \\ \mu_d \\ \mu_s \\ \mu_b \\ \mu_{Q_1} \\ \mu_{Q_2} \\ \mu_{Q_3} \\ \mu_\Phi
\end{pmatrix}
=
\begin{pmatrix}
1 & 0 & 0 & 0 \\
\sfrac{7}{481} & \sfrac{1}{13} & -\sfrac{29}{111} & \sfrac{8}{111} \\
\sfrac{7}{481} & \sfrac{1}{13} & \sfrac{8}{111} & -\sfrac{29}{111} \\
-\sfrac{415}{962} & -\sfrac{11}{26} & \sfrac{2}{37} & \sfrac{2}{37} \\
\sfrac{59}{962} & \sfrac{1}{26} & -\sfrac{35}{111} & \sfrac{2}{111} \\
\sfrac{59}{962} & \sfrac{1}{26} & \sfrac{2}{111} & -\sfrac{35}{111} \\
\sfrac{3}{37} & 0 & -\sfrac{1}{37} & -\sfrac{1}{37} \\
\sfrac{3}{37} & 0 & -\sfrac{1}{37} & -\sfrac{1}{37} \\
\sfrac{3}{37} & 0 & -\sfrac{1}{37} & -\sfrac{1}{37} \\
-\sfrac{6}{481} & \sfrac{1}{13} & \sfrac{3}{37} & \sfrac{3}{37} \\
-\sfrac{6}{481} & \sfrac{1}{13} & \sfrac{3}{37} & \sfrac{3}{37} \\
-\sfrac{6}{481} & \sfrac{1}{13} & \sfrac{3}{37} & \sfrac{3}{37} \\
\sfrac{33}{962} & \sfrac{1}{26} & \sfrac{1}{37} & \sfrac{1}{37} \\
\sfrac{33}{962} & \sfrac{1}{26} & \sfrac{1}{37} & \sfrac{1}{37} \\
\sfrac{33}{962} & \sfrac{1}{26} & \sfrac{1}{37} & \sfrac{1}{37} \\
\sfrac{45}{962} & -\sfrac{1}{26} & -\sfrac{2}{37} & -\sfrac{2}{37}
\end{pmatrix}
\begin{pmatrix}
\bar{\mu}_e \\ \bar{\mu}_{\Delta_e} \\ \bar{\mu}_{\Delta_\mu} \\\bar{\mu}_{\Delta_\tau}
\end{pmatrix} \,.
\label{eq:eqV}
\end{equation}


In our benchmark scenario, wash-in leptogenesis is supposed to occur in this last temperature regime.
The equilibrium solution in Eq.~\eqref{eq:eqV} therefore represents the initial conditions for wash-in leptogenesis, \textit{i.e.}, the chemical configuration of the SM thermal bath that it will act upon. 
Compared to the situation right after the end of axion inflation, the chemical equilibrium in Eq.~\eqref{eq:eqV} now features reduced baryon and lepton asymmetries in consequence of the continuous $B+L$ wash-out by weak sphalerons since reheating.
Combining Eq.~\eqref{eq:eqV} with the input data in Eq.~\eqref{eq:muini}, we find $\eta_B = \eta_L = -198/481\,\eta_\chi$, which indicates sphaleron-induced shifts in the global baryon and lepton charges of $1047/962\,\eta_\chi$, respectively [see panel (b) in Fig.~\ref{fig:bars}]. 
This observation concludes our detailed discussion of the chemical transport in the SM plasma from $T \sim 10^{14}\,\textrm{GeV}$ all the way down to $T \sim 10^{5\cdots6}\,\textrm{GeV}$ and sets the stage for our analysis of wash-in leptogenesis.


\subsection{Wash-in leptogenesis}
\label{subsec:washin}


Let us now add RHN interactions to the picture, especially, the typical RHN wash-out processes known from standard thermal leptogenesis, which are dominated by RHN inverse decays, $\ell_\alpha \Phi \rightarrow N_i$ and $\bar{\ell}_\alpha \Phi^* \rightarrow N_i$, in our benchmark scenario.
At $T \sim M_i$, these interactions are equilibrated, unless we choose exceptionally small RHN Yukawa couplings, which allows us to impose three more conditions on the SM chemical potentials,
\begin{equation}
\label{eq:washin}
\mu_{\ell_\alpha} + \mu_\Phi = 0 \,,
\end{equation}
where the zero on the right-hand side stems from the fact that the three RHNs correspond to heavy Majorana fermions with zero chemical potential at the time of their decay.
Next, we rewrite the \textit{wash-in condition} in Eq.~\eqref{eq:washin} with the help of Eq.~\eqref{eq:eqV} as a condition for the global charges $\bar{\mu}_e$ and $\bar{\mu}_{\Delta_\alpha}$ in matrix form,
\begin{equation}
\label{eq:winmatrix}
\begin{pmatrix}
\mu_{\ell_e} + \mu_\Phi \\ \mu_{\ell_\mu} + \mu_\Phi \\ \mu_{\ell_\tau} + \mu_\Phi \end{pmatrix}
=
\begin{pmatrix}
-\sfrac{5}{13} \\ \sfrac{4}{37} \\ \sfrac{4}{37}
\end{pmatrix} \bar{\mu}_e - 
\begin{pmatrix}
\sfrac{6}{13} & 0 & \\
0 & \sfrac{41}{111} & \sfrac{4}{111} \\
0 & \sfrac{4}{111} & \sfrac{41}{111} 
\end{pmatrix}
\begin{pmatrix}
\bar{\mu}_{\Delta_e} \\ \bar{\mu}_{\Delta_\mu} \\\bar{\mu}_{\Delta_\tau}
\end{pmatrix} = \begin{pmatrix} 0 \\ 0 \\ 0 \end{pmatrix}\,.
\end{equation}
This condition is imposed by the interactions of each of the three RHN species independently, which reflects the fact that all three RHNs decay in the same temperature regime in our benchmark scenario.
Thanks to Eq.~\eqref{eq:winmatrix}, it is now straightforward to determine the new chemical equilibrium in the presence of efficient RHN interactions.
The \textit{lepton-number-violating} (LNV) RHN interactions drive the system to a new chemical attractor and thereby \textit{wash in} nonzero $\Delta_\alpha$ charges whose size is controlled by the primordial input charge $\bar{\mu}_e$,
\begin{equation}
\label{eq:Deltaemutau}
\begin{pmatrix}
\bar{\mu}_{\Delta_e} \\ \bar{\mu}_{\Delta_\mu} \\\bar{\mu}_{\Delta_\tau}
\end{pmatrix}
=
\begin{pmatrix}
\sfrac{6}{13} & 0 & \\
0 & \sfrac{41}{111} & \sfrac{4}{111} \\
0 & \sfrac{4}{111} & \sfrac{41}{111} 
\end{pmatrix}^{-1}
\begin{pmatrix}
-\sfrac{5}{13} \\ \sfrac{4}{37} \\ \sfrac{4}{37}
\end{pmatrix} \bar{\mu}_e = 
\begin{pmatrix}
\sfrac{13}{6} & 0 & \\
0 & \sfrac{41}{15} & -\sfrac{4}{15} \\
0 & -\sfrac{4}{15} & \sfrac{41}{15} 
\end{pmatrix}
\begin{pmatrix}
-\sfrac{5}{13} \\ \sfrac{4}{37} \\ \sfrac{4}{37}
\end{pmatrix} \bar{\mu}_e = 
\begin{pmatrix}
-\sfrac{5}{6} \\ \sfrac{4}{15} \\ \sfrac{4}{15}
\end{pmatrix} \bar{\mu}_e \,.
\end{equation}


The stage of wash-in leptogenesis hence results in the generation of a total $B\!-\!L$ charge-to-photon ratio
\begin{equation}
\label{eq:winbl}
\eta_{B-L} = \left(\frac{5}{6} - \frac{4}{15} - \frac{4}{15}\right) \eta_\chi = c_{B-L}^{\rm win} \eta_\chi \,,\qquad c_{B-L}^{\rm win} = \frac{3}{10} \,, 
\end{equation}
where we used that $\eta_e = -\eta_\chi$ [see Eq.~\eqref{eq:etaC}] and where we introduced the coefficient $c_{B-L}^{\rm win} = \eta_{B-L}/\eta_\chi$ to record the outcome of wash-in leptogenesis.
This expression for the wash-in contribution to the primordial $B\!-\!L$ asymmetry is our main result in this section.
In view of this result, several comments are in order:


(i) First of all, we note that Eq.~\eqref{eq:winbl} is completely independent of the details of $CP$ violation in the RHN sector.
$CP$ violation in RHN decays is typically quantified by $CP$ asymmetry parameters $\varepsilon_{i\alpha}$~\cite{Covi:1996wh}; these parameters, however, do not appear in our analysis and are in any case severely suppressed for RHN masses as low as a few 100 TeV (barring a resonant enhancement).
This observation is in accord with our discussion in Sec.~\ref{sec:introduction} and reflects the fact that wash-in leptogenesis allows us to separate the scales of $CP$ and $B\!-\!L$ violation.
The RHN interactions at low temperatures only serve the purpose to violate lepton number; the violation of $CP$ invariance is delegated to higher temperatures and accomplished by axion inflation.
At the time of wash-in leptogenesis, the $CP$-violating initial conditions set by axion inflation are then encoded in the remaining conserved global charges, \textit{i.e.}, the global right-handed electron number in our benchmark example.


(ii) The RHN mass scale in our benchmark scenario is determined by the requirement that at least one of the primordial input charges generated during axion inflation, $\bar{\mu}_e$, has not yet been altered by the chemical transport in the SM plasma.
This implies an absolute lower bound on the RHN mass scale that applies to any RHN mass spectrum in the context of wash-in leptogenesis:
At least one RHN species needs to decay and hence reshuffle the chemical potentials in the thermal bath at temperatures above the equilibration temperature of the electron Yukawa interaction,
\begin{equation}
\label{eq:M3bound}
M_3 \gtrsim T_{y_e} \sim 10^5\,\textrm{GeV} \,,
\end{equation}
provided that the LNV interactions of the other RHN species do not become efficient before sphaleron freeze-out.%
\footnote{In the remainder of this paper, though, we will mostly focus on RHN mass spectra with $M_{1,2,3} \gtrsim 10^5\,\textrm{GeV}$, assuming that all three RHN species reach chemical equilibrium at one point or other, in which case the bound in Eq.~\eqref{eq:M3bound} turns into $M_1 \gtrsim T_{y_e} \sim 10^5\,\textrm{GeV}$.}
Wash-in leptogenesis thus successfully operates at RHN masses as low as a few 100 TeV, which is four orders of magnitude below the typical mass range of standard thermal leptogenesis, $M_i \gtrsim 10^9\,\textrm{GeV}$~\cite{Davidson:2002qv,Buchmuller:2002rq}.
Moreover, it does not require small mass splittings in the RHN mass spectrum as in the case of resonant leptogenesis~\cite{Pilaftsis:1997jf,Pilaftsis:2003gt}.
A hierarchical spectrum as in our benchmark scenario is indeed a perfectly viable option.


(iii) In order to ensure a sufficient efficiency of wash-in leptogenesis, one only needs to assume that the standard RHN decay parameters $K_{i\alpha}= \Gamma_{i\alpha}\left(T=0\right)/H\left(T=M_i\right)$, which normally quantify the efficiency of wash-out processes, are large enough.
In other words, wash-in leptogenesis is particularly efficient in regions of parameter space that are otherwise characterized by a large asymmetry wash-out.
We refer to this parameter regime, otherwise known as the strong wash-out regime, as the \textit{strong wash-in regime}.
In the type-I seesaw model, light-neutrino masses of $\mathcal{O}\left(0.1\right)\,\textrm{eV}$ imply RHN decay parameters as large as $K_{i\alpha} \sim 10\cdots 100$ and hence naturally point to this regime (see, \textit{e.g.}, Ref.~\cite{Buchmuller:2004nz}).
As shown in Ref.~\cite{Domcke:2020quw}, the result in Eq.~\eqref{eq:winbl} is in this case in excellent agreement with the exact solution that one may obtain by explicitly solving the Boltzmann equations for the three flavored $B\!-\!L$ charges at the time of RHN decay.
In the strong wash-in regime, any deviations from Eq.~\eqref{eq:winbl} are exponentially suppressed by factors of the form $e^{-c_{i\alpha}K_{i\alpha}}$ with some coefficients $c_{i\alpha}$.
This situation changes in the weak wash-in regime, \textit{i.e.}, if we choose weaker RHN couplings, such that only a fraction of the totally available asymmetry is washed into the plasma, requiring one to resort to a description in terms of Boltzmann equations~\cite{Domcke:2020quw}.
In the following, we will, however, ignore this possibility and focus on the strong wash-in regime, which is well motivated by the low-energy neutrino data.


(iv) As discussed in Sec.~\ref{sec:helicity}, a necessary condition for successful wash-in leptogenesis is that the primordial fermion charges generated during axion inflation, alongside the helicity stored in the hypermagnetic field, are not erased by magnetic diffusion or the chiral plasma instability.
This requirement constraints the parameter space of axion inflation, as we will investigate in more detail in Secs.~\ref{subsec:independent} and \ref{subsec:estimates}.
Here, we merely remark that the chemical equilibrium in Eq.~\eqref{eq:eqV} now allows us to precisely calculate the chiral chemical potential $\bar{\mu}_5$ in Eq.~\eqref{eq:mu5def}, which controls the temperature scale of the chiral plasma instability, $T_{\rm CPI}$, in Eq.~\eqref{eq:TCPI},%
\footnote{
The chiral plasma instability terminates as soon as $\bar{\mu}_5 = 0$.
From Eq.~\eqref{eq:mu5_w_washin}, we can now read off that the flavored $B\!-\!L$ charges generated during wash-in leptogenesis, $\bar{\mu}_{\Delta_\alpha} \neq 0$, have a nontrivial impact on this condition.
If the chiral plasma instability occurs at $T_{y_e} \lesssim T_{\rm CPI} \lesssim T_{B-L}$, it will not result in $\bar{\mu}_e,q_{\rm CS} = 0$ as usual, but leave behind nonvanishing charge asymmetries and hypermagnetic helicity.
This observation generalizes to other baryogenesis mechanisms (\textit{e.g.}, leptoflavorgenesis~\cite{Mukaida:2021sgv}, which is characterized by $\sum_\alpha \bar{\mu}_{\Delta_\alpha} = 0$ and at least two $\bar{\mu}_{\Delta_\alpha} \neq 0$) and can source nonvanishing $\bar \mu_e$ and helicity, even if these two quantities are initially zero~\cite{Domcke:2022uue}.}
\begin{equation}
\frac{\bar{\mu}_5}{T} = \frac{711}{481}\,\frac{\bar{\mu}_e}{T} + \frac{5}{13}\,\frac{\bar{\mu}_{\Delta_e}}{T} - \frac{4}{37} \left(\frac{\bar{\mu}_{\Delta_\mu}}{T} + \frac{\bar{\mu}_{\Delta_\tau}}{T}\right) \,.
\label{eq:mu5_w_washin}
\end{equation}
As evident from this relation, the numerical value of $\bar{\mu}_e$ changes in consequence of wash-in leptogenesis, even though the set of equilibrated SM interactions remains the same across the interval $T \sim 10^5\,\textrm{GeV} \cdots 10^6\,\textrm{GeV}$.
Before wash-in leptogenesis, the three flavored $B\!-\!L$ charges vanish, while after wash-in leptogenesis, we need to work with the nonzero flavored $B\!-\!L$ charges in Eq.~\eqref{eq:Deltaemutau}.
Together with Eq.~\eqref{eq:muini}, we thus obtain
\begin{align}
\label{eq:c5larger}
\left.\frac{\bar{\mu}_5}{T}\right|_{T > T_{B-L}} & = -6\,c_5^>\,\chi \,,\qquad c_5^> = \frac{711}{481} \,, \\
\label{eq:c5smaller}
\left.\frac{\bar{\mu}_5}{T}\right|_{T < T_{B-L}} & = -6\,c_5^<\,\chi \,,\qquad c_5^< = \frac{11}{10} \,.
\end{align}


(v) Similarly, we can study the evolution of the global baryon charge across the stage of wash-in leptogenesis.
Using again the solution in Eq.~\eqref{eq:eqV}, we find the following relation among the relevant asymmetries,
\begin{equation}
\label{eq:etaBv}
\eta_B = \frac{6}{481}\left[33\,\eta_e + 37\,\eta_{\Delta_e} + 26 \left(\eta_{\Delta_\mu} + \eta_{\Delta_\tau}\right) \right] \,,
\end{equation}
which results in $\eta_B = -198/481\,\eta_\chi$ at $T > T_{B-L}$, as already stated at the end of Sec.~\ref{subsec:transport} [see panel (b) of Fig.~\ref{fig:bars}].
In the course of wash-in leptogenesis, the baryon asymmetry, however, evolves into $\eta_B = -1/5\,\eta_\chi$, corresponding to a shift of $509/2405\, \eta_\chi$.
Together with our result in Eq.~\eqref{eq:winbl}, $\eta_{B-L} = 3/10\,\eta_\chi$, this implies a lepton asymmetry $\eta_L = -1/2\,\eta_\chi$ as well as a baryon-plus-lepton asymmetry $\eta_{B+L} = -7/10\,\eta_\chi$ after wash-in leptogenesis.
This observation tells us that the violation of baryon-minus-lepton number by RHN interactions during wash-in leptogenesis is in fact accompanied by the violation of baryon-plus-lepton number by weak sphalerons:
While $\eta_{B-L}$ changes from $0$ to $3/10\,\eta_\chi$, $\eta_{B+L}$ receives a shift of $593/4810\,\eta_\chi$ [see panel (c) of Fig.~\ref{fig:bars}]. 


(vi) We stress that the baryon asymmetry in Eq.~\eqref{eq:etaBv} corresponds to the global baryon charge in the temperature regime $T \sim 10^5\,\textrm{GeV} \cdots 10^6\,\textrm{GeV}$.
In order to compute the baryon asymmetry in the present Universe, we still need to track the chemical transport in the SM plasma all the way down to the time of sphaleron decoupling at $T \simeq 135\,\textrm{GeV}$~\cite{DOnofrio:2014rug}.
To this end, we first determine the chemical equilibrium at temperatures below $T_{y_e}$ and above the EWPT, where \textit{all} SM interactions are equilibrated, including the electron Yukawa interaction.
As before, we can write down a system of linear equations, this time consisting of four charge constraints (for $\Delta_\alpha$ and $Y$) and twelve equilibrium conditions, which gives rise to the following equilibrium solution,
\begin{equation}
\begin{pmatrix}
\mu_e \\ \mu_\mu \\ \mu_\tau \\ \mu_{\ell_e} \\ \mu_{\ell_\mu} \\ \mu_{\ell_\tau} \\ \mu_u \\ \mu_c \\ \mu_t \\ \mu_d \\ \mu_s \\ \mu_b \\ \mu_{Q_1} \\ \mu_{Q_2} \\ \mu_{Q_3} \\ \mu_\Phi
\end{pmatrix}
=
\begin{pmatrix}
-\sfrac{185}{711} & \sfrac{52}{711} & \sfrac{52}{711} \\
\sfrac{52}{711} & -\sfrac{185}{711} & \sfrac{52}{711} \\
\sfrac{52}{711} & \sfrac{52}{711} & -\sfrac{185}{711} \\
-\sfrac{221}{711} & \sfrac{16}{711} & \sfrac{16}{711} \\
\sfrac{16}{711} & -\sfrac{221}{711} & \sfrac{16}{711} \\
\sfrac{16}{711} & \sfrac{16}{711} & -\sfrac{211}{711} \\
-\sfrac{5}{237} & -\sfrac{5}{237} & -\sfrac{5}{237} \\
-\sfrac{5}{237} & -\sfrac{5}{237} & -\sfrac{5}{237} \\
-\sfrac{5}{237} & -\sfrac{5}{237} & -\sfrac{5}{237} \\
\sfrac{19}{237} & \sfrac{19}{237} & \sfrac{19}{237} \\
\sfrac{19}{237} & \sfrac{19}{237} & \sfrac{19}{237} \\
\sfrac{19}{237} & \sfrac{19}{237} & \sfrac{19}{237} \\
\sfrac{7}{237} & \sfrac{7}{237} & \sfrac{7}{237} \\
\sfrac{7}{237} & \sfrac{7}{237} & \sfrac{7}{237} \\
\sfrac{7}{237} & \sfrac{7}{237} & \sfrac{7}{237} \\
-\sfrac{4}{79} & -\sfrac{4}{79} & -\sfrac{4}{79} 
\end{pmatrix}
\begin{pmatrix}
\bar{\mu}_{\Delta_e} \\ \bar{\mu}_{\Delta_\mu} \\\bar{\mu}_{\Delta_\tau}
\end{pmatrix} \,.
\label{eq:eqVI}
\end{equation}
This solution coincides with the chemical equilibrium shortly above the EWPT in standard leptogenesis scenarios.
In particular, it implies the well-known relation between the global $B\!-\!L$ and $B$ charge asymmetries
\begin{equation}
\eta_B = c_{\rm sph} \left(\eta_{\Delta_e} + \eta_{\Delta_\mu} + \eta_{\Delta_\tau}\right) = c_{\rm sph}\,\eta_{B-L} \,,\qquad c_{\rm sph} = \frac{28}{79} \,,
\end{equation}
where $c_{\rm sph}$ is referred to as the sphaleron conversion factor (in the symmetric phase of the SM plasma).
In combination with the outcome of wash-in leptogenesis, $\eta_{B-L} = 3/10\,\eta_\chi$, we therefore obtain
\begin{equation}
\eta_B = c_{\rm sph}\,c_{B-L}^{\rm win}\,\eta_\chi = \frac{42}{395}\eta_\chi \,, \qquad \eta_L = \left(c_{\rm sph}-1\right) c_{B-L}^{\rm win}\,\eta_\chi = -\frac{153}{790}\eta_\chi \,,
\end{equation}
or equivalently,
\begin{equation}
\eta_{B-L} = c_{B-L}^{\rm win}\,\eta_\chi = \frac{3}{10}\eta_\chi \,, \qquad \eta_{B+L} = \left(2\,c_{\rm sph}-1\right) c_{B-L}^{\rm win}\,\eta_\chi = -\frac{69}{790}\eta_\chi \,.
\end{equation}
Compared to the situation in the previous temperature regime, $T \sim 10^5\,\textrm{GeV} \cdots 10^6\,\textrm{GeV}$, the global $B\!-\!L$ charge hence remains unchanged, while the global $B\!+\!L$ charge receives a shift of $242/395\,\eta_\chi$ [see panel (d) of Fig.~\ref{fig:bars}].


Finally, we need to account for the evolution of the SM chemical potentials through the EWPT, especially, from the onset of the phase transition at $T \simeq 160\,\textrm{GeV}$~\cite{DOnofrio:2015gop}, when the Higgs field begins to develop a nonzero vacuum expectation value, to the freeze-out of the weak sphaleron processes at $T \simeq 135\,\textrm{GeV}$~\cite{DOnofrio:2014rug}.
The chemical transport in this temperature regime, \textit{i.e.}, the electroweak-broken phase, differs in several aspects from the chemical transport in the electroweak-symmetric phase~\cite{Harvey:1990qw}:
Weak isospin symmetry is spontaneously broken, the global hypercharge is no longer conserved, and the chemical potential of the Higgs boson vanishes.
Instead, the global electric charge is conserved ($\bar{\mu}_Q = 0$), and the $W$ boson is allowed to pick up a nonzero chemical potential.
The vanishing chemical potential of the Higgs boson implies in particular that we no longer need to distinguish between the chemical potentials of left- and right-handed Weyl fermions.
Instead, all fermions except for the three SM neutrino species correspond to massive Dirac fermions whose left- and right-handed components possess the same chemical potential. 
Meanwhile, the three SM neutrino species can still be treated as massless left-handed Weyl fermions; neutrino oscillations caused by their tiny Majorana masses will only begin to affect their chemical potentials at much later times~\cite{Dolgov:2002ab,Wong:2002fa}.
Accounting for these aspects of the electroweak-broken phase, we are able to write down a system of linear equations, consisting of four charge constraints (for $\Delta_\alpha$ and $Q$) and nine equilibrium conditions, that is solved by 
\begin{equation}
\begin{pmatrix}
\mu_e \\ \mu_\mu \\ \mu_\tau \\ \mu_{\nu_e} \\ \mu_{\nu_\mu} \\ \mu_{\nu_\tau} \\ \mu_u \\ \mu_c \\ \mu_t \\ \mu_d \\ \mu_s \\ \mu_b \\ \mu_{W^+}
\end{pmatrix}
=
\begin{pmatrix}
-\sfrac{95}{333} & \sfrac{16}{333} & \sfrac{16}{333} \\
\sfrac{16}{333} & -\sfrac{95}{333} & \sfrac{16}{333} \\
\sfrac{16}{333} & \sfrac{16}{333} & -\sfrac{95}{333} \\
-\sfrac{107}{333} & \sfrac{4}{333} & \sfrac{4}{333} \\
\sfrac{4}{333} & -\sfrac{107}{333} & \sfrac{4}{333} \\
\sfrac{4}{333} & \sfrac{4}{333} & -\sfrac{107}{333} \\
\sfrac{1}{111} & \sfrac{1}{111} & \sfrac{1}{111} \\
\sfrac{1}{111} & \sfrac{1}{111} & \sfrac{1}{111} \\
\sfrac{1}{111} & \sfrac{1}{111} & \sfrac{1}{111} \\
\sfrac{5}{111} & \sfrac{5}{111} & \sfrac{5}{111} \\
\sfrac{5}{111} & \sfrac{5}{111} & \sfrac{5}{111} \\
\sfrac{5}{111} & \sfrac{5}{111} & \sfrac{5}{111} \\
-\sfrac{4}{111} & -\sfrac{4}{111} & -\sfrac{4}{111} 
\end{pmatrix}
\begin{pmatrix}
\bar{\mu}_{\Delta_e} \\ \bar{\mu}_{\Delta_\mu} \\\bar{\mu}_{\Delta_\tau}
\end{pmatrix} \,,
\label{eq:eqVII}
\end{equation}
where $\mu_e$, $\mu_\mu$, $\mu_\tau$, $\mu_u$, $\mu_c$, $\mu_t$, $\mu_d$, $\mu_s$, and $\mu_b$ now denote the chemical potentials of the corresponding Dirac fermions.
This chemical equilibrium implies the following relation between the global $B\!-\!L$ and $B$ charges,
\begin{equation}
\label{eq:etaBBL}
\eta_B = \bar{c}_{\rm sph} \left(\eta_{\Delta_e} + \eta_{\Delta_\mu} + \eta_{\Delta_\tau}\right) = \bar{c}_{\rm sph}\,\eta_{B-L} \,,\qquad \bar{c}_{\rm sph} = \frac{12}{37} \,,
\end{equation}
where $\bar{c}_{\rm sph}$ represents the sphaleron conversion factor in the broken phase of the SM plasma~\cite{Harvey:1990qw,Laine:1999wv}.


The baryon and lepton asymmetries originating from wash-in leptogenesis hence freeze out at
\begin{equation}
\label{eq:etaBetachi}
\eta_B = \bar{c}_{\rm sph}\,c_{B-L}^{\rm win}\,\eta_\chi = \frac{18}{185}\eta_\chi \,, \qquad \eta_L = \left(\bar{c}_{\rm sph}-1\right) c_{B-L}^{\rm win}\,\eta_\chi = -\frac{15}{74}\eta_\chi \,,
\end{equation}
or equivalently,
\begin{equation}
\label{eq:etaBLetachi}
\eta_{B-L} = c_{B-L}^{\rm win}\,\eta_\chi = \frac{3}{10}\eta_\chi \,, \qquad \eta_{B+L} = \left(2\,\bar{c}_{\rm sph}-1\right) c_{B-L}^{\rm win}\,\eta_\chi = -\frac{39}{370}\eta_\chi \,,
\end{equation}
which is displayed in panel (e) of Fig.~\ref{fig:bars}.
After the EWPT, the baryon asymmetry no longer evolves, apart from the trivial dilution of the baryon-to-photon ratio caused by the decreasing number of relativistic degrees of freedom in the thermal bath.
The wash-in contribution to the present-day baryon asymmetry thus reads
\begin{equation}
\label{eq:etaBwin}
\eta_B^{\rm win} = \frac{g_{*,s}\left(T_0\right)}{g_{*,s}\left(T_{B-L}\right)}\,c_B^{\rm win}\,\eta_\chi = \frac{g_{*,s}\left(T_0\right)}{g_{*,s}\left(T_{B-L}\right)}\,\frac{\pi^2}{\zeta\left(3\right)g_\gamma}c_B^{\rm win}\,\chi  \,, \qquad c_B^{\rm win} = \bar{c}_{\rm sph}\,c_{B-L}^{\rm win} = \frac{18}{185} \,,
\end{equation}
where we used Eq.~\eqref{eq:etachi} in the second step, and where $g_{*,s}\left(T_0\right) = 43/11$ and $g_{*,s}\left(T_{B-L}\right) = 427/4$ denote the effective numbers of entropic degrees of freedom today and at the time of wash-in leptogenesis, respectively.
This expression is the final result of our discussion of wash-in leptogenesis in this section.
We conclude that, in order to obtain a baryon asymmetry of the right order of magnitude, $\eta_B \sim 10^{-(9\cdots10)}$, the $CP$ asymmetry parameter $\chi$ needs to be of the order of $\chi \sim 10^{-(7\cdots8)}$, which is exactly what we anticipated in Fig.~\ref{fig:chi},
\begin{equation}
\label{eq:etaBwin2}
\eta_B^{\rm win} \simeq 0.15\,c_B^{\rm win}\,\chi \simeq 4.4 \times 10^{-10}\:\Big(\frac{c_B^{\rm win}}{18/185}\Big)\left(\frac{\chi}{3 \times 10^{-8}}\right) \,.
\end{equation}


\subsection{Baryogenesis from helicity decay}
\label{subsec:decay}


In the context of our benchmark scenario in this section, we shall assume that the helicity stored in the hypermagnetic field survives and the field itself remains fully helical all the way down to the EWPT.%
\footnote{If the hypermagnetic field is only partially helical, its coherence length and field strength at a fixed value of the total helicity are larger than in the fully helical case. 
This results in large baryon isocurvature perturbations and hence severe constraints from inhomogeneous big-bang nucleosynthesis~\cite{Kamada:2020bmb}.
In the parameter space of our interest, the hypermagnetic field is, however, in the turbulent regime and its coherence length is sufficiently short so that no constraint from baryon isocurvature perturbations arises.}
The necessary conditions for this to happen were outlined in Sec.~\ref{sec:helicity}; in Sec.~\ref{subsec:estimates}, we will identify the viable region in the parameter space of axion inflation where these conditions are indeed satisfied.
For the time being, let us, however, solely focus on the implications for the BAU if the primordial helicity of the gauge field is not erased before the EWPT.
The crucial observation in this case is that the decay of the hypermagnetic helicity during the EWPT will result in another contribution to the baryon asymmetry in accord with the chiral anomaly of the baryon-number current in Eq.~\eqref{eq:JBL} and the relation among the involved global charges in Eq.~\eqref{eq:DeltaBL}.
%
%
We refer to this mechanism as baryogenesis from helicity decay, which has been first proposed in Refs.~\cite{Giovannini:1997gp,Giovannini:1997eg} and then studied in more detail in Refs.~\cite{Fujita:2016igl,Kamada:2016eeb,Kamada:2016cnb} as well as in Refs.~\cite{Anber:2015yca,Jimenez:2017cdr,Domcke:2019mnd} in relation to axion inflation.


The helicity stored in the hypermagnetic field decreases for two reasons.
On the one hand, it decays in consequence of Ohmic dissipation because of the finite conductivity of the SM plasma~\cite{Giovannini:1997gp}.
On the other hand, it is driven to zero by electroweak symmetry breaking, which rotates the physical vector fields in the electroweak sector in a way such that the Abelian contribution to the chiral anomalies of the global $B$ and $L$ currents is removed.
In other words, baryon and lepton number are anomalously violated by the hypercharge gauge field, but they are preserved in electromagnetic interactions.
This change on the right-hand side of the anomaly equation~\eqref{eq:JBL} is reflected in a corresponding change on the left-hand side, \textit{i.e.}, the generation of an additional contribution to the baryon asymmetry.
The effect of electroweak symmetry breaking on the hypermagnetic helicity can be parametrized in terms of the temperature-dependent weak mixing angle $\theta_{\rm w}\left(T\right)$, which determines the unscreened mode among the electroweak gauge fields.
For the electroweak crossover found in the SM for a $125\,\textrm{GeV}$ Higgs boson, the angle $\theta_{\rm w}\left(T\right)$ vanishes before the onset of the EWPT and smoothly evolves to its low-temperature value during the EWPT~\cite{Kajantie:1996qd,DOnofrio:2015gop}.  


Around the time of sphaleron freeze-out at $T \simeq 135\,\textrm{GeV}$, the temperature dependence of the weak mixing angle represents the dominant effect on the evolution of the hypermagnetic helicity, which allows us to neglect Ohmic dissipation in our estimate of the baryon asymmetry.
In this approximation, the transport equation for the charge density $q_B$ in the broken phase obtains the following form~\cite{Kamada:2016cnb},
\begin{equation}
\label{eq:transport}
\left(\frac{d}{dt} + 3H\right)q_B = \frac{111}{34}\,\Gamma_{\rm ws}\left[c_B^{\rm dec}\,\chi\,T^3 - \left(q_B-\bar{c}_{\rm sph}\,q_{B-L}\right)\right] \,.
\end{equation}
Here, the first term inside the brackets represents the source term originating from the decaying hypermagnetic helicity, while the second term is the standard sphaleron term that is responsible for the conversion from $B\!-\!L$ to $B$.
The rate $\Gamma_{\rm ws}$ correspondingly denotes the rate of weak sphaleron processes~\cite{DOnofrio:2014rug},
\begin{equation}
\label{eq:gammaws}
\Gamma_{\rm ws} \simeq T\,\exp\left[-146.6 + 0.83 \left(\frac{T}{1\,\textrm{GeV}}\right)\right] \,,
\end{equation}
which is valid in the broken phase, $T \lesssim 160\,\textrm{GeV}$.
The coefficient $111/34$ in Eq.~\eqref{eq:transport} has also be found in Ref.~\cite{Kamada:2016cnb} and follows from the chemical equilibrium in the broken phase in the limit of slow sphaleron processes.
To see this, note that the sphaleron conversion term in Eq.~\eqref{eq:transport} must be proportional to the sum of the chemical potentials of the fermion fields that belong to $SU(2)_L$ doublets in the symmetric phase, 
\begin{equation}
\label{eq:sigma}
\Sigma = 3 \left[3\,\frac{\mu_u + \mu_d}{2} + 3\,\frac{\mu_c + \mu_s}{2} + 3\,\frac{\mu_t + \mu_b}{2} + \frac{\mu_e + \mu_{\nu_e}}{2} + \frac{\mu_\mu + \mu_{\nu_\mu}}{2} + \frac{\mu_\tau + \mu_{\nu_\tau}}{2}\right] \,.
\end{equation}
Here, the overall factor of $3$ counts the units of baryon charge generated per sphaleron transition; the factors of $3$ in front of the quark chemical potentials count color degrees of freedom; and the chemical potentials of species whose left-handed components originally belonged to the same $SU(2)_L$ doublets in the symmetric phase are averaged over.
This last factor of $1/2$ may also be regarded as an overall normalization that ensures a smooth matching with the corresponding linear combination of chemical potentials in the symmetric phase.
In the next step, we need to evaluate $\Sigma$ as a function of conserved and slowly violated charges, including baryon number, which is conserved by all SM processes except for the weak sphalerons.
To do so, we need to modify the result in Eq.~\eqref{eq:eqVII}, which describes the chemical equilibrium in the broken phase assuming fast sphalerons.
If we drop this assumption and treat the sphalerons as slow, we obtain a system of linear equations\,---\,five charge constraints (for $\Delta_\alpha$, $Q$, and $B$) and eight equilibrium conditions\,---\,that is solved by
\begin{equation}
\begin{pmatrix}
\mu_e \\ \mu_\mu \\ \mu_\tau \\ \mu_{\nu_e} \\ \mu_{\nu_\mu} \\ \mu_{\nu_\tau} \\ \mu_u \\ \mu_c \\ \mu_t \\ \mu_d \\ \mu_s \\ \mu_b \\ \mu_{W^+}
\end{pmatrix}
=
\begin{pmatrix}
-\sfrac{49}{153} & \sfrac{2}{153} & \sfrac{2}{153} & \sfrac{11}{102} \\
\sfrac{2}{153} & -\sfrac{49}{153} & \sfrac{2}{153} & \sfrac{11}{102} \\
\sfrac{2}{153} & \sfrac{2}{153} & -\sfrac{49}{153} & \sfrac{11}{102} \\
-\sfrac{55}{153} & -\sfrac{4}{153} & -\sfrac{4}{153} & \sfrac{2}{17} \\
-\sfrac{4}{153} & -\sfrac{55}{153} & -\sfrac{4}{153} & \sfrac{2}{17} \\
-\sfrac{4}{153} & -\sfrac{4}{153} & -\sfrac{55}{153} & \sfrac{2}{17} \\
-\sfrac{1}{51} & -\sfrac{1}{51} & -\sfrac{1}{51} & \sfrac{3}{34} \\
-\sfrac{1}{51} & -\sfrac{1}{51} & -\sfrac{1}{51} & \sfrac{3}{34} \\
-\sfrac{1}{51} & -\sfrac{1}{51} & -\sfrac{1}{51} & \sfrac{3}{34} \\
\sfrac{1}{51} & \sfrac{1}{51} & \sfrac{1}{51} & \sfrac{4}{51} \\
\sfrac{1}{51} & \sfrac{1}{51} & \sfrac{1}{51} & \sfrac{4}{51} \\
\sfrac{1}{51} & \sfrac{1}{51} & \sfrac{1}{51} & \sfrac{4}{51} \\
-\sfrac{2}{51} & -\sfrac{2}{51} & -\sfrac{2}{51} & \sfrac{1}{102}
\end{pmatrix}
\begin{pmatrix}
\bar{\mu}_{\Delta_e} \\ \bar{\mu}_{\Delta_\mu} \\\bar{\mu}_{\Delta_\tau} \\\bar{\mu}_B
\end{pmatrix} \,.
\label{eq:eqVIII}
\end{equation}
Evaluating the linear combination $\Sigma$ in Eq.~\eqref{eq:sigma} in this chemical equilibrium then gives in the desired result,
\begin{equation}
\Sigma = \frac{111}{34}\left(\bar{\mu}_B-\bar{c}_{\rm sph}\,\bar{\mu}_{B-L}\right) \,.
\end{equation}


The coefficient $c_B^{\rm sph}$ in Eq.~\eqref{eq:transport}, finally, accounts for the evolution of the weak mixing angle $\theta_{\rm w}\left(T\right)$~\cite{Kamada:2016cnb},
\begin{equation}
c_B^{\rm dec} = \frac{17}{37} \left(1 + \frac{g_L^2}{g_Y^2}\right) \frac{H\,\Theta}{\Gamma_{\rm ws}}\,, \qquad \Theta =  -T\,\frac{d \theta_{\rm w}}{dT}\,\sin\left(2\theta_{\rm w}\right) \,,
\end{equation}
and hence parametrizes the source term in units of $\chi T^3$.
Here, the factor $17/37$ simply ensures that the source term enters the transport equation with an overall numerical prefactor of $17/37 \times 111/34 = 3/2$~\cite{Kamada:2016cnb}.
In order to estimate the temperature dependence of $\theta_{\rm w}\left(T\right)$, we will use the analytical one-loop result in Ref.~\cite{Kajantie:1996qd}, which is based on the dimensionally reduced effective thermal field theory description of the Standard Model,
\begin{equation}
\label{eq:thetaw}
\cos^2 \theta_{\rm w}\left(T\right) = \cos^2 \theta_{\rm w}\left(T=0\right) \left[1 + \frac{11}{6\pi}\,\sin^2 \theta_{\rm w}\left(T=0\right)\,\frac{g_L T}{v\left(T\right)}\right] \,.
\end{equation}
Here, $v\left(T\right)$ denotes the temperature-dependent Higgs vacuum expectation value, $\left<\Phi^\dagger\Phi\right> = v^2\left(T\right)/2$ during the EWPT, which has been studied in a numerical lattice simulation in Ref.~\cite{DOnofrio:2015gop}.
A simple fit formula describing these numerical lattice results, valid at $130\,\textrm{GeV} \lesssim T \lesssim 160\,\textrm{GeV}$, has been worked out in Ref.~\cite{Kamada:2016eeb},
\begin{equation}
v\left(T\right) \approx 0.23\,T \sqrt{162 - \frac{T}{1\,\textrm{GeV}}} \,.
\end{equation}
This expression, together with $g_L \simeq 0.64$, $g_Y \simeq 0.35$, $\cos^2 \theta_{\rm w}\left(T=0\right) \simeq 0.78$ and $\sin^2 \theta_{\rm w}\left(T=0\right) \simeq 0.22$, then allows us to evaluate $c_B^{\rm dec}$ at the time of sphaleron freeze-out, when the final value of the BAU is determined,%
\footnote{In doing so, we take into account the one-loop running of $g_L$ and $g_Y$ from the $Z$ pole to energy scale of sphaleron freeze-out.}
\begin{equation}
T_{\rm sph} = 135\,\textrm{GeV} \,: \qquad c_B^{\rm dec} \simeq 0.05 \left(\frac{H/\Gamma_{\rm ws}}{0.19}\right)\left(\frac{\Theta}{0.14}\right) \,.
\end{equation}


For definiteness, we will work with $c_B^{\rm dec} \simeq 0.05$ in the following.
We, however, caution that this estimate comes with a considerable numerical uncertainty of up to three orders of magnitude.
On the one hand, the ratio $H/\Gamma_{\rm ws}$ is exponentially sensitive to the chosen freeze-out temperature, simply because $\Gamma_{\rm ws}$ decreases exponentially fast during the EWPT [see Eq.~\eqref{eq:gammaws}].
On the other hand, the accuracy of the one-loop result in Eq.~\eqref{eq:thetaw} is limited, which is, \textit{e.g.}, reflected in the fact that it only roughly agrees with the numerical lattice results in Ref.~\cite{DOnofrio:2015gop} (for more details, see also the discussion in Refs.~\cite{Kamada:2016cnb,Kamada:2020bmb}).
A more accurate determination of the temperature dependence of the weak mixing angle during the EWPT is, however, not available at present, which is why we will content ourselves with $c_B^{\rm dec} \simeq 0.05$ as a representative benchmark value in our analysis.


In order to solve the transport equation \eqref{eq:transport}, we can make use of the fact that it is linear in the baryon charge density.
This allows us to split the total charge density into two contributions, $q_B = q_B^{\rm dec} + q_B^{\rm lep}$, where $q_B^{\rm dec}$ denotes the outcome of baryogenesis from helicity decay and $q_B^{\rm lep}$ is the combined contribution to the baryon asymmetry from wash-in and standard thermal leptogenesis.
These partial asymmetries then satisfy
\begin{equation}
\left(\frac{d}{dt} + 3H\right)q_B^{\rm dec} = \frac{111}{34}\,\Gamma_{\rm ws}\left(c_B^{\rm dec}\,\chi\,T^3 - q_B^{\rm dec}\right) \,, \qquad \left(\frac{d}{dt} + 3H\right)q_B^{\rm lep} = \frac{111}{34}\,\Gamma_{\rm ws}\left(\bar{c}_{\rm sph}\,q_{B-L}-q_B^{\rm lep}\right) \,,
\end{equation}
which immediately allows us to read off their freeze-out values at the time of sphaleron decoupling.
The contribution originating from leptogenesis clearly agrees with the result in Eq.~\eqref{eq:etaBBL}, \textit{i.e.}, it simply follows from the standard sphaleron conversion formula, $q_B^{\rm lep} = \bar{c}_{\rm sph}\,q_{B-L}$, while the baryogenesis contribution from helicity decay is directly determined by the dimensionless helicity density $\chi$ produced during axion inflation,
\begin{equation}
q_B^{\rm dec} = c_B^{\rm dec}\,\chi\,T^3 \,.
\end{equation}
In terms of charge-to-photon ratios, we thus obtain the following additional asymmetries from helicity decay,
\begin{equation}
\eta_B = c_B^{\rm dec}\,\eta_\chi \,, \qquad \eta_L = c_B^{\rm dec}\,\eta_\chi \,, \qquad \eta_{B-L} = 0 \,, \qquad \eta_{B+L} = 2c_B^{\rm dec}\,\eta_\chi \,,
\end{equation}
where we used the fact that baryogenesis from helicity decay preserves baryon-minus-lepton number.


In addition to the extra baryon charge density $q_B^{\rm dec}$, we therefore also obtain an extra contribution to the lepton charge density of equal size, $q_L^{\rm dec} = q_B^{\rm dec}$.
At the same time, the hypermagnetic field is transformed into the usual magnetic field of electromagnetism, which means that no hypermagnetic helicity that would be capable of sourcing a baryon asymmetry survives after the EWPT. 
Meanwhile, the magnetic field emanating from the EWPT does remain helical.
But instead of \textit{hypermagnetic helicity}, it now features \textit{magnetic helicity}, which is of phenomenological interest in its own right,%
\footnote{Similar to hypermagnetic fields, maximally helical magnetic fields experience an ``inverse cascade'' evolution in the turbulent regime (see also Sec.~\ref{sec:helicity}), which can lead to strong present-day magnetic fields with large coherence length in cosmic voids~\cite{Brandenburg:1996fc,Banerjee:2004df,Brandenburg:2016odr}.
The search for and a possible hint of helical intergalactic magnetic fields, based on the parity-odd correlation functions of diffuse gamma rays emitted by blazars, is discussed, \textit{e.g.}, in Refs.~\cite{Tashiro:2013bxa,Tashiro:2013ita,Tashiro:2014gfa,Chen:2014qva} (see also Ref.~\cite{MAGIC:2022piy}).
The $\eta_\chi$ values required for successful baryogenesis after axion inflation, however, lead to magnetic fields that are not strong enough to explain these blazar observations~\cite{Jimenez:2017cdr,Domcke:2019mnd}.}
even if it is no longer relevant for the evolution of the baryon asymmetry.
In panel (e) of Fig.~\ref{fig:bars}, we indicate this behavior of the helicity density by setting the CS charge-to-photon ratio $\eta_{\rm CS}$ to zero towards the end of the EWPT, which completes our discussion of Fig.~\ref{fig:bars}.


The only remaining step, as far as the discussion of our benchmark scenario in this section is concerned, thus consists in relating $q_B^{\rm dec}$ to the present-day baryon asymmetry.
In analogy to Eq.~\eqref{eq:etaBwin}, we have
\begin{equation}
\eta_B^{\rm dec} = \frac{g_{*,s}\left(T_0\right)}{g_{*,s}\left(T_{\rm sph}\right)}\,c_B^{\rm dec}\,\eta_\chi = \frac{g_{*,s}\left(T_0\right)}{g_{*,s}\left(T_{\rm sph}\right)}\,\frac{\pi^2}{\zeta\left(3\right)g_\gamma}c_B^{\rm dec}\,\chi \,,
\end{equation}
which happens to be of the same order of magnitude as the wash-in contribution in Eq.~\eqref{eq:etaBwin2},
\begin{equation}
\eta_B^{\rm dec} \simeq 0.15\,c_B^{\rm dec}\,\chi \simeq 2.3 \times 10^{-10}\:\Big(\frac{c_B^{\rm dec}}{0.05}\Big)\left(\frac{\chi}{3 \times 10^{-8}}\right) \,.
\end{equation}
We therefore conclude that it is indeed necessary to account for both contributions to the final baryon asymmetry.
This statement applies in particular to earlier studies of baryogenesis from helicity decay, whose outcome can receive important corrections as soon as RHNs are added to the theory.
In summary, we find that wash-in leptogenesis after axion inflation in combination with baryogenesis from helicity decay leads to 
\begin{equation}
\label{eq:etaBtot}
\eta_B^{\rm tot} = \eta_B^{\rm win} + \eta_B^{\rm dec} \simeq 0.15\left(c_B^{\rm win} + c_B^{\rm dec}\right)\chi \simeq 6.6 \times 10^{-10}\:\Big(\frac{c_B^{\rm win} + c_B^{\rm dec}}{18/185 + 0.05}\Big)\left(\frac{\chi}{3 \times 10^{-8}}\right) \,,
\end{equation}
which is one of our main results in this paper.
As expected the final asymmetry is controlled by the dimensionless helicity density $\chi$, which quantifies the amount of $CP$ violation during axion inflation and which needs to take a value of around $\chi \sim 10^{-(7\cdots8)}$ in order to set the stage for successful baryogenesis.


\section{Range of viable scenarios}
\label{sec:scenarios}


Our extensive discussion of the specific benchmark scenario in the previous section now enables us to readily generalize our analysis and map out the full range of viable scenarios without much additional effort.
To this end, we will first extend our investigation of wash-in leptogenesis to larger RHN masses, $M_i \gg 10^5\,\textrm{GeV}$, and correspondingly larger leptogenesis temperature scales $T_{B-L}$ (see Sec.~\ref{subsec:regimes}).
In a second step, we will then combine all of our results obtained in Secs.~\ref{sec:helicity}, \ref{sec:bau}, and \ref{subsec:regimes} and identify the viable region in parameter space that allows to produce the BAU while avoiding the constraints from magnetic diffusion and the chiral plasma instability.
First, we will do this in a slightly more model-independent way in Sec.~\ref{subsec:independent}, which will provide us with general results that can also be applied to alternative mechanisms of primordial magnetogenesis beyond our scenario of axion inflation.
Then, in Sec.~\ref{subsec:estimates}, we will finally turn to the main case our interest, primordial magnetogenesis during axion inflation, and discuss the constraints on its parameter space. 


\subsection{Temperature regimes}
\label{subsec:regimes}


In Sec.~\ref{subsec:washin}, we saw that the action of wash-in leptogenesis in the temperature regime $T \sim 10^5\cdots10^6\,\textrm{GeV}$ can be parametrized in terms of three dimensionless coefficients:
(i) $c_{B-L}^{\rm win}$, which relates the dimensionless helicity density $\chi$ to the primordial $B\!-\!L$ asymmetry produced during wash-in leptogenesis [see Eq.~\eqref{eq:winbl}];
equivalently, one may also work in terms of the coefficient $c_B^{\rm win} = \bar{c}_{\rm sph}\,c_{B-L}^{\rm win}$ where $\bar{c}_{\rm sph} = 12/37$, which relates $\chi$ to the primordial baryon asymmetry produced during wash-in leptogenesis [see Eq.~\eqref{eq:etaBwin}];
(ii) $c_5^>$, which relates $\chi$ to the chiral chemical potential shortly before wash-in leptogenesis [see Eq.~\eqref{eq:c5larger}]; and
(iii) $c_5^<$, which relates $\chi$ to the chiral chemical potential shortly after wash-in leptogenesis [see Eq.~\eqref{eq:c5smaller}].


In the temperature window $T \sim 10^5\cdots10^6\,\textrm{GeV}$, we found $c_{B-L}^{\rm win} = 3/10$, $c_B^{\rm win} = 18/185$, $c_5^> = 711/481$, and $c_5^< = 11/10$.
By contrast, if we attempted to realize wash-in leptogenesis at lower RHN masses, $M_i \lesssim 10^5\,\textrm{GeV}$, all of these coefficients would turn out to vanish, $c_{B-L}^{\rm win} = c_B^{\rm win} = c_5^> = c_5^< = 0$, for the following reason:
Around $T \sim 10^5\,\textrm{GeV}$, the electron Yukawa interaction enters thermal equilibrium, such that, from this point on all the way down to the EWPT, all SM interactions are equilibrated.
Thus, if no lepton asymmetry has been generated by the time the temperature has reached $T \sim 10^5\,\textrm{GeV}$, all chemical potentials in the thermal bath will vanish, as immediately follows from Eq.~\eqref{eq:eqVI} with $\bar{\mu}_{\Delta_\alpha} = 0$ on the right-hand side.
This means that no primordial global charges survive in the plasma that could be reprocessed by the RHN interactions around $T \sim M_i \lesssim 10^5\,\textrm{GeV}$.
Similarly, no chiral chemical potential survives that could trigger the chiral plasma instability.


At higher temperatures, the situation is more interesting.
To see this, let us now consider increasingly larger values of the leptogenesis temperature scale $T_{B-L}$, which we are going to estimate as follows,
\begin{equation}
\label{eq:TBL}
T_{B-L} \sim \min\left\{T_{N_i},T_{\Delta L = 2}\right\} \,, \qquad T_{N_i} \simeq M_i \,, \qquad T_{\Delta L = 2} \simeq 6 \times 10^{12}\,\textrm{GeV} \left(\frac{0.05\,\textrm{eV}}{m_\nu}\right)^2 \,.
\end{equation}
Here, $T_{N_i}$ denotes the decay temperature of the respective RHN species $N_i$, and $T_{\Delta L = 2}$ is the freeze-out temperature of the dimension-5 Weinberg operator in the seesaw extension of the Standard Model~\cite{Domcke:2020kcp}, with $m_\nu \sim 0.05\,\textrm{GeV}$ representing the mass scale of the light SM neutrinos.
Interactions mediated by the Weinberg operator violate total lepton number by two units, $\Delta L = 2$, and can thus play the role of the LNV interactions required for wash-in leptogenesis in the limit of very large RHN masses.
That is, for $M_i \gtrsim 10^{13}\,\textrm{GeV}$, LNV processes do not yet freeze out at the time of RHN decay, but only when lepton--Higgs scatterings with heavy off-shell RHNs in the intermediate state become inefficient at $T \sim T_{\Delta L = 2}$.%
\footnote{Renormalizable interactions such as the Yukawa interactions of the SM fermions begin to \textit{enter} thermal equilibrium as the temperature of the thermal bath steadily decreases.
The electron Yukawa interaction is, \textit{e.g.}, inefficient at temperatures above $T_{y_e}$ and efficient at temperatures below $T_{y_e}$. 
By contrast, the interactions mediated by the Weinberg operator scale differently with temperature because of the dimensionful coupling constant and thus \textit{leave} thermal equilibrium at temperatures around $T_{\Delta L = 2}$.}
The estimate in Eq.~\eqref{eq:TBL} tells us that we can easily realize wash-in leptogenesis at higher temperatures simply by increasing the RHN masses $M_i$. 
Wash-in leptogenesis can in particular occur in the temperature regimes (iv), (iii) and (ii), which we introduced in Sec.~\ref{subsec:transport}, if we choose the lightest RHN mass, $M_1$, to lie in one of the intervals $\left(10^6,10^9\right)\,\textrm{GeV}$, $\left(10^9,10^{11\cdots12}\right)\,\textrm{GeV}$, and $\left(10^{11\cdots12},10^{13}\right)\,\textrm{GeV}$, respectively.
For completeness, we will also consider wash-in leptogenesis in temperature regime (i), $T \in \left(10^{13},10^{15}\right)\,\textrm{GeV}$, in the following.
In this way, we will be able to account for the uncertainty in the freeze-out temperature of the Weinberg operator $T_{\Delta L = 2}$, which, as can be seen from Eq.~\eqref{eq:TBL}, may not be too different from the lower boundary of temperature regime (i). 


First, we consider RHN masses in the range $M_i \sim 10^6\cdots10^9\,\textrm{GeV}$.
In this case, we can evaluate the strong wash-in condition in Eq.~\eqref{eq:washin} by making use of the chemical equilibrium solution in Eq.~\eqref{eq:eqIV}
\begin{equation}
\begin{pmatrix}
\mu_{\ell_e} + \mu_\Phi \\ \mu_{\ell_\mu} + \mu_\Phi \\ \mu_{\ell_\tau} + \mu_\Phi \end{pmatrix}
=
\begin{pmatrix}
-\sfrac{37}{716} & -\sfrac{265}{716} \\
-\sfrac{13}{179} & \sfrac{23}{179}   \\
-\sfrac{13}{179} & \sfrac{23}{179} 
\end{pmatrix}
\begin{pmatrix}
\bar{\mu}_{u-d} \\ \bar{\mu}_e 
\end{pmatrix}
- 
\begin{pmatrix}
\sfrac{339}{716} & \sfrac{3}{179} & \sfrac{3}{179}  \\
\sfrac{3}{179} & \sfrac{211}{537} & \sfrac{32}{537} \\
\sfrac{3}{179} & \sfrac{32}{537} & \sfrac{211}{537}
\end{pmatrix}
\begin{pmatrix}
\bar{\mu}_{\Delta_e} \\ \bar{\mu}_{\Delta_\mu} \\\bar{\mu}_{\Delta_\tau}
\end{pmatrix} = \begin{pmatrix} 0 \\ 0 \\ 0 \end{pmatrix}\,.
\end{equation}
This equilibrium condition results in the following flavored $B\!-\!L$ charges after wash-in leptogenesis,
\begin{equation}
\label{eq:washinIV}
\begin{pmatrix}
\eta_{\Delta_e} \\ \eta_{\Delta_\mu} \\ \eta_{\Delta_\tau}
\end{pmatrix} = 
\begin{pmatrix}
-\sfrac{41}{51} & -\sfrac{5}{51} \\
 \sfrac{16}{51} & -\sfrac{8}{51}   \\
 \sfrac{16}{51} & -\sfrac{8}{51} 
\end{pmatrix}
\begin{pmatrix}
\eta_{u-d} \\ \eta_e 
\end{pmatrix} = 
\begin{pmatrix}
\sfrac{46}{51} \\ -\sfrac{8}{51} \\ -\sfrac{8}{51}
\end{pmatrix} \eta_\chi \,,
\end{equation}
where we used in the last step the primordial input charges after axion inflation in Eq.~\eqref{eq:etaC}.
The sum of the three flavored $B\!-\!L$ charges yields the total $B\!-\!L$ asymmetry and hence the coefficient $c_{B-L}^{\rm win}$,
\begin{equation}
\eta_{B-L} = -\frac{7}{17}\,\eta_{u-d} - \frac{3}{17}\,\eta_e = c_{B-L}^{\rm win} \eta_\chi \,,\qquad c_{B-L}^{\rm win} = \frac{10}{17} \,,
\end{equation}
which immediately translates to a coefficient for the primordial baryon asymmetry of $c_B^{\rm win} = 120/629$.
Similarly, we can use the chemical equilibrium in Eq.~\eqref{eq:eqIV} in order to determine the chiral chemical potential,
\begin{equation}
\label{eq:mu5iv}
\frac{\bar{\mu}_5}{T} = \frac{173}{1074}\,\frac{\bar{\mu}_{u-d}}{T} + \frac{513}{358}\,\frac{\bar{\mu}_e}{T} + \frac{151}{358}\,\frac{\bar{\mu}_{\Delta_e}}{T} - \frac{10}{179} \left(\frac{\bar{\mu}_{\Delta_\mu}}{T} + \frac{\bar{\mu}_{\Delta_\tau}}{T}\right) \,.
\end{equation}
Making use of the primordial chemical potentials in Eq.~\eqref{eq:muini} and the flavored $B\!-\!L$ charges in Eq.~\eqref{eq:washinIV}, we are able to explicitly evaluate this expression right before as well as right after wash-in leptogenesis, which provides us with the coefficients $c_5^>$ and $c_5^<$ in temperature regime (iv), $c_5^> = 856/537$ and $c_5^< = 61/51$.


Next, we repeat the analysis for temperature regime (iii), \textit{i.e.}, for RHN masses $M_i \sim 10^9\cdots10^{11\cdots12}\,\textrm{GeV}$.
If we evaluate the strong wash-in condition in Eq.~\eqref{eq:washin} based on the equilibrium in Eq.~\eqref{eq:eqIII}, we obtain
\begin{equation}
\label{eq:washinIII}
\begin{pmatrix}
\mu_{\ell_e} + \mu_\Phi \\ \mu_{\ell_\mu} + \mu_\Phi \\ \mu_{\ell_\tau} + \mu_\Phi \end{pmatrix}
=
\begin{pmatrix}
 \sfrac{84}{589}   & -\sfrac{123}{1178}  & -\sfrac{421}{1178} \\
-\sfrac{421}{1178} & -\sfrac{123}{1178}  & \sfrac{84}{589}   \\
 \sfrac{86}{589}   & -\sfrac{84}{589}    & \sfrac{86}{589} 
\end{pmatrix}
\begin{pmatrix}
\bar{\mu}_\mu \\ \bar{\mu}_{u-d} \\ \bar{\mu}_e 
\end{pmatrix}
- 
\begin{pmatrix}
 \sfrac{585}{1178} & -\sfrac{2}{589}    & \sfrac{26}{589}  \\
-\sfrac{2}{589}    &  \sfrac{585}{1178} & \sfrac{26}{589}  \\
 \sfrac{26}{589}   &  \sfrac{26}{589}   & \sfrac{251}{589}
\end{pmatrix}
\begin{pmatrix}
\bar{\mu}_{\Delta_e} \\ \bar{\mu}_{\Delta_\mu} \\\bar{\mu}_{\Delta_\tau}
\end{pmatrix} \,.
\end{equation}
However, unlike in the previous case, it is now no longer guaranteed that all three linear combinations of chemical potentials on the left-hand side of this relation must vanish in the course of wash-in leptogenesis.
The crucial difference compared to temperature regime (iv) is that, at temperatures above $T \sim 10^9\,\textrm{GeV}$, the muon Yukawa interaction has not yet equilibrated, which effectively reduces the lepton sector to a two-flavor system.
In temperature regime (iii), the SM interactions are only able to probe the tau-flavor content of a given lepton state; coherent superpositions of electron- and muon-flavor states remain unperturbed.
For our purposes, this means that the labels $\alpha = e$ and $\alpha = \mu$ in Eq.~\eqref{eq:washinIII} are meaningless to some extent.
They merely denote \textit{a possible} basis of the two-dimensional $e$\,--\,$\mu$ flavor space; but this basis does not necessarily need to coincide with the physical electron and muon flavors at lower temperatures.
As far as the SM interactions are concerned, any orthonormal basis of the $e$\,--\,$\mu$ flavor space is as good as any other at $T \gtrsim 10^9\,\textrm{GeV}$.
Meanwhile, the same statement does not hold true with regard to the RHN interactions during wash-in leptogenesis.
\textit{A priori}, each RHN species interacts with one specific linear combination of lepton flavors, \textit{i.e.}, along one specific direction in the three-dimensional $e$\,--\,$\mu$\,--\,$\tau$ flavor space, which follows from the relation among its Yukawa couplings to the SM lepton-Higgs pairs. 
In temperature regime (iii), the tau Yukawa interaction then probes the tau-flavor content of these states; but the coherence in the $e$\,--\,$\mu$ flavor space remains preserved. 
For each RHN species, there is hence a particular direction in $e$\,--\,$\mu$ flavor space along which it interacts with SM lepton-Higgs pairs as well as an orthogonal direction in $e$\,--\,$\mu$ flavor space along which it does not interact.


In view of this situation, we must now distinguish between different scenarios.
First, let us assume that wash-in leptogenesis is driven by only one RHN species, $N_1$, which is, \textit{e.g.}, possible if the heavier RHN species $N_2$ and $N_3$ have masses above the reheating temperature, $M_{2,3} \gtrsim T_{\rm rh}$, such that they are never produced after inflation. 
In this case, which we will refer to as $N_1$-dominated wash-in leptogenesis in the following, we only have to deal with two relevant directions in flavor space: the tau-flavor direction and the direction in $e$\,--\,$\mu$ flavor space along which the $N_1$ interactions are active.
Therefore, given the basis freedom in $e$\,--\,$\mu$ flavor space from the SM perspective, we are able to identify, \textit{w.l.o.g.}, $\alpha = e$ in Eq.~\eqref{eq:washinIII} with the $N_1$ wash-in direction and $\alpha = \mu$ with its orthogonal complement.
In the strong wash-in regime, we thus have to impose
\begin{equation}
\begin{pmatrix}
\mu_{\ell_e} + \mu_\Phi\\ \mu_{\ell_\tau} + \mu_\Phi \end{pmatrix}
=
\begin{pmatrix}
 \sfrac{84}{589}   & -\sfrac{123}{1178}  & -\sfrac{421}{1178} \\
 \sfrac{86}{589}   & -\sfrac{84}{589}    & \sfrac{86}{589} 
\end{pmatrix}
\begin{pmatrix}
\bar{\mu}_\mu \\ \bar{\mu}_{u-d} \\ \bar{\mu}_e 
\end{pmatrix}
- 
\begin{pmatrix}
 \sfrac{585}{1178} & -\sfrac{2}{589}    & \sfrac{26}{589}  \\
 \sfrac{26}{589}   &  \sfrac{26}{589}   & \sfrac{251}{589}
\end{pmatrix}
\begin{pmatrix}
\bar{\mu}_{\Delta_e} \\ \bar{\mu}_{\Delta_\mu} \\\bar{\mu}_{\Delta_\tau}
\end{pmatrix} = \begin{pmatrix} 0 \\ 0  \end{pmatrix} \,.
\end{equation}
Solving these two conditions for $\bar{\mu}_{\Delta_e}$ and $\bar{\mu}_{\Delta_\tau}$ results in the following asymmetries after $N_1$ freeze-out,
\begin{equation}
\label{eq:washinIIIa}
\begin{pmatrix}
\eta_{\Delta_e} \\ \eta_{\Delta_\tau}
\end{pmatrix} = 
\begin{pmatrix}
  \sfrac{4}{247} & \sfrac{64}{247} & -\sfrac{45}{247} & -\sfrac{187}{247} \\
- \sfrac{2}{19} &  \sfrac{6}{19}   & -\sfrac{6}{19}   &  \sfrac{8}{19}    \\
\end{pmatrix}
\begin{pmatrix}
\eta_{\Delta_\mu} \\ \eta_\mu \\ \eta_{u-d} \\ \eta_e 
\end{pmatrix} = 
\begin{pmatrix}
\sfrac{168}{247} \\ -\sfrac{8}{19}
\end{pmatrix} \eta_\chi \,,
\end{equation}
where $\eta_{\Delta_\mu}$ (\textit{i.e.}, the asymmetry along the flavor direction that is orthogonal to the $N_1$ wash-in direction in $e$\,--\,$\mu$ flavor space) is now treated as yet another conserved charge.
In the second step in Eq.~\eqref{eq:washinIIIa}, we replaced the primordial charges $\eta_\mu$, $\eta_{u-d}$, and $\eta_e$ by their input values listed in Eq.~\eqref{eq:etaC} and set $\eta_{\Delta_\mu} = 0$.
This is in line with our assumption of $N_1$-dominated wash-in leptogenesis, where no pre-existing asymmetries (possibly caused by $N_2$ or $N_3$ wash-in leptogenesis) are present.
The two nonzero flavored $B\!-\!L$ charges sum to
\begin{equation}
\label{eq:etaBLiii}
\eta_{B-L} = \sum_{\alpha = e,\tau} \eta_{\Delta_\alpha} = -\frac{22}{247}\,\eta_{\Delta_\mu} + \frac{142}{247}\,\eta_\mu -\frac{123}{247}\,\eta_{u-d} - \frac{83}{247}\,\eta_e = c_{B-L}^{\rm win} \eta_\chi \,,\qquad c_{B-L}^{\rm win} = \frac{64}{247} \,,
\end{equation}
and correspondingly $c_B^{\rm win} = 768/9139$.
In addition, we use the chemical equilibrium in Eq.~\eqref{eq:eqIII} to find $\bar{\mu}_5$,
\begin{equation}
\frac{\bar{\mu}_5}{T} = \frac{828}{589}\left(\frac{\bar{\mu}_e}{T} + \frac{\bar{\mu}_\mu}{T}\right) + \frac{109}{589}\,\frac{\bar{\mu}_{u-d}}{T} + \frac{188}{589}\left(\frac{\bar{\mu}_{\Delta_e}}{T} + \frac{\bar{\mu}_{\Delta_\mu}}{T}\right) - \frac{88}{589}\,\frac{\bar{\mu}_{\Delta_\tau}}{T} \,,
\end{equation}
which, together with $\bar{\mu}_{\Delta_\mu} = 0$ and the attractor solution in Eq.~\eqref{eq:washinIIIa}, yields $c_5^> = 1765/589$ and $c_5^< = 671/247$.


Let us now discuss these results.
First, recall that, in our derivation, we identified $\alpha = e$ with the $N_1$ wash-in direction in $e$\,--\,$\mu$ flavor space and $\alpha = \mu$ with its orthogonal complement.
In passing, we mention that we would have obtained the same coefficients for the opposite identification.
This immediately follows from the flavor-blind initial conditions after inflation (in particular, $\bar{\mu}_e = \bar{\mu}_\mu$) and the fact that the system of equations in Eq.~\eqref{eq:washinIII} is invariant under the exchange of all $e$ and $\mu$ indices. 
In fact, given the symmetric initial conditions after axion inflation, we can relate $\bar{\mu}_e$ and $\bar{\mu}_\mu$ to the trace over chemical potentials in $e$\,--\,$\mu$ flavor space and write $\bar{\mu}_e = \bar{\mu}_\mu = 1/2\left(\bar{\mu}_e + \bar{\mu}_\mu\right)$.
This relation allows us to symmetrize the result in Eq.~\eqref{eq:etaBLiii},
\begin{equation}
\eta_{B-L} = \sum_{\alpha = e,\tau} \eta_{\Delta_\alpha} = \frac{59}{494}\left(\eta_e + \eta_\mu\right) -\frac{123}{247}\,\eta_{u-d} - \frac{22}{247}\,\eta_{\Delta_\mu} \,,
\end{equation}
which is consistent with the discussion in the appendix of Ref.~\cite{Domcke:2020quw}. 


Next, let us relax our assumption regarding the role of the $N_2$ and $N_3$ RHNs and assume that both species are able to generate primordial lepton asymmetries at temperatures $T \gtrsim 10^{11\cdots12}\,\textrm{GeV}$, either via wash-in or standard thermal leptogenesis.
In this case, $N_1$ wash-in leptogenesis in temperature regime (iii) will be subject to heavy-neutrino flavor effects.
That is, if the $N_2$ and $N_3$ RHNs should be responsible for a first stage of leptogenesis at high temperatures, we will no longer be able to assume $\eta_{\Delta_\mu} = 0$ in our calculation.
Instead, we have to keep $\eta_{\Delta_\mu}$ throughout our analysis and include it in our sum over the flavored $B\!-\!L$ charges,
\begin{equation}
\eta_{B-L} = \sum_{\alpha = e,\mu,\tau} \eta_{\Delta_\alpha} = \frac{59}{494}\left(\eta_e + \eta_\mu\right) -\frac{123}{247}\,\eta_{u-d} + \frac{225}{247}\,\eta_{\Delta_\mu} \,,
\end{equation}
which results in the following modification of the four coefficients $c_{B-L}^{\rm win}$, $c_B^{\rm win}$, $c_5^>$, and $c_5^<$,
\begin{align}
\label{eq:coeffiii}
c_{B-L}^{\rm win} & = \frac{64}{247} + \frac{225}{247}\,c_\perp^* \,, \qquad c_B^{\rm win} = \frac{768}{9139} + \frac{2700}{9139}\,c_\perp^* \,, \\
\label{eq:coeffiii2}
c_5^> & = \frac{1765}{589} - \frac{188}{589}\left(c_{\Delta_e}^* + c_{\Delta_\mu}^*\right) + \frac{88}{589}\,c_{\Delta_\tau}^* \,, \qquad c_5^< = \frac{671}{247} - \frac{84}{247}\,c_\perp^* \,.
\end{align}
Here, the coefficients $c_{\Delta_\alpha}^* = \eta_{\Delta_\alpha}/\eta_\chi$ quantify the three pre-existing flavored $B\!-\!L$ asymmetries generated during $N_{2,3}$ leptogenesis, while $c_\perp^*$ specifically measures the pre-existing asymmetry along the flavor direction in $e$\,--\,$\mu$ flavor space that is immune to $N_1$ wash-in leptogenesis.
Given our convention chosen above, $c_\perp^* = \eta_{\Delta_\mu}/\eta_\chi$, while more generally, one may give the protected flavor direction a new name and write $c_\perp^* = \eta_{\Delta_\perp}/\eta_\chi$ (see also the notation and conventions in Ref.~\cite{Domcke:2020quw}).
The explicit value of $c_\perp^*$ is model-dependent but calculable. 
Given a specification of all parameters in the RHN sector, one is able to use standard results for thermal leptogenesis or the formalism for wash-in leptogenesis developed here and in Ref.~\cite{Domcke:2020quw} to compute the coefficient $c_\perp^*$. 


The coefficients in Eqs.~\eqref{eq:coeffiii} and \eqref{eq:coeffiii2} encode our results for $N_1$ wash-in leptogenesis at temperatures $T \in \left(10^9, 10^{11\cdots12}\right)\,\textrm{GeV}$ in the two-flavor regime, both for scenarios with ($c_\perp^* \neq 0$) and without ($c_\perp^* = 0$) a pre-existing asymmetry along the blind flavor direction.
In addition, it is possible to construct a three-flavor scenario in temperature regime (iii), which can be realized when at least two RHN species contribute to wash-in leptogenesis.
In this case, if the two lepton states interacting with the active RHN species plus the tau flavor are linearly independent, all three dimensions in flavor space can be accessed and strong wash-in leads to 
\begin{equation}
\begin{pmatrix}
\mu_{\ell_e} + \mu_\Phi \\ \mu_{\ell_\mu} + \mu_\Phi \\ \mu_{\ell_\tau} + \mu_\Phi \end{pmatrix}
=
\begin{pmatrix}
 \sfrac{84}{589}   & -\sfrac{123}{1178}  & -\sfrac{421}{1178} \\
-\sfrac{421}{1178} & -\sfrac{123}{1178}  & \sfrac{84}{589}   \\
 \sfrac{86}{589}   & -\sfrac{84}{589}    & \sfrac{86}{589} 
\end{pmatrix}
\begin{pmatrix}
\bar{\mu}_\mu \\ \bar{\mu}_{u-d} \\ \bar{\mu}_e 
\end{pmatrix}
- 
\begin{pmatrix}
 \sfrac{585}{1178} & -\sfrac{2}{589}    & \sfrac{26}{589}  \\
-\sfrac{2}{589}    &  \sfrac{585}{1178} & \sfrac{26}{589}  \\
 \sfrac{26}{589}   &  \sfrac{26}{589}   & \sfrac{251}{589}
\end{pmatrix}
\begin{pmatrix}
\bar{\mu}_{\Delta_e} \\ \bar{\mu}_{\Delta_\mu} \\\bar{\mu}_{\Delta_\tau}
\end{pmatrix} = \begin{pmatrix} 0 \\ 0 \\ 0 \end{pmatrix} \,,
\end{equation}
after all.
Wash-in leptogenesis at $T \in \left(10^9, 10^{11\cdots12}\right)\,\textrm{GeV}$ in the three-flavor regime hence yields 
\begin{equation}
\begin{pmatrix}
\eta_{\Delta_e} \\ \eta_{\Delta_\mu} \\ \eta_{\Delta_\tau}
\end{pmatrix} = 
\begin{pmatrix}
\sfrac{20}{81} & -\sfrac{5}{27} & -\sfrac{61}{81}  \\
-\sfrac{61}{81} & -\sfrac{5}{27} & \sfrac{20}{81}  \\
\sfrac{32}{81}  & -\sfrac{8}{27} &  \sfrac{32}{81} \\
\end{pmatrix}
\begin{pmatrix}
\eta_\mu \\ \eta_{u-d} \\ \eta_e 
\end{pmatrix} = 
\begin{pmatrix}
\sfrac{56}{81} \\ \sfrac{56}{81} \\ -\sfrac{40}{81}
\end{pmatrix} \eta_\chi \,,
\end{equation}
and correspondingly the following total $B\!-\!L$ asymmetry and coefficients for the chiral chemical potential,
\begin{equation}
\eta_{B-L} = -\frac{1}{9}\left(\eta_e + \eta_\mu\right) -\frac{2}{3}\,\eta_{u-d} = c_{B-L}^{\rm win} \eta_\chi \,,\qquad c_{B-L}^{\rm win} = \frac{8}{9} \,, \qquad c_B^{\rm win} = \frac{32}{111} \,, \qquad c_5^> = \frac{1765}{589} \,, \qquad c_5^< = \frac{67}{27} \,.
\end{equation}


Moving on to temperature regime (ii), $T \in \left(10^{11\cdots12}, 10^{13}\right)\,\textrm{GeV}$, we have to keep paying attention to coherence\,/\,decoherence as well as heavy-neutrino flavor effects.
Similarly as before, we first naively evaluate the relevant chemical potentials, this time making use of the chemical equilibrium in Eq.~\eqref{eq:eqII},
\begin{equation}
\label{eq:washinii0}
\begin{pmatrix}
\mu_{\ell_e} + \mu_\Phi \\ \mu_{\ell_\mu} + \mu_\Phi \\ \mu_{\ell_\tau} + \mu_\Phi \end{pmatrix}
=
\begin{pmatrix}
\sfrac{1}{6} & \sfrac{7}{46} & \sfrac{7}{46} & -\sfrac{9}{23} & -\sfrac{8}{23} \\
\sfrac{1}{6} & \sfrac{7}{46} & -\sfrac{8}{23} & -\sfrac{9}{23} & \sfrac{7}{46} \\
\sfrac{1}{6} & -\sfrac{8}{23} & \sfrac{7}{46} & -\sfrac{9}{23} & \sfrac{7}{46} \\
\end{pmatrix}
\begin{pmatrix}
\bar{\mu}_B \\ \bar{\mu}_\tau \\ \bar{\mu}_\mu \\ \bar{\mu}_{u-d} \\ \bar{\mu}_e 
\end{pmatrix}
- 
\begin{pmatrix}
\sfrac{15}{23} & \sfrac{7}{46}  & \sfrac{7}{46}  \\
\sfrac{7}{46}  & \sfrac{15}{23} & \sfrac{7}{46}  \\
\sfrac{7}{46}  & \sfrac{7}{46}  & \sfrac{15}{23}
\end{pmatrix}
\begin{pmatrix}
\bar{\mu}_{\Delta_e} \\ \bar{\mu}_{\Delta_\mu} \\\bar{\mu}_{\Delta_\tau}
\end{pmatrix} \,.
\end{equation}
This system of equations is invariant under the exchange of any two lepton flavor indices ($e \leftrightarrow \mu$, $e \leftrightarrow \tau$, $\mu \leftrightarrow \tau$), which reflects the fact that the SM interactions do not distinguish between the three lepton flavors at temperatures above the equilibration temperature of the tau Yukawa interaction, $T_{y_\tau} \sim 10^{12}\,\textrm{GeV}$.
The SM interactions rather preserve the coherence of lepton flavor states at $T \gtrsim 10^{12}\,\textrm{GeV}$, which leads us to consider three different scenarios: wash-in leptogenesis in the one-flavor, two-flavor, and three-flavor regime.


First, let us investigate the one-flavor regime, assuming that the $N_2$ and $N_3$ RHN species are not active in temperature regime (ii). 
In this case, wash-in leptogenesis only occurs along one direction in $e$\,--\,$\mu$\,--\,$\tau$ flavor space. 
As before, we can use the basis freedom in flavor space to identify this direction, in a slightly abusive notation but \textit{w.l.o.g.}, with $\alpha = e$, while $\alpha = \mu$ and $\alpha = \tau$ now span the two-dimensional co-space that is immune to $N_1$ wash-in.
In the one-flavor regime, there is hence only one strong wash-in condition,
\begin{equation}
\label{eq:washinii}
\mu_{\ell_e} + \mu_\Phi = \frac{1}{6}\,\bar{\mu}_B + \frac{7}{46}\left(\bar{\mu}_\mu + \bar{\mu}_\tau\right) - \frac{9}{23}\,\bar{\mu}_{u-d} - \frac{8}{23}\,\bar{\mu}_e - \frac{15}{23}\,\bar{\mu}_{\Delta_e} - \frac{7}{46}\left(\bar{\mu}_{\Delta_\mu} + \bar{\mu}_{\Delta_\tau}\right) = 0 \,,
\end{equation}
with $\bar{\mu}_{\Delta_\mu} + \bar{\mu}_{\Delta_\tau} = \bar{\mu}_{\Delta_\perp}$ quantifying the pre-existing asymmetry in the two-dimensional flavor subspace that $N_1$ wash-in has no access to.
Solving Eq.~\eqref{eq:washinii} for $\bar{\mu}_{\Delta_e}$ then immediately provides us with the total $B\!-\!L$ charge,
\begin{align}
\label{eq:washiniiA}
\eta_{\Delta_e} & = \frac{23}{90}\,\eta_B + \frac{7}{30}\left(\eta_\mu + \eta_\tau\right) - \frac{3}{5}\,\eta_{u-d} - \frac{8}{15}\,\eta_e - \frac{7}{30}\,\eta_{\Delta_\perp} = \frac{17}{60}\,\eta_\chi - \frac{7}{30}\,\eta_{\Delta_\perp} \,,\\
\eta_{B-L} = \eta_{\Delta_e} + \eta_{\Delta_\perp} & = \frac{23}{90}\,\eta_B + \frac{7}{30}\left(\eta_\mu + \eta_\tau\right) - \frac{3}{5}\,\eta_{u-d} - \frac{8}{15}\,\eta_e + \frac{23}{30}\,\eta_{\Delta_\perp} = c_{B-L}^{\rm win}\,\eta_\chi \,,
\end{align}
where we used again Eq.~\eqref{eq:etaC} in order to relate the various input asymmetries to the reference asymmetry $\eta_\chi$.
Hence, employing the notation $c_\perp^* = \eta_{\Delta_\perp}/\eta_\chi$ introduced above, the coefficients $c_{B-L}^{\rm win}$ and $c_B^{\rm win}$ now read
\begin{equation}
c_{B-L}^{\rm win} = \frac{17}{60} + \frac{23}{30}\,c_\perp^* \,, \qquad c_B^{\rm win} = \frac{17}{185} + \frac{46}{185}\,c_\perp^* \,.
\end{equation}
Here, $c_\perp^* = 0$ corresponds to the case of $N_1$-dominated wash-in leptogenesis, while $c_\perp^* \neq 0$ occurs in scenarios where the $N_2$ and $N_3$ RHNs are responsible for the generation of pre-existing asymmetries at $T \gtrsim 10^{13}\,\textrm{GeV}$.
Finally, we can use the chemical equilibrium in Eq.~\eqref{eq:eqII} to evaluate $\bar{\mu}_5$ in temperature regime (ii),
\begin{equation}
\label{eq:mu5ii}
\frac{\bar{\mu}_5}{T} = \frac{1}{8}\frac{\bar{\mu}_B}{T} + \frac{121}{92}\left(\frac{\bar{\mu}_e}{T} + \frac{\bar{\mu}_\mu}{T} + \frac{\bar{\mu}_\tau}{T}\right) + \frac{6}{23}\,\frac{\bar{\mu}_{u-d}}{T} + \frac{17}{92}\left(\frac{\bar{\mu}_{\Delta_e}}{T} + \frac{\bar{\mu}_{\Delta_\mu}}{T} + \frac{\bar{\mu}_{\Delta_\tau}}{T}\right) \,,
\end{equation}
which allows us to deduce the coefficients
\begin{equation}
c_5^> = \frac{1617}{368} - \frac{17}{92}\,c_{B-L}^* \,, \qquad c_5^< = \frac{521}{120} - \frac{17}{120}\,c_\perp^* \,,
\end{equation}
where $c_{B-L}^* = c_{\Delta_e}^* + c_{\Delta_\mu}^* + c_{\Delta_\tau}^*$ accounts for the possible pre-existing $B-L$ asymmetry from $N_{2,3}$ leptogenesis.


Similarly as in temperature regime (iii), we would have obtained the same results for $c_{B-L}^{\rm win}$, $c_B^{\rm win}$, $c_5^>$, and $c_5^<$, if we had picked a different convention in Eq.~\eqref{eq:washinii} and used $\mu_{\ell_\mu} + \mu_\Phi = 0$ or $\mu_{\ell_\tau} + \mu_\Phi = 0$ as our strong wash-in condition in the single-flavor regime.
Moreover, we can symmetrize our result for the total $B\!-\!L$ charge by introducing the trace over chemical potentials in $e$\,--\,$\mu$\,--\,$\tau$ flavor space, $\bar{\mu}_e = \bar{\mu}_\mu = \bar{\mu}_\tau = 1/3\left(\bar{\mu}_e + \bar{\mu}_\mu + \bar{\mu}_\tau\right)$,
\begin{equation}
\label{eq:washiniiX}
\eta_{B-L} = \frac{23}{90}\,\eta_B - \frac{1}{45}\left(\eta_e + \eta_\mu + \eta_\tau\right) - \frac{3}{5}\,\eta_{u-d} + \frac{23}{30}\,\eta_{\Delta_\perp} \,,
\end{equation}
which is again consistent with the discussion in the appendix of Ref.~\cite{Domcke:2020quw}. 


If the interactions of two RHN species, $N_1$ and $N_2$, are efficient at $T \in \left(10^{11\cdots12}, 10^{13}\right)\,\textrm{GeV}$ and if we assume that these two RHN species interact with linearly independent combinations of lepton flavor states, wash-in leptogenesis will operate in the two-flavor regime.
In this case, strong wash-in results in two conditions,
\begin{equation}
\label{eq:washiniia}
\begin{pmatrix}
\mu_{\ell_e} + \mu_\Phi \\ \mu_{\ell_\mu} + \mu_\Phi \end{pmatrix}
=
\begin{pmatrix}
\sfrac{1}{6} & \sfrac{7}{46} & \sfrac{7}{46} & -\sfrac{9}{23} & -\sfrac{8}{23} \\
\sfrac{1}{6} & \sfrac{7}{46} & -\sfrac{8}{23} & -\sfrac{9}{23} & \sfrac{7}{46}
\end{pmatrix}
\begin{pmatrix}
\bar{\mu}_B \\ \bar{\mu}_\tau \\ \bar{\mu}_\mu \\ \bar{\mu}_{u-d} \\ \bar{\mu}_e 
\end{pmatrix}
- 
\begin{pmatrix}
\sfrac{15}{23} & \sfrac{7}{46}  & \sfrac{7}{46}  \\
\sfrac{7}{46}  & \sfrac{15}{23} & \sfrac{7}{46}
\end{pmatrix}
\begin{pmatrix}
\bar{\mu}_{\Delta_e} \\ \bar{\mu}_{\Delta_\mu} \\\bar{\mu}_{\Delta_\tau}
\end{pmatrix} \,,
\end{equation}
where we now identify, \textit{w.lo.g.}, $\alpha = e$ and $\alpha = \mu$ with the basis of the two-dimensional flavor space that accommodates the $N_1$ and $N_2$ wash-in directions, while the $\alpha = \tau$ direction remains immune to wash-in, \textit{i.e.}, $\bar{\mu}_{\Delta_\tau} = \bar{\mu}_{\Delta_\perp}$ in the two-flavor regime.
As before, we shall allow for the possibility of a nonvanishing initial value, $\bar{\mu}_{\Delta_\perp} \neq 0$, which may originate from $N_3$ leptogenesis at $T \gtrsim 10^{13}\,\textrm{GeV}$.
The solution of Eq.~\eqref{eq:washiniia} then reads,
\begin{equation}
\label{eq:washiniiB}
\begin{pmatrix}
\eta_{\Delta_e} \\ \eta_{\Delta_\mu}
\end{pmatrix} =
\begin{pmatrix}
-\sfrac{7}{37} & \sfrac{23}{111} & \sfrac{7}{37} & \sfrac{14}{37} & -\sfrac{18}{37} & -\sfrac{23}{37} \\
-\sfrac{7}{37} & \sfrac{23}{111} & \sfrac{7}{37} & -\sfrac{23}{37} & -\sfrac{18}{37} & \sfrac{14}{37}
\end{pmatrix}
\begin{pmatrix}
\eta_{\Delta_\tau} \\ \eta_B \\ \eta_\tau \\ \eta_\mu \\ \eta_{u-d} \\ \eta_e 
\end{pmatrix} = 
\begin{pmatrix}
\sfrac{17}{74} - \sfrac{7}{37}\,c_\perp^*\\ \sfrac{17}{74} - \sfrac{7}{37}\,c_\perp^*
\end{pmatrix} \eta_\chi \,,
\end{equation}
which translates to the following outcome for the total $B\!-\!L$ asymmetry, 
\begin{align}
\eta_{B-L} = \eta_{\Delta_e} + \eta_{\Delta_\mu} + \eta_{\Delta_\perp} & = \frac{46}{111}\,\eta_B + \frac{14}{37}\,\eta_\tau - \frac{9}{37}\left(\eta_e + \eta_\mu\right) - \frac{36}{37}\,\eta_{u-d} + \frac{23}{37}\,\eta_{\Delta_\tau} \\
& = \frac{46}{111}\,\eta_B - \frac{4}{111}\left(\eta_e + \eta_\mu + \eta_\tau\right) - \frac{36}{37}\,\eta_{u-d} + \frac{23}{37}\,\eta_{\Delta_\perp} = c_{B-L}^{\rm win}\,\eta_\chi \,.
\end{align}
where, in the second line, we set $\eta_{\Delta_\tau} = \eta_{\Delta_\perp}$ and symmetrized the result using $\eta_e = \eta_\mu = \eta_\tau = 1/3\left(\eta_e + \eta_\mu + \eta_\tau\right)$.
Together with Eq.~\eqref{eq:mu5ii}, we thus obtain the following coefficients in the two-flavor regime,
\begin{equation}
\label{eq:washiniiY}
c_{B-L}^{\rm win} = \frac{17}{37} + \frac{23}{37}\,c_\perp^* \,, \qquad c_B^{\rm win} = \frac{204}{1369} + \frac{276}{1369}\,c_\perp^* \,, \qquad c_5^> = \frac{1617}{368} - \frac{17}{92}\,c_{B-L}^* \,, \qquad c_5^< = \frac{2551}{592} - \frac{17}{148}\,c_\perp^* \,. 
\end{equation}
Again, this result is independent of our concrete identification of the various directions in flavor space.


Finally, let us turn to wash-in leptogenesis at $T \in \left(10^{11\cdots12}, 10^{13}\right)\,\textrm{GeV}$ in the three-flavor regime, which is realized when all three RHN species are active in temperature regime (ii) and operate along three linearly independent directions in flavor space.
In this case, we deduce three strong wash-in conditions from Eq.~\eqref{eq:washinii0},
\begin{equation}
\begin{pmatrix}
\mu_{\ell_e} + \mu_\Phi \\ \mu_{\ell_\mu} + \mu_\Phi \\ \mu_{\ell_\tau} + \mu_\Phi \end{pmatrix}
=
\begin{pmatrix}
\sfrac{1}{6} & \sfrac{7}{46} & \sfrac{7}{46} & -\sfrac{9}{23} & -\sfrac{8}{23} \\
\sfrac{1}{6} & \sfrac{7}{46} & -\sfrac{8}{23} & -\sfrac{9}{23} & \sfrac{7}{46} \\
\sfrac{1}{6} & -\sfrac{8}{23} & \sfrac{7}{46} & -\sfrac{9}{23} & \sfrac{7}{46} \\
\end{pmatrix}
\begin{pmatrix}
\bar{\mu}_B \\ \bar{\mu}_\tau \\ \bar{\mu}_\mu \\ \bar{\mu}_{u-d} \\ \bar{\mu}_e 
\end{pmatrix}
- 
\begin{pmatrix}
\sfrac{15}{23} & \sfrac{7}{46}  & \sfrac{7}{46}  \\
\sfrac{7}{46}  & \sfrac{15}{23} & \sfrac{7}{46}  \\
\sfrac{7}{46}  & \sfrac{7}{46}  & \sfrac{15}{23}
\end{pmatrix}
\begin{pmatrix}
\bar{\mu}_{\Delta_e} \\ \bar{\mu}_{\Delta_\mu} \\\bar{\mu}_{\Delta_\tau}
\end{pmatrix} = \begin{pmatrix}
0 \\ 0 \\ 0
\end{pmatrix}\,,
\end{equation}
which we solve for the three flavored $B\!-\!L$ asymmetries,
\begin{equation}
\label{eq:washiniiC}
\begin{pmatrix}
\eta_{\Delta_e} \\ \eta_{\Delta_\mu} \\ \eta_{\Delta_\tau}
\end{pmatrix} = 
\begin{pmatrix}
\sfrac{23}{132} & \sfrac{7}{22} & \sfrac{7}{22} & -\sfrac{9}{22} & -\sfrac{15}{22} \\
\sfrac{23}{132} & \sfrac{7}{22} & -\sfrac{15}{22} & -\sfrac{9}{22} & \sfrac{7}{22} \\
\sfrac{23}{132} & -\sfrac{15}{22} & \sfrac{7}{22} & -\sfrac{9}{22} & \sfrac{7}{22} \\
\end{pmatrix}
\begin{pmatrix}
\eta_B \\ \eta_\tau \\ \eta_\mu \\ \eta_{u-d} \\ \eta_e 
\end{pmatrix} =
\begin{pmatrix}
\sfrac{17}{88} \\ \sfrac{17}{88} \\ \sfrac{17}{88} 
\end{pmatrix} \eta_\chi \,,
\end{equation}
and which in turn yield the following total $B\!-\!L$ charge and coefficients for the chiral chemical potential,
\begin{align}
\eta_{B-L} & = \frac{23}{44}\,\eta_B -\frac{1}{22}\left(\eta_e + \eta_\mu + \eta_\tau\right) -\frac{27}{22}\,\eta_{u-d} \,, \\
c_{B-L}^{\rm win} & = \frac{51}{88} \,, \qquad c_B^{\rm win} = \frac{153}{814} \,, \qquad c_5^> = \frac{1617}{368} \,, \qquad c_5^< = \frac{1509}{352} \,. \label{eq:washiniiZ}
\end{align}


Last but not least, we turn to wash-in leptogenesis in temperature regime (i), $T \in \left(10^{13}, 10^{15}\right)\,\textrm{GeV}$, where, qualitatively, the analysis proceeds in exactly the same way as in temperature regime (ii).
Therefore, in order to quote our results in regime (i), it will suffice if we merely state how the various equations that we encountered in our discussion of regime (ii) need to be updated.
We begin with Eq.~\eqref{eq:washinii0}, which gets replaced by
\begin{equation}
\begin{pmatrix}
\mu_{\ell_e} + \mu_\Phi \\ \mu_{\ell_\mu} + \mu_\Phi \\ \mu_{\ell_\tau} + \mu_\Phi \end{pmatrix}
=
\begin{pmatrix}
\sfrac{2}{9} & \sfrac{1}{9} & \sfrac{1}{6} & \sfrac{1}{6} & -\sfrac{5}{9} & -\sfrac{1}{3} \\
\sfrac{2}{9} & \sfrac{1}{9} & \sfrac{1}{6} & -\sfrac{1}{3} & -\sfrac{5}{9} & \sfrac{1}{6} \\
\sfrac{2}{9} & \sfrac{1}{9} & -\sfrac{1}{3} & \sfrac{1}{6} & -\sfrac{5}{9} & \sfrac{1}{6} \\
\end{pmatrix}
\begin{pmatrix}
\bar{\mu}_u \\ \bar{\mu}_B \\ \bar{\mu}_\tau \\ \bar{\mu}_\mu \\ \bar{\mu}_{u-d} \\ \bar{\mu}_e 
\end{pmatrix}
- 
\begin{pmatrix}
\sfrac{2}{3} & \sfrac{1}{6} & \sfrac{1}{6} \\
\sfrac{1}{6} & \sfrac{2}{3} & \sfrac{1}{6} \\
\sfrac{1}{6} & \sfrac{1}{6} & \sfrac{2}{3}
\end{pmatrix}
\begin{pmatrix}
\bar{\mu}_{\Delta_e} \\ \bar{\mu}_{\Delta_\mu} \\\bar{\mu}_{\Delta_\tau}
\end{pmatrix} \,.
\end{equation}
In the one-flavor regime, we need to update the numbers in Eqs.~\eqref{eq:washiniiA} to \eqref{eq:washiniiX}, which leads us to
\begin{align}
\eta_{\Delta_e} & = \frac{1}{3}\,\eta_u + \frac{1}{6}\,\eta_B + \frac{1}{4}\left(\eta_\mu + \eta_\tau\right) - \frac{5}{6}\,\eta_{u-d} - \frac{1}{2}\,\eta_e - \frac{1}{4}\,\eta_{\Delta_\perp} = \frac{5}{36}\,\eta_\chi - \frac{1}{4}\,\eta_{\Delta_\perp} \,,\\
\eta_{B-L} = \eta_{\Delta_e} + \eta_{\Delta_\perp} & = \frac{1}{3}\,\eta_u + \frac{1}{6}\,\eta_B - \frac{5}{6}\,\eta_{u-d} + \frac{3}{4}\,\eta_{\Delta_\perp} \,, \label{eq:etaBione}\\
\frac{\bar{\mu}_5}{T} & = \frac{229}{162}\,\frac{\bar{\mu}_u}{T} - \frac{37}{162}\frac{\bar{\mu}_B}{T} + \frac{38}{27}\left(\frac{\bar{\mu}_e}{T} + \frac{\bar{\mu}_\mu}{T} + \frac{\bar{\mu}_\tau}{T}\right) - \frac{127}{162}\,\frac{\bar{\mu}_{u-d}}{T} + \frac{5}{54}\left(\frac{\bar{\mu}_{\Delta_e}}{T} + \frac{\bar{\mu}_{\Delta_\mu}}{T} + \frac{\bar{\mu}_{\Delta_\tau}}{T}\right) \,,
\end{align}
and
\begin{equation}
c_{B-L}^{\rm win} = \frac{5}{36} + \frac{3}{4}\,c_\perp^* \,, \qquad c_B^{\rm win} = \frac{5}{111} + \frac{9}{37}\,c_\perp^* \,, \qquad c_5^> = \frac{4841}{972} - \frac{5}{54}\,c_{B-L}^* \,, \qquad c_5^< =\frac{1073}{216} - \frac{5}{72}\,c_\perp^* \,. 
\end{equation}
Here, we used the relation $\eta_e = \eta_\mu = \eta_\tau = \frac{1}{3}\left(\eta_e + \eta_\mu + \eta_\tau\right)$ in Eq.~\eqref{eq:etaBione}, which results in the cancellation of all three chemical potentials in $\eta_{B-L}$.
In the two-flavor regime, we update Eqs.~\eqref{eq:washiniiB} to \eqref{eq:washiniiY} and thus find
\begin{equation}
\begin{pmatrix}
\eta_{\Delta_e} \\ \eta_{\Delta_\mu}
\end{pmatrix} =
\begin{pmatrix}
-\sfrac{1}{5} & \sfrac{4}{15} & \sfrac{2}{15} & \sfrac{1}{5} & \sfrac{2}{5} & -\sfrac{2}{3} & -\sfrac{3}{5} \\
-\sfrac{1}{5} & \sfrac{4}{15} & \sfrac{2}{15} & \sfrac{1}{5} & -\sfrac{3}{5} & -\sfrac{2}{3} & \sfrac{2}{5}
\end{pmatrix}
\begin{pmatrix}
\eta_{\Delta_\tau} \\ \eta_u \\ \eta_B \\ \eta_\tau \\ \eta_\mu \\ \eta_{u-d} \\ \eta_e 
\end{pmatrix} = 
\begin{pmatrix}
\sfrac{1}{9} - \sfrac{1}{5}\,c_\perp^*\\ \sfrac{1}{9} - \sfrac{1}{5}\,c_\perp^*
\end{pmatrix} \eta_\chi \,,
\end{equation}
alongside the total $B\!-\!L$ asymmetry
\begin{align}
\eta_{B-L} = \eta_{\Delta_e} + \eta_{\Delta_\mu} + \eta_{\Delta_\perp} & = \frac{8}{15}\,\eta_u + \frac{4}{15}\,\eta_B + \frac{2}{5}\,\eta_\tau - \frac{1}{5}\left(\eta_e + \eta_\mu\right) - \frac{4}{3}\,\eta_{u-d} + \frac{3}{5}\,\eta_{\Delta_\tau} \\
& = \frac{8}{15}\,\eta_u + \frac{4}{15}\,\eta_B - \frac{4}{3}\,\eta_{u-d} + \frac{3}{5}\,\eta_{\Delta_\perp} \,,
\end{align}
and the coefficients
\begin{equation}
c_{B-L}^{\rm win} = \frac{2}{9} + \frac{3}{5}\,c_\perp^* \,, \qquad c_B^{\rm win} = \frac{8}{111} + \frac{36}{185}\,c_\perp^* \,, \qquad c_5^> = \frac{4841}{972} - \frac{5}{54}\,c_{B-L}^* \,, \qquad c_5^< = \frac{1607}{324} - \frac{1}{18}\,c_\perp^* \,. 
\end{equation}
In the three-flavor regime, finally, we update Eqs.~\eqref{eq:washiniiC} to \eqref{eq:washiniiZ} and thus obtain
\begin{equation}
\begin{pmatrix}
\eta_{\Delta_e} \\ \eta_{\Delta_\mu} \\ \eta_{\Delta_\mu}
\end{pmatrix} =
\begin{pmatrix}
\sfrac{2}{9} & \sfrac{1}{9} & \sfrac{1}{3} & \sfrac{1}{3} & -\sfrac{5}{9} & -\sfrac{2}{3} \\
\sfrac{2}{9} & \sfrac{1}{9} & \sfrac{1}{3} & -\sfrac{2}{3} & -\sfrac{5}{9} & \sfrac{1}{3} \\
\sfrac{2}{9} & \sfrac{1}{9} & -\sfrac{2}{3} & \sfrac{1}{3} & -\sfrac{5}{9} & \sfrac{1}{3}
\end{pmatrix}
\begin{pmatrix}
\eta_u \\ \eta_B \\ \eta_\tau \\ \eta_\mu \\ \eta_{u-d} \\ \eta_e 
\end{pmatrix} = 
\begin{pmatrix}
\sfrac{5}{54} \\ \sfrac{5}{54} \\ \sfrac{5}{54}
\end{pmatrix} \eta_\chi \,,
\end{equation}
together with 
\begin{equation}
\eta_{B-L} = \frac{2}{3}\,\eta_u + \frac{1}{3}\,\eta_B - \frac{5}{3}\,\eta_{u-d} \,, \qquad
c_{B-L}^{\rm win} = \frac{5}{18} \,, \qquad c_B^{\rm win} = \frac{10}{111} \,, \qquad c_5^> = \frac{4841}{972} \,, \qquad c_5^< = \frac{1204}{243} \,.
\end{equation}


These results complete our discussion of wash-in leptogenesis in temperature regimes (i) to (v). 
An overview of all the numerical coefficients $c_{B-L}^{\rm win}$, $c_B^{\rm win}$, $c_5^>$, and $c_5^<$ that we derived in this section can be found in Tab.~\ref{tab:coefficients}.
Our results for $c_B^{\rm win}$ in Tab.~\ref{tab:coefficients} can in particular be used in Eq.~\eqref{eq:etaBtot}, repeated here for convenience,
\begin{equation*}
\eta_B^{\rm tot} = \eta_B^{\rm win} + \eta_B^{\rm dec} \simeq 0.15\left(c_B^{\rm win} + c_B^{\rm dec}\right)\chi \simeq 6.6 \times 10^{-10}\:\Big(\frac{c_B^{\rm win} + c_B^{\rm dec}}{18/185 + 0.05}\Big)\left(\frac{\chi}{3 \times 10^{-8}}\right) \,,
\end{equation*}
to evaluate the total BAU that originates from fermion and gauge-field production during axion inflation.
At the same time, our results for $c_5^>$ and $c_5^<$ can be used in Eqs.~\eqref{eq:c5larger} and \eqref{eq:c5smaller} to evaluate $T_{\rm CPI}$ in Eq.~\eqref{eq:TCPI}.


\begin{table}
\begin{center}
\caption{Numerical coefficients describing the outcome of wash-in leptogenesis.
$c_{B-L}^{\rm win}$ and $c_B^{\rm win}$ relate the $B\!-\!L$ and $B$ asymmetries generated during wash-in leptogenesis to the $\eta_\chi$ reference asymmetry, $\eta_{B-L} = c_{B-L}^{\rm win}\,\eta_\chi$ and $\eta_B = c_B^{\rm win}\,\eta_\chi$ [see Eqs.~\eqref{eq:etaBetachi} and \eqref{eq:etaBLetachi}], which is valid at the time of sphaleron freeze-out, \textit{i.e.}, before the entropy injection in consequence of the decreasing number of relativistic degrees of freedom.
$c_5^>$ and $c_5^<$ can be used in Eqs.~\eqref{eq:c5larger} and \eqref{eq:c5smaller} to evaluate our estimate of $T_{\rm CPI}$ in Eq.~\eqref{eq:TCPI}.
The three different flavor regimes correspond to realizations of wash-in leptogenesis along one, two, or three linearly independent directions in the $e$\,--\,$\mu$\,--\,$\tau$ flavor space, respectively.
In the two-flavor scenario, $c_\perp^*$ measures the possibly nonzero pre-existing flavor asymmetry along the direction in $e$\,--\,$\mu$ flavor space that is immune to RHN wash-in, while in the one-flavor scenario, $c_\perp^*$ measures the possibly nonzero pre-existing flavor asymmetry in the two-dimensional flavor space that is orthogonal to the active flavor direction.
Correspondingly, the coefficients $c_{\Delta_{\alpha}}^*$ measure the possibly nonzero pre-existing flavor asymmetries along the respective directions in flavor space, $\alpha = e,\mu,\tau$, where we write $c_{\Delta_{e + \mu}}^* = c_{\Delta_e}^* + c_{\Delta_\mu}^*$ for the ease of notation and where $c_{B-L}^*$ stands for $c_{B-L}^* = c_{\Delta_e}^* + c_{\Delta_\mu}^* + c_{\Delta_\tau}^*$.}
\label{tab:coefficients}
\renewcommand{\arraystretch}{1.8}
\begin{tabular}{|c||cccccc|}
\hline
$T\,\left[\textrm{GeV}\right] $ & $\left(0,10^5\right)$ & $\left(10^5,10^6\right)$ & $\left(10^6,10^9\right)$ & $\left(10^9,10^{11-12}\right)$ & $\left(10^{11-12},10^{13}\right)$ & $\left(10^{13},10^{15}\right)$ \\
\hline\hline
\multicolumn{7}{|c|}{Three-flavor regime} \\
\hline\hline
$c_{B-L}^{\rm win}$ & $0$ & $\frac{3}{10}$ & $\frac{10}{17}$ & $\frac{8}{9}$ & $\frac{51}{88}$ & $\frac{5}{18}$ \\
\hline
$c_B^{\rm win}$ & $0$ & $\frac{18}{185}$ & $\frac{120}{629}$ & $\frac{32}{111}$ & $\frac{153}{814}$ & $\frac{10}{111}$ \\
\hline
$c_5^>$ & $0$ & $\frac{711}{481}$ & $\frac{856}{537}$ & $\frac{1765}{589}$ & $\frac{1617}{368}$ & $\frac{4841}{972}$ \\
\hline
$c_5^<$  & $0$ & $\frac{11}{10}$ & $\frac{61}{51}$ & $\frac{67}{27}$ & $\frac{1509}{352}$ & $\frac{1204}{243}$ \\
\hline\hline
\multicolumn{7}{|c|}{Two-flavor regime} \\
\hline\hline
$c_{B-L}^{\rm win}$ &   &   &   & $\frac{64}{247} + \frac{225}{247}\,c_\perp^*$ & $\frac{17}{37} + \frac{23}{37}\,c_\perp^*$ & $\frac{2}{9} + \frac{3}{5}\,c_\perp^*$ \\
\hline
$c_B^{\rm win}$ &   &   &   & $\frac{768}{9139} + \frac{2700}{9139}\,c_\perp^*$ & $\frac{204}{1369} + \frac{276}{1369}\,c_\perp^*$ & $\frac{8}{111} + \frac{36}{185}\,c_\perp^*$ \\
\hline
$c_5^>$ &   &   &   & $\frac{1765}{589} - \frac{188}{589}\,c_{\Delta_{e+\mu}}^* + \frac{88}{589}\,c_{\Delta_\tau}^*$ & $\frac{1617}{368} - \frac{17}{92}\,c_{B-L}^*$ & $\frac{4841}{972} - \frac{5}{54}\,c_{B-L}^*$ \\
\hline
$c_5^<$  &   &   &   & $\frac{671}{247} - \frac{84}{247}\,c_\perp^*$ & $\frac{2551}{592} - \frac{17}{148}\,c_\perp^*$ & $\frac{1607}{324} - \frac{1}{18}\,c_\perp^*$ \\
\hline\hline
\multicolumn{7}{|c|}{One-flavor regime} \\
\hline\hline
$c_{B-L}^{\rm win}$ &   &   &   &   & $\frac{17}{60} + \frac{23}{30}\,c_\perp^*$ & $\frac{5}{36} + \frac{3}{4}\,c_\perp^*$ \\
\hline
$c_B^{\rm win}$ &   &   &   &   & $\frac{17}{185} + \frac{46}{185}\,c_\perp^*$ & $\frac{5}{111} + \frac{9}{37}\,c_\perp^*$ \\
\hline
$c_5^>$ &   &   &   &   & $\frac{1617}{368} - \frac{17}{92}\,c_{B-L}^*$ & $\frac{4841}{972} - \frac{5}{54}\,c_{B-L}^*$ \\
\hline
$c_5^<$  &   &   &   &   & $\frac{521}{120} - \frac{17}{120}\,c_\perp^*$ & $\frac{1073}{216} - \frac{5}{72}\,c_\perp^*$\\
\hline
\end{tabular}
\end{center}
\end{table}


\subsection{Model-independent results}
\label{subsec:independent}


We are now able to combine all results derived in the previous sections and identify the viable regions in parameter space.
In doing so, let us first be slightly more general and discuss the implications of our analysis for a broader class of models of primordial magnetogenesis.
Thus far, our main focus has been on primordial magnetogenesis during axion inflation, which comes with two distinct advantages:
(i) First of all, the dual production of helical gauge fields and fermionic charge asymmetries during axion inflation is not impeded by plasma effects. 
The electric currents in a thermal plasma induce additional friction in the equation of motion of the gauge field, which renders gauge-field production less efficient. 
This problem is avoided if primordial magnetogenesis occurs during inflation.
(ii) A second advantage is that, during inflation, the axion field continuously rolls in the same direction in field space.
The sign of its velocity is hence fixed, which in turn leads to the amplification of only one helicity (negative or positive) in the gauge field.
This needs to be compared to scenarios in which an oscillating axion field is responsible for gauge-field production after inflation.
In such scenarios, the axion velocity repeatedly flips its sign as the axion oscillates around the minimum of its potential, which results in the amplification of both helicities and hence a reduced total helicity.


Nonetheless, if the axion--vector coupling is large enough, it might become feasible to generate a sufficiently large hypermagnetic helicity, alongside a corresponding set of fermionic charge asymmetries, in models of postinflationary axion evolution. 
An important aspect in this case consists in the fact that an oscillating axion field is also subject to Hubble friction, which leads to damped oscillations around the potential minimum.
It is therefore possible to generate a nonvanishing net helicity from asymmetric oscillations; see, \textit{e.g.}, Refs.~\cite{Fujita:2015iga,Adshead:2016iae}, which discuss axion-driven magnetogenesis at the time of inflaton oscillations during reheating. 
Meanwhile, it is challenging to realize efficient axion-driven magnetogenesis at later times in the cosmological evolution, \textit{e.g.}, in scenarios where the axion field begins to oscillate during the radiation-dominated era, without requiring a prohibitively large axion--vector coupling. 
To overcome this problem, one could imagine that the onset of axion oscillations is delayed by an additional time-dependent contribution to the axion mass, or one could consider axion oscillations during an early stage of matter domination.
A third option would be to replace the axion field by a complex field rolling in the complex plane~\cite{Kamada:2019uxp,Co:2019wyp,Co:2020xlh,Co:2020jtv}, similarly as in the Affleck--Dine mechanism~\cite{Affleck:1984fy,Dine:1995uk,Dine:1995kz}, such that the sign of the axion velocity remains unchanged during magnetogenesis and helicity production becomes more efficient.


Precisely estimating the resultant helicity and fermion asymmetries in these alternative cases is more involved than in our scenario based on axion inflation. 
We therefore do not make an attempt at such an estimate, leaving a more detailed investigation for future work, but simply remark that we expect the parametrization introduced in Sec.~\ref{sec:ic} to be useful for alternative scenarios of magnetogenesis as well.
That is, we expect that, also in postinflationary scenarios, it should be possible to characterize the outcome of magnetogenesis in terms of (i) the Hubble rate at the end of magnetogenesis, $H_{\rm end}$; (ii) a typical length scale, $c_\lambda H_{\rm end}^{-1}$; (iii) a typical time scale, $c_\tau H_{\rm end}^{-1}$; and (iv) a dimensionless helicity yield parameter, $\chi$.
In addition, helicity production after inflation will also be accompanied by the creation of fermionic charge asymmetries.
During axion inflation, the relation among the various charges is determined by Eq.~\eqref{eq:qiend}; in other scenarios, the precise relations are going to depend on the set of equilibrated SM interactions at the time of magnetogenesis.


If magnetogenesis occurs during radiation domination, the Hubble rate $H_{\rm end}$ can be readily related to the corresponding temperature scale,
\begin{equation}
T_{\rm end} = \left(\frac{90}{\pi^2 g_*}\right)^{1/4} \left(H_{\rm end}\,M_{\rm Pl}\right)^{1/2} \,.
\end{equation}
In the case of axion inflation, the end of magnetogenesis coincides with the end of inflation. 
In the approximation of instantaneous reheating, which we have been relying upon thus far, the temperature $T_{\rm end}$ thus coincides with the reheating temperature $T_{\rm rh}$.
Together with the parameter $\chi$, this temperature scale spans the two-dimensional parameter space that we are going to be interested in now.
In the following, we shall identify the viable regions in the $\chi$\,--\,$T_{\rm rh}$ plane (or equivalently, $\chi$\,--\,$T_{\rm end}$ plane in postinflationary scenarios) that are safe from magnetic diffusion and the chiral plasma instability and that at the same time yield the correct baryon asymmetry.
In doing so, we will first assume $c_\lambda \sim c_\tau$ for concreteness, which implies the following simple relation between the magnitudes of the electric and the magnetic field at the end of magnetogenesis,
\begin{equation}
\label{eq:EBindependent}
\left<\bm{E}^2\right> \sim \left<\bm{B}^2\right> \sim \left|\left<\bm{E}\cdot\bm{B}\right>\right| \,,
\end{equation}
and allows us to rewrite the magnetic Reynolds numbers in Eqs.~\eqref{eq:Rm_max} and \eqref{eq:Rm_visc} as functions of $\chi$, where
\begin{equation}
\rho_B = \frac{\left<\bm{B}^2\right>}{2a^4} \sim \frac{3\pi}{2\alpha_Y}\,\chi\,H_{\rm end}^{\vphantom{3}}T_{\rm end}^3 \,.
\end{equation}
Up to possible modifications by the coefficients $c_\lambda$ and $c_\tau$, we expect this rough estimate to be representative of a broader class of magnetogenesis models that goes beyond the specific case of axion inflation.


\begin{figure}
\begin{center}
\includegraphics[width=0.95\textwidth]{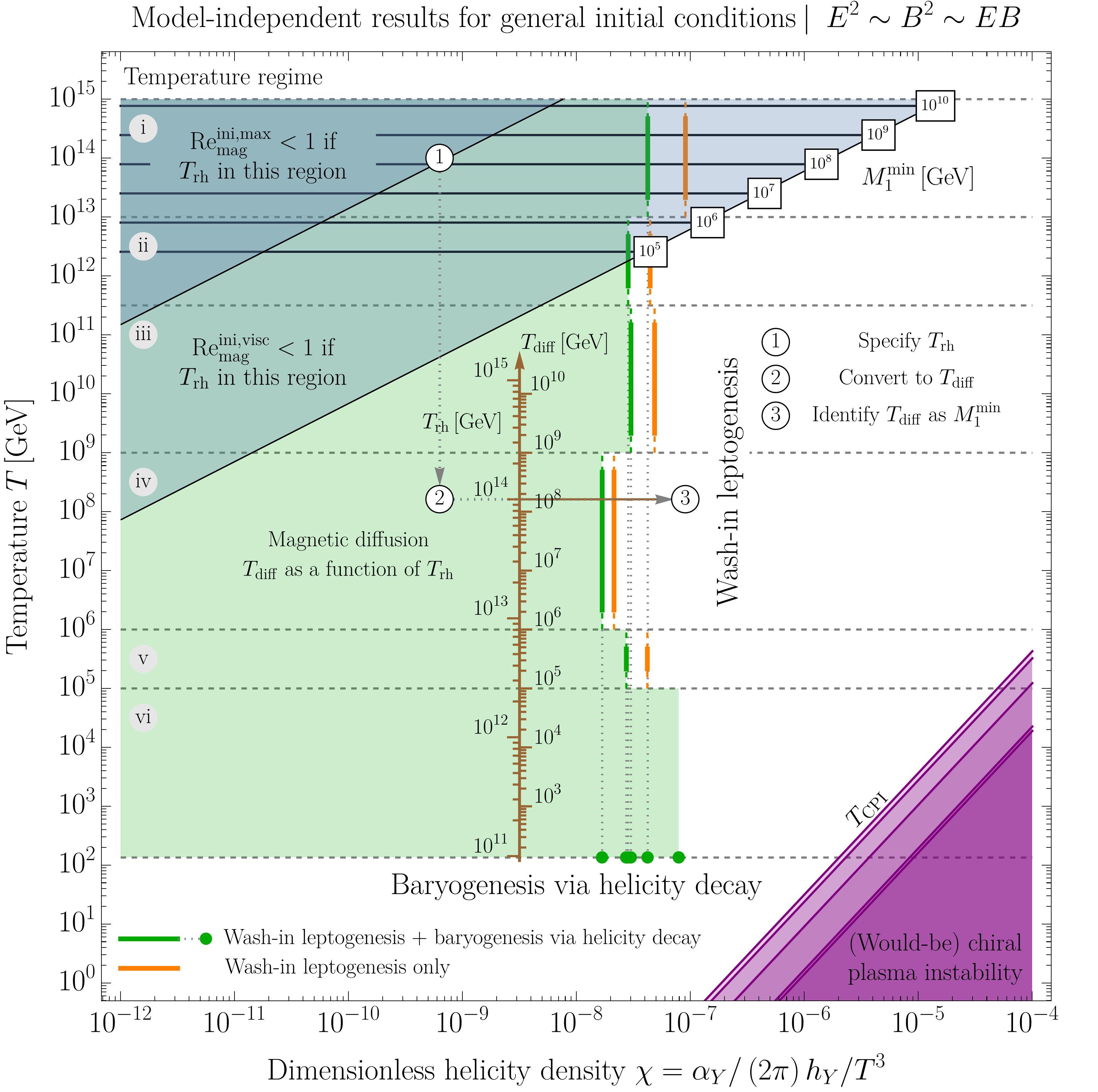}
\caption{Viable parameter space for successful baryogenesis after primordial hypermagnetogenesis.
In each temperature regime, (i) to (v), the vertical bars respectively indicate the required values of the dimensionless helicity density $\chi$ if the BAU receives contributions from wash-in leptogenesis \textit{and} baryogenesis via helicity (green bars) or from wash-in leptogenesis \textit{only} (orange bars).
In temperature regime (vi), $T_{\rm sph}\lesssim T \lesssim 10^5\,\textrm{GeV}$, we indicate the $\chi$ value that leads to correct baryon asymmetry if baryogenesis via helicity decay yields the only contribution to the BAU, $\chi \simeq 7.9 \times 10^{-8}$.
Smaller $\chi$ values are typically harmless, while larger values will lead to the overproduction of baryon number, barring cancellations with other contributions of different origin.
The shaded region in the lower right corner highlights where the chiral plasma instability \textit{would} occur, around $T \sim T_{\rm CPI}$ [see Eq.~\eqref{eq:TCPI}], if it were not already rendered ineffective at higher temperatures by the electron Yukawa interaction.
If the initial conditions after the end of hypermagnetogenesis are located in the shaded region in the top left corner, the initial magnetic Reynolds number (either $\textrm{Re}_{\rm mag}^{\rm ini, max}$ or $\textrm{Re}_{\rm mag}^{\rm ini, visc}$) is expected to be smaller than unity [see Eqs.~\eqref{eq:Rm_max} and \eqref{eq:Rm_visc}], which results in magnetic diffusion setting in at temperatures $T \sim T_{\rm diff}$ [see Eq.~\eqref{eq:Tdiff}].
The temperature $T_{\rm diff}$ can be read off from the brown axis in dependence of the initial temperature $T_{\rm rh}$ (or $T_{\rm end}$).
The requirement $T_{B-L} > T_{\rm diff}$ can moreover be turned into a lower bound $M_1^{\rm min}$ on the $N_1$ mass.
See text for more details.}
\label{fig:modelindependent}
\end{center} 
\end{figure}


We now have all ingredients at our disposal to discuss the constraints on the $\chi$\,--\,$T_{\rm rh}$ plane (see Fig.~\ref{fig:modelindependent}).
For illustrative purpose, Fig.~\ref{fig:modelindependent} highlights the $\chi$ values that are necessary to obtain the correct BAU in temperature regimes (i) to (v) assuming wash-in leptogenesis to be driven exclusively by $N_1$ RHNs. 
That is, in temperature regimes (i) and (ii), we assume one active wash-in direction in flavor space; in temperature regime (iii), two active wash-in directions in flavor space; and in temperature regimes (iv) and (v), three active wash-in directions in flavor space.
Furthermore, we indicate the required $\chi$ values for wash-in leptogenesis in isolation [see Eq.~\eqref{eq:etaBwin2}] and for wash-in leptogenesis in combination with baryogenesis from helicity decay [see Eq.~\eqref{eq:etaBtot}].
In both cases, we use the respective coefficients $c_B^{\rm win}$ listed in Tab.~\ref{tab:coefficients}, and in the latter case, we set $c_B^{\rm dec} = 0.05$.
The temperature regime in which $\chi$ needs to be read off from Fig.~\ref{fig:modelindependent} is determined by the value of $T_{B-L}$.


In order to assess whether or not the different baryogenesis scenarios presented in Fig.~\ref{fig:modelindependent} are actually viable, we need to evaluate our estimates for the CPI temperature in Eq.~\eqref{eq:TCPI} and the magnetic Reynolds number in Eqs.~\eqref{eq:Rm_max} and \eqref{eq:Rm_visc}.
As for the CPI temperature, we use the coefficients $c_5^<$ listed in Tab.~\ref{tab:coefficients}, setting the contributions from any pre-existing asymmetries to zero, $c_\perp^* = 0$.
The resulting values of $T_{\rm CPI}$ as functions of $\chi$ are shown by the solid purple lines in the lower right corner of Fig.~\ref{fig:modelindependent}.
In view of these estimates of the CPI temperature, several comments are in order.


(i) First of all, we note that using $c_5^>$ rather than $c_5^<$ would result in very similar results.
All of these coefficients are of the same order of magnitude $c_5^> \sim c_5^< \sim 1$; the precise value of the numerical coefficient entering our estimate of $T_{\rm CPI}$ is therefore less relevant for our qualitative and quantitative conclusions.


(ii) In almost the entire $\chi$ range displayed in Fig.~\ref{fig:modelindependent}, $T_{\rm CPI}$ turns out to be smaller than $T_{y_e}$.
This means that, when the temperature of the thermal bath drops to $T \sim T_{\rm CPI}$, the chiral plasma instability will actually \textit{not} occur because, at $T \sim T_{y_e}$, the electron Yukawa interaction will have already erased the entire chiral chemical potential $\bar{\mu}_5$.
The corresponding region in Fig.~\ref{fig:modelindependent}, $T \leq T_{\rm CPI}$, is therefore labeled ``(would-be) chiral plasma instability'', in order to indicate that baryogenesis in this region \textit{would} be endangered if the chiral plasma instability were not already rendered ineffective by the electron Yukawa interaction at earlier times.


(iii) Successful baryogenesis after axion inflation requires $\chi$ values in the range $\chi \sim 10^{-\left(7\cdots8\right)}$.
For such small values, our estimate of the (would-be) CPI temperature turns out to be negligibly small.
The scenarios we are interested in therefore always feature a strong hierarchy of temperature scales, $T_{\rm CPI} \ll T_{B-L}$.%
\footnote{This hierarchy is also the reason why we evaluate $T_{\rm CPI}$ using the coefficients $c_5^<$ in Fig.~\ref{fig:modelindependent}; see, however, also comment (i) above.}


(iv) For completeness, let us also comment on the parameter region at very large $\chi$ values where $T_{\rm CPI} > T_{y_e}$ and the chiral plasma instability can actually occur.
In this case, we need to distinguish between two possible scenarios, $T_{y_e} < T_{B-L} < T_{\rm CPI}$ and $T_{y_e} < T_{\rm CPI} < T_{B-L}$.
The former scenario is less interesting, as it simply leads to the erasure of the hypermagnetic helicity and all fermion asymmetries, so that no baryon asymmetry is created.
However, in the latter scenario, $T_{y_e} < T_{\rm CPI} < T_{B-L}$, we observe an interesting interplay between wash-in leptogenesis, baryogenesis from helicity decay, and the chiral plasma instability.
Our first observation is that, for the large $\chi$ values that are necessary to realize this scenario, wash-in leptogenesis drastically overproduces the baryon asymmetry.
Since $B\!-\!L$ remains preserved during the chiral plasma instability, this contribution to the final BAU notably survives down to low temperatures.
At same time, the chiral $B\!-\!L$ charges generated during wash-in leptogenesis affect the outcome of the chiral plasma instability in a nontrivial way.
Consider, \textit{e.g.}, a scenario where $T_{\rm CPI}$ falls into temperature regime (iv), $T \in \left(10^6,10^9\right)\,\textrm{GeV}$, such that the chiral plasma instability strives to erase the chiral chemical potential in Eq.~\eqref{eq:mu5iv}.
In the presence of nonvanishing $B\!-\!L$ asymmetries, the condition $\bar{\mu}_5 = 0$ then implies nonvanishing chemical potentials $\bar{\mu}_e$ and $\bar{\mu}_{u-d}$ in dependence of the three $\bar{\mu}_{\Delta_\alpha}$, even after the completion of the chiral plasma instability.
This result deviates from the standard outcome of the chiral plasma instability, which typically results in the erasure of all fermionic charge asymmetries.
Furthermore, the conservation law in Eq.~\eqref{eq:qihlaw}, which applies to the charges $q_e$ and $q_{u-d}$ in temperature regime (iv), tells us that an incomplete erasure of $\bar{\mu}_e$ and $\bar{\mu}_{u-d}$ directly translates to an incomplete erasure of the hypermagnetic helicity during the chiral plasma instability.
A fraction of the helicity stored in the hypermagnetic field thus survives down to the electroweak phase transition and causes the generation of a second contribution to the BAU, on top of the already-too-large contribution from wash-in leptogenesis, via baryogenesis from helicity decay.
In passing, we mention that similar arguments can be used to derive new constraints on the size of primordial lepton flavor asymmetries in the early Universe~\cite{Domcke:2022uue}.
 

Next, we turn to our estimates of the magnetic Reynolds number based on Eqs.~\eqref{eq:Rm_max} and \eqref{eq:Rm_visc}.
In Fig.~\ref{fig:modelindependent}, we indicate the regions where these estimates return a magnetic Reynolds number smaller than unity.
If the initial conditions for the further evolution after magnetogenesis fall into these regions, the hypermagnetic field is not strong enough to develop a turbulent regime before the temperature reaches $T_{\rm diff}$ in Eq.~\eqref{eq:Tdiff}. 
In this case, magnetic diffusion will set in at $T \sim T_{\rm diff}$ and erase the helicity stored in the hypermagnetic field as well as the fermionic charge asymmetries.
If wash-in leptogenesis already occurs at temperatures above this threshold, $T_{B-L} > T_{\rm diff}$, the corresponding contribution to the BAU will survive down to low temperatures, whereas there will be no contribution to the final asymmetry from baryogenesis via helicity decay.
Therefore, if magnetic diffusion occurs and $T_{B-L} > T_{\rm diff}$, the relevant $\chi$ values resulting in the correct BAU correspond to the red vertical bars in Fig.~\ref{fig:modelindependent} that are labeled ``wash-in leptogenesis only''.
Conversely, if magnetic diffusion can be avoided thanks to turbulence, the relevant $\chi$ values resulting in the correct BAU correspond to the green vertical bars in Fig.~\ref{fig:modelindependent} that are labeled ``wash-in leptogenesis $+$ baryogenesis via helicity decay''.


Whether or not wash-in leptogenesis takes place early enough, before the onset of magnetic diffusion, depends on the RHN mass scale [see Eq.~\eqref{eq:TBL}].
The requirement that $T_{B-L}$ be larger than $T_{\rm diff}$ in scenarios with small initial magnetic Reynolds number can therefore be formulated as a lower bound on $M_1$, if we assume that wash-in leptogenesis is exclusively driven by $N_1$ RHNs. 
This bound can be determined according to the following algorithm:
(i) Choose an initial temperature $T_{\rm rh}$ (or $T_{\rm end}$) at the end of magnetogenesis.
(ii) Use the brown axis in Fig.~\ref{fig:modelindependent} to read off the corresponding value of $T_{\rm diff}$. 
That is, focus on the ticks on the left-hand side of this axis and find the location that corresponds to the chosen value of $T_{\rm rh}$.
Then, switch to the ticks on the opposite (right-hand) side of the axis and read off the desired value of $T_{\rm diff}$. 
(iii) Identify this value of $T_{\rm diff}$ as the corresponding lower bound on $M_1$ that follows from the requirement $T_{B-L} > T_{\rm diff}$.
The lower bound on the $N_1$ mass found in this way is well approximated by [see also Eqs.~\eqref{eq:Treh} and \eqref{eq:Tdiff}]
\begin{equation}
M_1^{\rm min} \sim T_{\rm diff} \sim 10^8\,\textrm{GeV} \:\left(\frac{T_{\rm rh}}{10^{14}\,\textrm{GeV}}\right)^2 \,.
\end{equation}


In Fig.~\ref{fig:modelindependent}, we explicitly indicate the lower bounds on $M_1$ found in this way in the top part of the plot, \textit{i.e.}, in the regions with small initial magnetic Reynolds number.
In view of these results, we conclude that scenarios with $\chi \sim 10^{-\left(7\cdots8\right)}$ and $T_{\rm rh} \lesssim 10^{12}\,\textrm{GeV}$ manage to successfully generate the BAU without any disturbance from magnetic diffusion or the chiral plasma instability. 
In temperature regimes (ii) to (v), it is therefore possible to obtain the correct BAU from the combination of wash-in leptogenesis and baryogenesis via helicity decay.
For $\chi \sim 10^{-\left(7\cdots8\right)}$ and larger values of $T_{\rm rh}$, the chiral plasma instability still plays no role; but magnetic diffusion may become relevant.
This is not the case as long as we estimate the magnetic Reynolds number in terms of $\textrm{Re}_{\rm mag}^{\rm ini, max}$ [see Eq.~\eqref{eq:Rm_max}], but needs to be taken into account when we estimate it instead in terms of $\textrm{Re}_{\rm mag}^{\rm ini, visc}$ [see Eq.~\eqref{eq:Rm_visc}].
More precisely, for $T_{\rm rh} \simeq 10^{13}\,\textrm{GeV}$, magnetic diffusion sets in around $T_{\rm diff} \simeq 10^6\,\textrm{GeV}$, if we rely on our estimate in Eq.~\eqref{eq:Rm_visc}.
Thus, there will be no contribution to the BAU from baryogenesis via helicity decay; the $N_1$ mass needs to be at least as large as $M_1^{\rm min} \simeq 10^6\,,\textrm{GeV}$; and wash-in leptogenesis can only be realized in temperature regimes (ii) to (iv).
Similarly, for $T_{\rm rh} \simeq 10^{14}\,\textrm{GeV}$ and again estimating the magnetic Reynolds number in terms of Eq.~\eqref{eq:Rm_visc}, magnetic diffusion sets in around $T_{\rm diff} \simeq 10^8\,\textrm{GeV}$.
The only contribution to the final BAU therefore then stems again from wash-in leptogenesis; the $N_1$ mass needs to be at least as large as $M_1^{\rm min} \simeq 10^8\,\textrm{GeV}$; and wash-in leptogenesis can only be realized in temperature regimes (i) to (iv).


\subsection{Estimates for axion inflation}
\label{subsec:estimates}


\begin{figure}
\begin{center}
\includegraphics[width=0.49\textwidth]{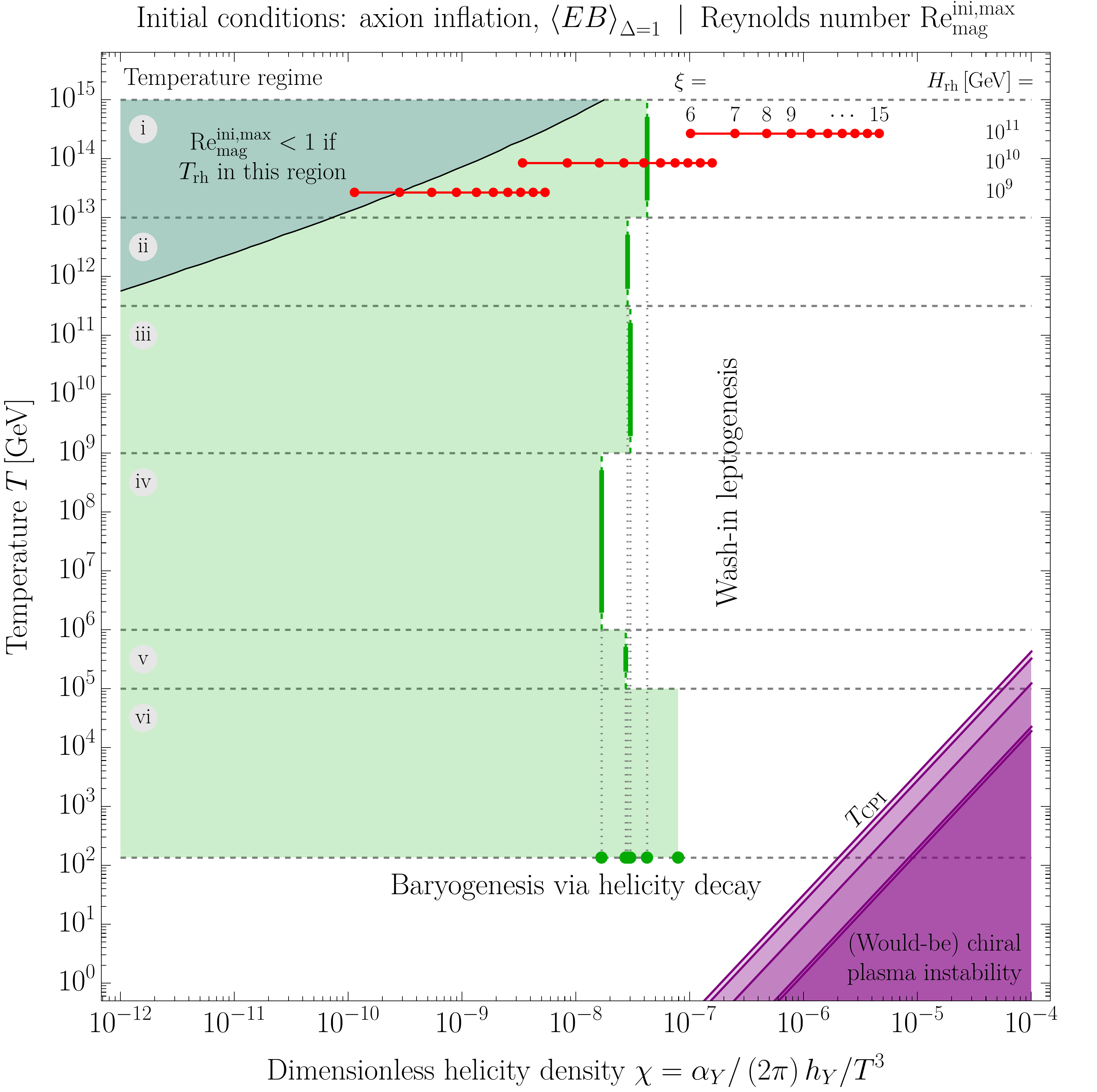}\quad
\includegraphics[width=0.49\textwidth]{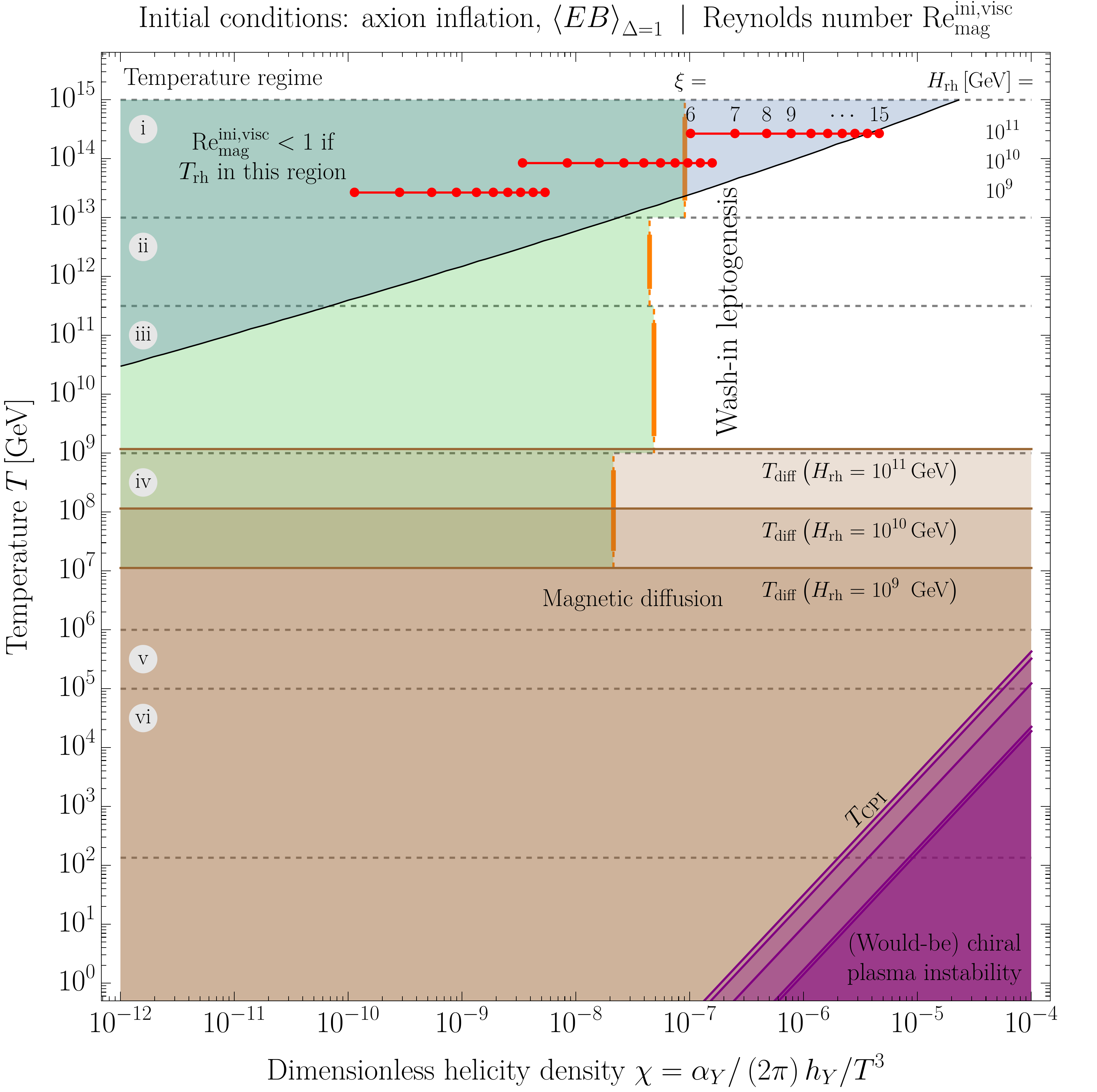}

\bigskip

\includegraphics[width=0.49\textwidth]{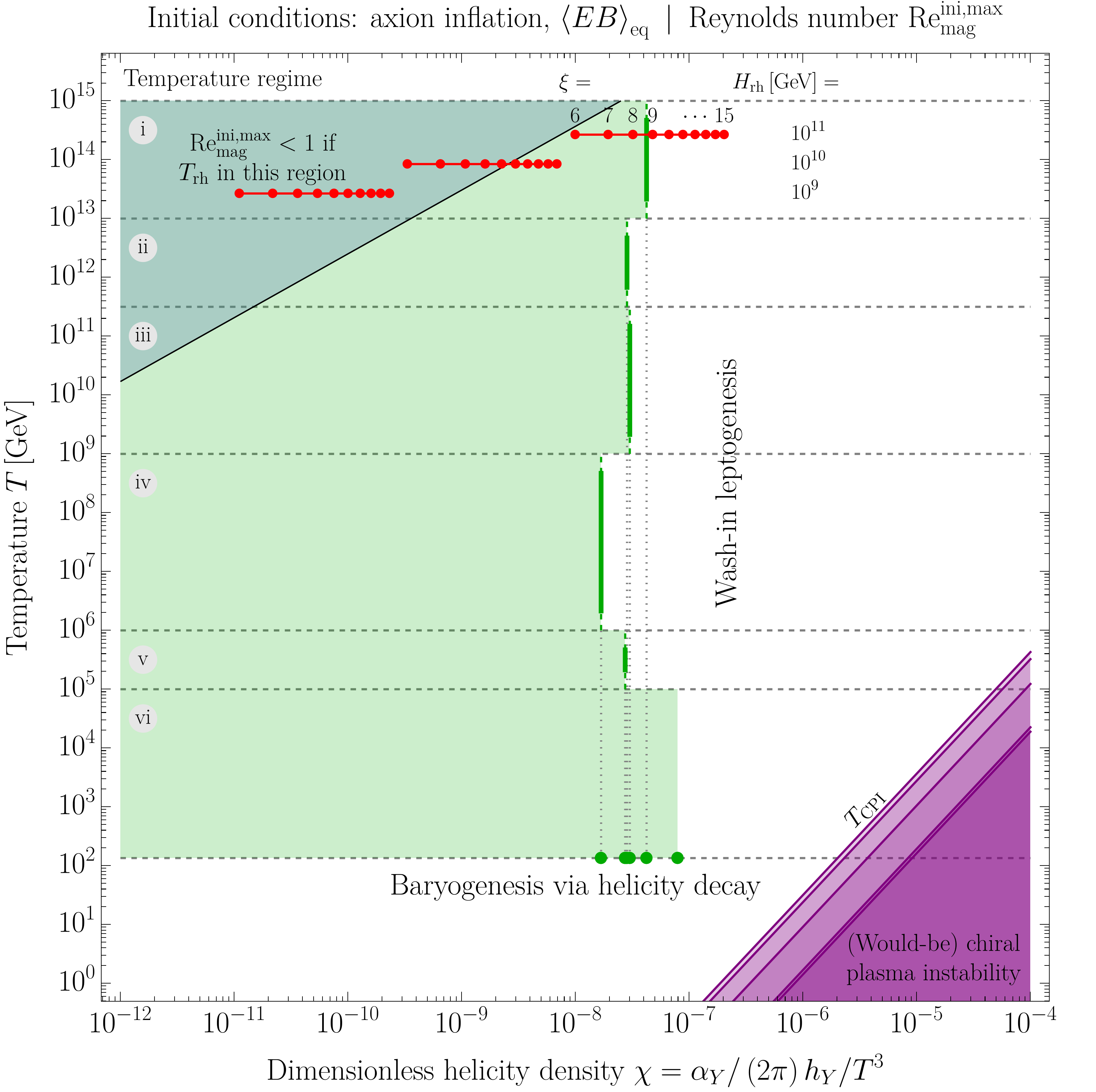}\quad
\includegraphics[width=0.49\textwidth]{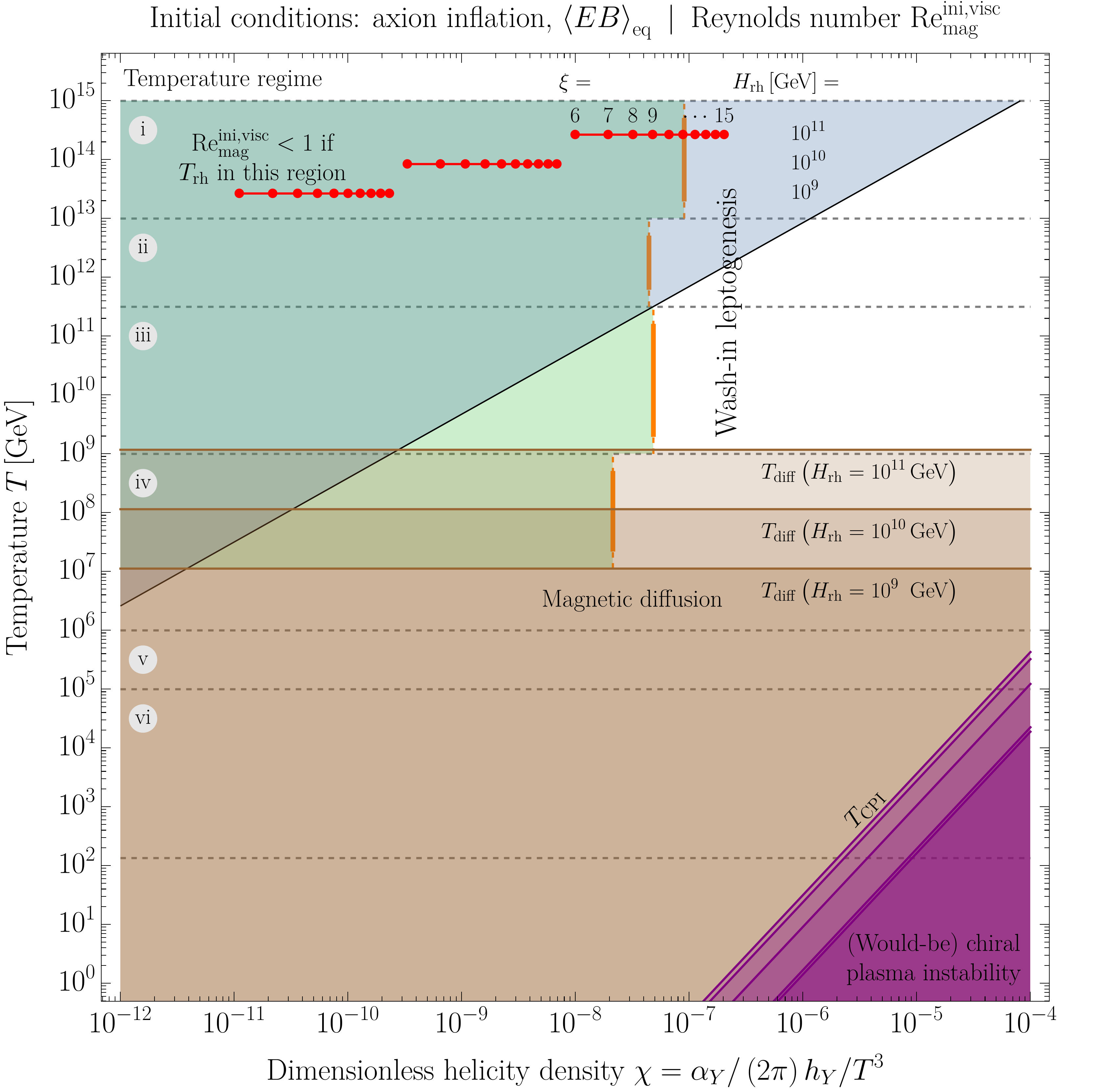}
\caption{Viable parameter space for successful baryogenesis after axion inflation.
The four panels differ from each other in terms of the estimate of the magnetic Reynolds number, $\textrm{Re}_{\rm mag}^{\rm ini, max}$ (left column) or $\textrm{Re}_{\rm mag}^{\rm ini, visc}$ (right column), as well as in terms of the method to relate the magnetic energy density to the dimensionless helicity density $\chi$, GEF estimate with $\Delta = 1$ in the electric picture (top row) or equilibrium estimate in the magnetic picture (bottom row) [see Eqs.~\eqref{eq:rhoBchi} and \eqref{eq:fit}].
The red points and red solid lines indicate possible values of $\chi$ and $T_{\rm rh}$ that can be achieved at the end of axion inflation in dependence of the parameters $\xi$ and $H_{\rm rh}$ (see Fig.~\ref{fig:chi}).
In contrast to Fig.~\ref{fig:modelindependent}, we no longer show the brown axis that allows one to convert from $T_{\rm rh}$ to $T_{\rm diff}$.
Instead, we explicitly mark the regions that are ruled out by magnetic diffusion if the initial magnetic Reynolds number is expected to be smaller than unity.
All other elements shown in the four plots in this figure are equivalent to the corresponding elements in Fig.~\ref{fig:modelindependent}.
See text for more details.}
\label{fig:viable}
\end{center}
\end{figure}
 

In the previous section, we made the simplifying assumption that $E_{\rm end}^2 \sim B_{\rm end}^2 \sim E_{\rm end}B_{\rm end}$ at the end of hypermagnetogenesis [see Eq.~\eqref{eq:EBindependent}].
Let us now focus on axion inflation, in which case these relations can be made slightly more precise.
For concreteness, we shall consider two of the approaches that we discussed in Sec.~\ref{subsec:estimates}: (i) the equilibrium estimate in the magnetic picture [see Eq.~\eqref{eq:EBeq}] and (ii) the GEF estimate in the electric picture with the damping factor $\Delta$ set to $\Delta = 1$ [see Eq.~\eqref{eq:delta}].
In both cases, our numerical results for the electric and magnetic field strengths allow us to derive simple fit formulas that relate the magnetic energy density $\rho_B$ to the dimensionless helicity density $\chi$ and hence allow us to evaluate Eqs.~\eqref{eq:Rm_max} and \eqref{eq:Rm_visc},
\begin{equation}
\label{eq:rhoBchi}
\rho_B = \frac{\left<\bm{B}^2\right>}{2a^4} \,, \qquad \log_{10}\left(\frac{\left<\bm{B}^2\right>}{a^4H^4}\right) = F\left(x\right) \,, \qquad x = \log_{10}\left|\frac{\left<\bm{E}\cdot\bm{B}\right>}{a^4H^4}\right| = \log_{10}\left(\frac{3\pi}{\alpha_Y}\,\chi\,\frac{T^3}{H^3}\right)_{\rm end} \,, 
\end{equation}
where 
\begin{equation}
\label{eq:fit}
F\left(x\right) \simeq
\begin{cases}
-0.35 + 0.96\,x & \quad \textrm{;\quad equilibrium estimate} \\
\phantom{-}0.30 + 0.68\,x + 0.039\,x^2 & \quad \textrm{;\quad GEF estimate $\left(\Delta = 1\right)$} 
\end{cases} \,.
\end{equation}
Both fit functions manage to reproduce our exact numerical results with excellent accuracy.
For more details on semianalytical fit functions describing the outcome of the GEF approach, see the analysis in Ref.~\cite{Gorbar:2021zlr}.


In addition to our two different estimates of $\left<\bm{E}\cdot\bm{B}\right>$, we will also work again with our two estimates of the magnetic Reynolds number in Eqs.~\eqref{eq:Rm_max} and \eqref{eq:Rm_visc}.
In total, this results in four different combinations, for which we collectively analyze the viable parameter space in Fig.~\ref{fig:viable}.
The four panels in Fig.~\ref{fig:viable} contain similar information as Fig.~\ref{fig:modelindependent} for the most part.
However, on top, we also indicate typical initial conditions after axion inflation, in terms of $\chi$ and $T_{\rm rh}$ values that follow from typical values of the parameters $\xi$ and $H_{\rm rh}$ in Fig.~\ref{fig:chi}.
Based on this information, we are then able to draw the following conclusions.


If our optimistic estimate of the magnetic Reynolds number, $\textrm{Re}_{\rm mag}^{\rm ini, max}$, can be trusted, baryogenesis after axion inflation can proceed without any interference from magnetic diffusion or the chiral plasma instability.
This means that both wash-in leptogenesis and baryogenesis via helicity decay contribute to the final baryon asymmetry, and remarkably enough, they manage to produce the observed BAU in exactly the strip of parameter space that is left unaffected by these potentially dangerous phenomena. 
The GEF estimate of the efficiency of hypermagnetogenesis then points to $\xi \sim 10$ and $H_{\rm rh} \sim 10^{10}\,\textrm{GeV}$, whereas the equilibrium estimate in the magnetic picture signals a preference for a slightly larger Hubble rate, \textit{e.g.}, $\xi \sim 9$ and $H_{\rm rh} \sim 10^{11}\,\textrm{GeV}$.


On the other hand, if we trust our less optimistic estimate for the magnetic Reynolds number, $\textrm{Re}_{\rm mag}^{\rm ini, visc}$, we arrive at the conclusion that the initial conditions after axion inflation are not sufficient to avoid the onset of magnetic diffusion at later times.
Baryogenesis via helicity decay does not contribute to the final baryon asymmetry in this case, so that the entire BAU solely originates from wash-in leptogenesis.
Using the GEF estimate of $\left<\bm{E}\cdot\bm{B}\right>$, we observe that $\xi \sim 6$ and $H_{\rm rh} \sim 10^{11}\,\textrm{GeV}$ promises to result in favorable initial conditions after axion inflation.
Magnetic diffusion will then set in around $T_{\rm diff} \sim 10^9\,\textrm{GeV}$, which still leaves enough room for successful wash-in leptogenesis in temperature regimes (i) to (iii).
Alternatively, one may go to slightly smaller values of the Hubble rate and larger $\xi$ values, \textit{e.g.}, $\xi \sim 13$ and $H_{\rm rh} \sim 10^{10}\,\textrm{GeV}$, which lowers the diffusion temperature by an order of magnitude and hence also allows for wash-in leptogenesis in temperature regime (iv). 
In both cases, the mass of the RHN $N_1$ needs to be large enough in order to trigger wash-in leptogenesis before the onset of magnetic diffusion, which amounts to a lower bound of $M_1^{\rm min} \sim 10^{8\cdots9}\,\textrm{GeV}$.
Finally, for the equilibrium estimate of $\left<\bm{E}\cdot\bm{B}\right>$ in the magnetic picture, the situation becomes even more restricted.
In this case, we have to work with $\xi \sim 11$ and $H_{\rm rh} \sim 10^{11}\,\textrm{GeV}$, which eliminates the possibility of wash-in leptogenesis in temperature regime (iv) and raises the lower bound on the $N_1$ mass to $M_1^{\rm min} \sim 10^9\,\textrm{GeV}$.


Despite these bounds, Fig.~\ref{fig:viable} clearly illustrates that wash-in leptogenesis after axion inflation is a viable option across large regions of parameter space, even if we estimate the magnetic Reynolds number in a slightly less optimistic way.
Wash-in leptogenesis can especially rescue the successful generation of the BAU in parameter regions where baryogenesis via helicity decay is spoiled by magnetic diffusion.
In this sense, wash-in leptogenesis manages to restore parameter regions that were already deemed unviable in earlier work that exclusively focused on baryogenesis via helicity decay and did not consider the existence of RHNs~\cite{Domcke:2019mnd}.


\section{Conclusions}
\label{sec:conclusions}


The addition of RHNs to the SM particle content has nontrivial implications for the chemical transport in the primordial plasma in the early Universe.
Similar to weak sphaleron processes, which can wash in a baryon asymmetry in the presence of a primordial lepton asymmetry, RHN processes can wash in a $B\!-\!L$ asymmetry in the presence of primordial input charges.
This observation is the basis for the mechanism of wash-in leptogenesis, which we introduced in Ref.~\cite{Domcke:2020quw} and which generalizes standard thermal leptogenesis to situations featuring a nontrivial chemical background induced by new $CP$-violating dynamics at higher temperatures.
In scenarios of wash-in leptogenesis, the energy scales of $CP$ and $B\!-\!L$ violation are hence separated from each other, which offers a wealth of opportunities for model building.
The RHN sector is no longer burdened with the requirement of sufficiently large $CP$ violation, and $CP$ violation in general no longer has to be tied to the generation of $B\!-\!L$.
Instead, it suffices if a dynamical $CP$-violating process at high energies, which may be referred to as \textit{chargegenesis}, results in any of the conserved global charges that are present in the SM thermal bath at high temperatures.
In the extreme case of temperatures in the range $T \sim 10^{13\cdots15}\,\textrm{GeV}$, this means that chargegenesis simply needs to produce a subset of the in total 15 available conserved charges,
\begin{equation}
q_e \,,\quad q_{2 B_1 - B_2 - B_3} \,,\quad q_{u-d} \,,\quad q_{d-s} \,,\quad q_{B_1 - B_2} \,,\quad q_\mu \,,\quad q_{u-c} \,,\quad q_\tau \,,\quad q_{d-b} \,,\quad q_B \,,\quad q_u \,.
\end{equation}
The RHN interactions at lower temperatures will then act upon the nontrivial chemical background induced by chargegenesis, which will lead to $B\!-\!L$ violation and hence \textit{baryogenesis via leptogenesis via chargegenesis}.


One can imagine a variety of possible chargegenesis scenarios (see also Ref.~\cite{Domcke:2020quw}). 
Scenarios that predominantly result in the production of a charge asymmetry in right-handed electrons or muons may, \textit{e.g.}, be referred to as \textit{electrogenesis} and \textit{muogenesis}, and so on and so forth.
Among these various possibilities, we focused on a particularly attractive option in this paper: axion inflation coupled to the SM hypercharge gauge sector.
The nonperturbative dynamics of this model lead to the exponential amplification of the hypercharge gauge field in one of its two helicity states, which in turn results in fermion production from the strong gauge-field background.
As a consequence, axion inflation coupled to the hypercharge gauge sector leads to the dual production of maximally helical hypermagnetic fields and a set of fermionic charge asymmetries,
\begin{equation}
\label{eq:axionq}
q_{\rm CS} \,, \qquad q_e \,, \qquad q_{u-d} \,, \qquad q_\mu, \qquad q_\tau \,, \qquad q_B \,, \qquad q_u \,,
\end{equation}
where the CS density of the hypercharge gauge field $q_{CS}$ is a measure of the helicity stored in it.
As we were able to show, all of these primordial charges are controlled by a single effective and dimensionless parameter, $\chi$, which quantifies the amount of $CP$ violation induced in the system at the end of inflation,
\begin{equation}
\chi \sim - \left.\frac{\alpha_Y}{3\pi}\frac{\left<\bm{E}\cdot\bm{B}\right>}{a^4HT^3} \right|_{\rm end} \,.
\end{equation}
Baryogenesis after axion inflation requires a $\chi$ value of the order of $\chi \sim 10^{-\left(7\cdots8\right)}$ (see Fig.~\ref{fig:viable}).
The actual value of the BAU, $\eta_B^{\rm obs} \sim 10^{-9}$, then follows from the product of $\chi$, a dilution factor related to the change in the number of relativistic degrees of freedom in the early Universe, and a conversion factor $c_B = c_B^{\rm win} + c_B^{\rm dec} \sim 0.1$, which accounts for the contributions to the final baryon asymmetry from wash-in leptogenesis as well as from baryogenesis via helicity decay around the time of the EWPT.


In this paper, we calculated the conversion factor $c_B^{\rm win}$ as a function of the leptogenesis temperature $T_{B-L}$, in five separate temperature regimes, and in dependence of the number of active lepton flavors participating in the dynamics of wash-in leptogenesis (see Tab.~\ref{tab:coefficients}).
These results significantly extend our previous results in Ref.~\cite{Domcke:2020quw} and provide a consistent treatment of coherence\,/\,decoherence as well as heavy-neutrino flavor effects.
Equipped with the coefficients listed in Tab.~\ref{tab:coefficients}, we were then able to identify the viable regions in the two-dimensional parameter space spanned by $\chi$ and the reheating temperature $T_{\rm rh}$ (see Fig.~\ref{fig:viable}).
\textit{A priori}, one may worry that the global charges in Eq.~\eqref{eq:axionq} could be erased by magnetic diffusion or the chiral plasma instability soon after inflation.
Based on rough estimates of the diffusion temperature $T_{\rm diff}$ [see Eq.~\eqref{eq:Tdiff}], the magnetic Reynolds number [see Eqs.~\eqref{eq:Rm_max} and \eqref{eq:Rm_visc}], and the CPI temperature $T_{\rm CPI}$ [see Eq.~\eqref{eq:TCPI}], we, however, argued that baryogenesis after axion inflation typically takes place in parameter regions that are spared from these two phenomena.
Magnetic diffusion might become an issue, according to our less optimistic estimate of the magnetic Reynolds number; but even in this case, a large temperature window for successful wash-in leptogenesis remains (see the two plots in the right column of Fig.~\ref{fig:viable}).
In the absence of heavy RHNs, one would have to conclude in this case that baryogenesis after axion inflation does not succeed in explaining the observed BAU, \textit{i.e.}, there is no contribution to the BAU from baryogenesis via helicity decay.
However, adding RHNs with masses above $M_1^{\rm min} \sim 10^{8\cdots9}\,\textrm{GeV}$ to the SM particle content can rescue these scenarios.


Baryogenesis after axion inflation is tightly related to the rich phenomenology of axion inflation coupled to the hypercharge gauge sector, which in addition to the production of primordial hypermagnetic fields also includes the generation of primordial perturbations.
Primordial scalar perturbations generated during axion inflation can, \textit{e.g.}, lead to the production of primordial black holes, while primordial tensor perturbations generated during axion inflation manifest themselves as a stochastic gravitational-wave background in the present Universe.
In the future, it will therefore be important to advance our understanding of nonperturbative particle production during axion inflation (see Sec.~\ref{subsec:efficiency}) in order to arrive at a clear quantitative picture of the relation between baryogenesis after axion inflation on the one hand and the phenomenological predictions for hypermagnetic fields, primordial black holes, and gravitational waves on the other hand.
Similarly, it will be important to better understand the evolution of the hypermagnetic helicity and chiral charge asymmetries after inflation, which, if one wants to go beyond the simple estimates used in the present paper, will require dedicated numerical MHD simulations.
Such simulations will then hopefully also shine more light on the validity of our estimates in Eqs.~\eqref{eq:Tdiff}, \eqref{eq:Rm_max}, \eqref{eq:Rm_visc}, and \eqref{eq:TCPI}.
Finally, the treatment of wash-in leptogenesis itself needs to be refined, so as to obtain a formalism that would allow one to compute the final BAU at any given value of $T_{B-L}$, irrespective of the rough assumptions that went into the definition of our five temperature regimes (i) to (v).
The best approach in this regard will likely be a full-fledged density matrix formalism that would be capable of treating the transition regimes between our individual temperature regimes.


We leave these tasks for future work and conclude that wash-in leptogenesis after axion inflation, in combination with baryogenesis via helicity decay, represents an exciting early-Universe scenario\,---\,not only does it tackle the mystery of the BAU from a new angle, it also acts as a well-motivated benchmark model for upcoming searches for relic magnetic fields, black holes, and gravitational waves from the early Universe.


\subsection*{Acknowledgements}


The authors would like to thank Oleksandr Sobol for helpful discussions.
K.\,K.\ was supported by JSPS KAKENHI Grant-in-Aid for Scientific Research (C) JP19K03842. 
K.\,M.\ was supported by MEXT Leading Initiative for Excellent Young Researchers Grant No.\ JPMXS0320200430 and by JSPS KAKENHI Grant No.\ JP22K14044.
M.\,Y.\ was supported by MEXT Leading Initiative for Excellent Young Researchers and by JSPS KAKENHI Grants No.\ 20H05851 and 21K13910.


\small
\bibliographystyle{utphys}
\bibliography{arxiv_2}

\providecommand{\href}[2]{#2}\begingroup\raggedright\begin{thebibliography}{100}

\bibitem{Aghanim:2018eyx}
{\bfseries Planck} Collaboration, N.~Aghanim {\em et~al.}, ``{Planck 2018
  results. VI. Cosmological parameters},''
  \href{http://dx.doi.org/10.1051/0004-6361/201833910}{{\em Astron. Astrophys.}
  {\bfseries 641} (2020) A6}, \href{http://arxiv.org/abs/1807.06209}{{\ttfamily
  arXiv:1807.06209 [astro-ph.CO]}}.

\bibitem{Zyla:2020zbs}
{\bfseries Particle Data Group} Collaboration, P.~A. Zyla {\em et~al.},
  ``{Review of Particle Physics},''
  \href{http://dx.doi.org/10.1093/ptep/ptaa104}{{\em PTEP} {\bfseries 2020}
  (2020) 083C01}.

\bibitem{Fukugita:1986hr}
M.~Fukugita and T.~Yanagida, ``{Baryogenesis Without Grand Unification},''
  \href{http://dx.doi.org/10.1016/0370-2693(86)91126-3}{{\em Phys. Lett. B}
  {\bfseries 174} (1986) 45--47}.

\bibitem{Minkowski:1977sc}
P.~Minkowski, ``{$\mu \to e\gamma$ at a Rate of One Out of $10^{9}$ Muon
  Decays?},'' \href{http://dx.doi.org/10.1016/0370-2693(77)90435-X}{{\em Phys.
  Lett. B} {\bfseries 67} (1977) 421--428}.

\bibitem{Yanagida:1979as}
T.~Yanagida, ``{Horizontal gauge symmetry and masses of neutrinos},'' {\em
  Conf. Proc. C} {\bfseries 7902131} (1979) 95--99.

\bibitem{Yanagida:1980xy}
T.~Yanagida, ``{Horizontal Symmetry and Masses of Neutrinos},''
  \href{http://dx.doi.org/10.1143/PTP.64.1103}{{\em Prog. Theor. Phys.}
  {\bfseries 64} (1980) 1103}.

\bibitem{GellMann:1980vs}
M.~Gell-Mann, P.~Ramond, and R.~Slansky, ``{Complex Spinors and Unified
  Theories},'' {\em Conf. Proc. C} {\bfseries 790927} (1979) 315--321,
  \href{http://arxiv.org/abs/1306.4669}{{\ttfamily arXiv:1306.4669 [hep-th]}}.

\bibitem{Mohapatra:1979ia}
R.~N. Mohapatra and G.~Senjanovic, ``{Neutrino Mass and Spontaneous Parity
  Nonconservation},'' \href{http://dx.doi.org/10.1103/PhysRevLett.44.912}{{\em
  Phys. Rev. Lett.} {\bfseries 44} (1980) 912}.

\bibitem{Schechter:1980gr}
J.~Schechter and J.~W.~F. Valle, ``{Neutrino Masses in $SU(2) \times U(1)$
  Theories},'' \href{http://dx.doi.org/10.1103/PhysRevD.22.2227}{{\em Phys.
  Rev. D} {\bfseries 22} (1980) 2227}.

\bibitem{Schechter:1981cv}
J.~Schechter and J.~W.~F. Valle, ``{Neutrino Decay and Spontaneous Violation of
  Lepton Number},'' \href{http://dx.doi.org/10.1103/PhysRevD.25.774}{{\em Phys.
  Rev. D} {\bfseries 25} (1982) 774}.

\bibitem{Chun:2017spz}
E.~J. Chun {\em et~al.}, ``{Probing Leptogenesis},''
  \href{http://dx.doi.org/10.1142/S0217751X18420058}{{\em Int. J. Mod. Phys. A}
  {\bfseries 33} no.~05n06, (2018) 1842005},
  \href{http://arxiv.org/abs/1711.02865}{{\ttfamily arXiv:1711.02865
  [hep-ph]}}.

\bibitem{Bodeker:2020ghk}
D.~Bodeker and W.~Buchmuller, ``{Baryogenesis from the weak scale to the grand
  unification scale},''
  \href{http://dx.doi.org/10.1103/RevModPhys.93.035004}{{\em Rev. Mod. Phys.}
  {\bfseries 93} no.~3, (2021) 035004},
  \href{http://arxiv.org/abs/2009.07294}{{\ttfamily arXiv:2009.07294
  [hep-ph]}}.

\bibitem{Dasgupta:2021ies}
B.~Dasgupta and J.~Kopp, ``{Sterile Neutrinos},''
  \href{http://dx.doi.org/10.1016/j.physrep.2021.06.002}{{\em Phys. Rept.}
  {\bfseries 928} (2021) 63}, \href{http://arxiv.org/abs/2106.05913}{{\ttfamily
  arXiv:2106.05913 [hep-ph]}}.

\bibitem{Akhmedov:1998qx}
E.~K. Akhmedov, V.~A. Rubakov, and A.~Y. Smirnov, ``{Baryogenesis via neutrino
  oscillations},'' \href{http://dx.doi.org/10.1103/PhysRevLett.81.1359}{{\em
  Phys. Rev. Lett.} {\bfseries 81} (1998) 1359--1362},
  \href{http://arxiv.org/abs/hep-ph/9803255}{{\ttfamily arXiv:hep-ph/9803255}}.

\bibitem{Kuzmin:1985mm}
V.~A. Kuzmin, V.~A. Rubakov, and M.~E. Shaposhnikov, ``{On the Anomalous
  Electroweak Baryon Number Nonconservation in the Early Universe},''
  \href{http://dx.doi.org/10.1016/0370-2693(85)91028-7}{{\em Phys. Lett. B}
  {\bfseries 155} (1985) 36}.

\bibitem{Yoshimura:1978ex}
M.~Yoshimura, ``{Unified Gauge Theories and the Baryon Number of the
  Universe},'' \href{http://dx.doi.org/10.1103/PhysRevLett.41.281}{{\em Phys.
  Rev. Lett.} {\bfseries 41} (1978) 281--284}. [Erratum: Phys.Rev.Lett. 42, 746
  (1979)].

\bibitem{Dimopoulos:1978kv}
S.~Dimopoulos and L.~Susskind, ``{On the Baryon Number of the Universe},''
  \href{http://dx.doi.org/10.1103/PhysRevD.18.4500}{{\em Phys. Rev. D}
  {\bfseries 18} (1978) 4500--4509}.

\bibitem{Toussaint:1978br}
D.~Toussaint, S.~B. Treiman, F.~Wilczek, and A.~Zee, ``{Matter - Antimatter
  Accounting, Thermodynamics, and Black Hole Radiation},''
  \href{http://dx.doi.org/10.1103/PhysRevD.19.1036}{{\em Phys. Rev. D}
  {\bfseries 19} (1979) 1036--1045}.

\bibitem{Weinberg:1979bt}
S.~Weinberg, ``{Cosmological Production of Baryons},''
  \href{http://dx.doi.org/10.1103/PhysRevLett.42.850}{{\em Phys. Rev. Lett.}
  {\bfseries 42} (1979) 850--853}.

\bibitem{Barr:1979ye}
S.~M. Barr, G.~Segre, and H.~A. Weldon, ``{The Magnitude of the Cosmological
  Baryon Asymmetry},'' \href{http://dx.doi.org/10.1103/PhysRevD.20.2494}{{\em
  Phys. Rev. D} {\bfseries 20} (1979) 2494}.

\bibitem{Domcke:2020quw}
V.~Domcke, K.~Kamada, K.~Mukaida, K.~Schmitz, and M.~Yamada, ``{Wash-In
  Leptogenesis},'' \href{http://dx.doi.org/10.1103/PhysRevLett.126.201802}{{\em
  Phys. Rev. Lett.} {\bfseries 126} no.~20, (2021) 201802},
  \href{http://arxiv.org/abs/2011.09347}{{\ttfamily arXiv:2011.09347
  [hep-ph]}}.

\bibitem{Sakharov:1967dj}
A.~D. Sakharov, ``{Violation of CP Invariance, C asymmetry, and baryon
  asymmetry of the universe},''
  \href{http://dx.doi.org/10.1070/PU1991v034n05ABEH002497}{{\em Pisma Zh. Eksp.
  Teor. Fiz.} {\bfseries 5} (1967) 32--35}.

\bibitem{Campbell:1992jd}
B.~A. Campbell, S.~Davidson, J.~R. Ellis, and K.~A. Olive, ``{On the baryon,
  lepton flavor and right-handed electron asymmetries of the universe},''
  \href{http://dx.doi.org/10.1016/0370-2693(92)91079-O}{{\em Phys. Lett.}
  {\bfseries B297} (1992) 118--124},
\href{http://arxiv.org/abs/hep-ph/9302221}{{\ttfamily arXiv:hep-ph/9302221
  [hep-ph]}}.

\bibitem{Cline:1993vv}
J.~M. Cline, K.~Kainulainen, and K.~A. Olive, ``{On the erasure and
  regeneration of the primordial baryon asymmetry by sphalerons},''
  \href{http://dx.doi.org/10.1103/PhysRevLett.71.2372}{{\em Phys. Rev. Lett.}
  {\bfseries 71} (1993) 2372--2375},
  \href{http://arxiv.org/abs/hep-ph/9304321}{{\ttfamily arXiv:hep-ph/9304321}}.

\bibitem{Cline:1993bd}
J.~M. Cline, K.~Kainulainen, and K.~A. Olive, ``{Protecting the primordial
  baryon asymmetry from erasure by sphalerons},''
  \href{http://dx.doi.org/10.1103/PhysRevD.49.6394}{{\em Phys. Rev.} {\bfseries
  D49} (1994) 6394--6409},
\href{http://arxiv.org/abs/hep-ph/9401208}{{\ttfamily arXiv:hep-ph/9401208
  [hep-ph]}}.

\bibitem{Fukugita:2002hu}
M.~Fukugita and T.~Yanagida, ``{Resurrection of grand unified theory
  baryogenesis},'' \href{http://dx.doi.org/10.1103/PhysRevLett.89.131602}{{\em
  Phys. Rev. Lett.} {\bfseries 89} (2002) 131602},
  \href{http://arxiv.org/abs/hep-ph/0203194}{{\ttfamily arXiv:hep-ph/0203194}}.

\bibitem{Fong:2015vna}
C.~S. Fong, ``{Baryogenesis from Symmetry Principle},''
  \href{http://dx.doi.org/10.1016/j.physletb.2015.11.055}{{\em Phys. Lett. B}
  {\bfseries 752} (2016) 247},
  \href{http://arxiv.org/abs/1508.03648}{{\ttfamily arXiv:1508.03648
  [hep-ph]}}.

\bibitem{Fong:2021xmi}
C.~S. Fong, ``{Cosmic evolution of lepton flavor charges},''
  \href{http://dx.doi.org/10.1103/PhysRevD.105.043004}{{\em Phys. Rev. D}
  {\bfseries 105} no.~4, (2022) 043004},
  \href{http://arxiv.org/abs/2109.04478}{{\ttfamily arXiv:2109.04478
  [hep-ph]}}.

\bibitem{Anber:2015yca}
M.~M. Anber and E.~Sabancilar, ``{Hypermagnetic Fields and Baryon Asymmetry
  from Pseudoscalar Inflation},''
  \href{http://dx.doi.org/10.1103/PhysRevD.92.101501}{{\em Phys. Rev. D}
  {\bfseries 92} no.~10, (2015) 101501},
  \href{http://arxiv.org/abs/1507.00744}{{\ttfamily arXiv:1507.00744
  [hep-th]}}.

\bibitem{Adshead:2016iae}
P.~Adshead, J.~T. Giblin, T.~R. Scully, and E.~I. Sfakianakis,
  ``{Magnetogenesis from axion inflation},''
  \href{http://dx.doi.org/10.1088/1475-7516/2016/10/039}{{\em JCAP} {\bfseries
  1610} (2016) 039},
\href{http://arxiv.org/abs/1606.08474}{{\ttfamily arXiv:1606.08474
  [astro-ph.CO]}}.

\bibitem{Jimenez:2017cdr}
D.~Jim\'enez, K.~Kamada, K.~Schmitz, and X.-J. Xu, ``{Baryon asymmetry and
  gravitational waves from pseudoscalar inflation},''
  \href{http://dx.doi.org/10.1088/1475-7516/2017/12/011}{{\em JCAP} {\bfseries
  1712} no.~12, (2017) 011},
\href{http://arxiv.org/abs/1707.07943}{{\ttfamily arXiv:1707.07943 [hep-ph]}}.

\bibitem{Domcke:2019mnd}
V.~Domcke, B.~von Harling, E.~Morgante, and K.~Mukaida, ``{Baryogenesis from
  axion inflation},''
  \href{http://dx.doi.org/10.1088/1475-7516/2019/10/032}{{\em JCAP} {\bfseries
  10} (2019) 032}, \href{http://arxiv.org/abs/1905.13318}{{\ttfamily
  arXiv:1905.13318 [hep-ph]}}.

\bibitem{Turner:1987bw}
M.~S. Turner and L.~M. Widrow, ``{Inflation Produced, Large Scale Magnetic
  Fields},''
\href{http://dx.doi.org/10.1103/PhysRevD.37.2743}{{\em Phys. Rev.} {\bfseries
  D37} (1988) 2743}.

\bibitem{Garretson:1992vt}
W.~D. Garretson, G.~B. Field, and S.~M. Carroll, ``{Primordial magnetic fields
  from pseudoGoldstone bosons},''
  \href{http://dx.doi.org/10.1103/PhysRevD.46.5346}{{\em Phys. Rev.} {\bfseries
  D46} (1992) 5346--5351},
\href{http://arxiv.org/abs/hep-ph/9209238}{{\ttfamily arXiv:hep-ph/9209238
  [hep-ph]}}.

\bibitem{Anber:2006xt}
M.~M. Anber and L.~Sorbo, ``{N-flationary magnetic fields},''
  \href{http://dx.doi.org/10.1088/1475-7516/2006/10/018}{{\em JCAP} {\bfseries
  0610} (2006) 018},
\href{http://arxiv.org/abs/astro-ph/0606534}{{\ttfamily arXiv:astro-ph/0606534
  [astro-ph]}}.

\bibitem{Domcke:2018eki}
V.~Domcke and K.~Mukaida, ``{Gauge Field and Fermion Production during Axion
  Inflation},'' \href{http://dx.doi.org/10.1088/1475-7516/2018/11/020}{{\em
  JCAP} {\bfseries 11} (2018) 020},
  \href{http://arxiv.org/abs/1806.08769}{{\ttfamily arXiv:1806.08769
  [hep-ph]}}.

\bibitem{Gorbar:2021rlt}
E.~V. Gorbar, K.~Schmitz, O.~O. Sobol, and S.~I. Vilchinskii, ``{Gauge-field
  production during axion inflation in the gradient expansion formalism},''
  \href{http://dx.doi.org/10.1103/PhysRevD.104.123504}{{\em Phys. Rev. D}
  {\bfseries 104} no.~12, (2021) 123504},
  \href{http://arxiv.org/abs/2109.01651}{{\ttfamily arXiv:2109.01651
  [hep-ph]}}.

\bibitem{Gorbar:2021zlr}
E.~V. Gorbar, K.~Schmitz, O.~O. Sobol, and S.~I. Vilchinskii,
  ``{Hypermagnetogenesis from axion inflation: Model-independent estimates},''
  \href{http://dx.doi.org/10.1103/PhysRevD.105.043530}{{\em Phys. Rev. D}
  {\bfseries 105} no.~4, (2022) 043530},
  \href{http://arxiv.org/abs/2111.04712}{{\ttfamily arXiv:2111.04712
  [hep-ph]}}.

\bibitem{Adler:1969gk}
S.~L. Adler, ``{Axial vector vertex in spinor electrodynamics},''
  \href{http://dx.doi.org/10.1103/PhysRev.177.2426}{{\em Phys. Rev.} {\bfseries
  177} (1969) 2426--2438}.
[,241(1969)].

\bibitem{Bell:1969ts}
J.~S. Bell and R.~Jackiw, ``{A PCAC puzzle: pi0 $\rightarrow$ gamma gamma in
  the sigma model},''
\href{http://dx.doi.org/10.1007/BF02823296}{{\em Nuovo Cim.} {\bfseries A60}
  (1969) 47--61}.

\bibitem{Pouquet:1976zz}
A.~Pouquet, U.~Frisch, and J.~Leorat, ``{Strong MHD helical turbulence and the
  nonlinear dynamo effect},''
\href{http://dx.doi.org/10.1017/S0022112076002140}{{\em J. Fluid Mech.}
  {\bfseries 77} (1976) 321--354}.

\bibitem{Kahniashvili:2012uj}
T.~Kahniashvili, A.~G. Tevzadze, A.~Brandenburg, and A.~Neronov, ``{Evolution
  of Primordial Magnetic Fields from Phase Transitions},''
  \href{http://dx.doi.org/10.1103/PhysRevD.87.083007}{{\em Phys. Rev.}
  {\bfseries D87} no.~8, (2013) 083007},
\href{http://arxiv.org/abs/1212.0596}{{\ttfamily arXiv:1212.0596
  [astro-ph.CO]}}.

\bibitem{Banerjee:2004df}
R.~Banerjee and K.~Jedamzik, ``{The Evolution of cosmic magnetic fields: From
  the very early universe, to recombination, to the present},''
  \href{http://dx.doi.org/10.1103/PhysRevD.70.123003}{{\em Phys. Rev.}
  {\bfseries D70} (2004) 123003},
\href{http://arxiv.org/abs/astro-ph/0410032}{{\ttfamily arXiv:astro-ph/0410032
  [astro-ph]}}.

\bibitem{Joyce:1997uy}
M.~Joyce and M.~E. Shaposhnikov, ``{Primordial magnetic fields, right-handed
  electrons, and the Abelian anomaly},''
  \href{http://dx.doi.org/10.1103/PhysRevLett.79.1193}{{\em Phys. Rev. Lett.}
  {\bfseries 79} (1997) 1193--1196},
\href{http://arxiv.org/abs/astro-ph/9703005}{{\ttfamily arXiv:astro-ph/9703005
  [astro-ph]}}.

\bibitem{Boyarsky:2011uy}
A.~Boyarsky, J.~Fr{\"o}hlich, and O.~Ruchayskiy, ``{Self-consistent evolution
  of magnetic fields and chiral asymmetry in the early Universe},''
  \href{http://dx.doi.org/10.1103/PhysRevLett.108.031301}{{\em Phys. Rev.
  Lett.} {\bfseries 108} (2012) 031301},
\href{http://arxiv.org/abs/1109.3350}{{\ttfamily arXiv:1109.3350
  [astro-ph.CO]}}.

\bibitem{Akamatsu:2013pjd}
Y.~Akamatsu and N.~Yamamoto, ``{Chiral Plasma Instabilities},''
  \href{http://dx.doi.org/10.1103/PhysRevLett.111.052002}{{\em Phys. Rev.
  Lett.} {\bfseries 111} (2013) 052002},
\href{http://arxiv.org/abs/1302.2125}{{\ttfamily arXiv:1302.2125 [nucl-th]}}.

\bibitem{Hirono:2015rla}
Y.~Hirono, D.~Kharzeev, and Y.~Yin, ``{Self-similar inverse cascade of magnetic
  helicity driven by the chiral anomaly},''
  \href{http://dx.doi.org/10.1103/PhysRevD.92.125031}{{\em Phys. Rev.}
  {\bfseries D92} no.~12, (2015) 125031},
\href{http://arxiv.org/abs/1509.07790}{{\ttfamily arXiv:1509.07790 [hep-th]}}.

\bibitem{Yamamoto:2016xtu}
N.~Yamamoto, ``{Scaling laws in chiral hydrodynamic turbulence},''
  \href{http://dx.doi.org/10.1103/PhysRevD.93.125016}{{\em Phys. Rev.}
  {\bfseries D93} no.~12, (2016) 125016},
\href{http://arxiv.org/abs/1603.08864}{{\ttfamily arXiv:1603.08864 [hep-th]}}.

\bibitem{Rogachevskii:2017uyc}
I.~Rogachevskii, O.~Ruchayskiy, A.~Boyarsky, J.~Fr{\"o}hlich, N.~Kleeorin,
  A.~Brandenburg, and J.~Schober, ``{Laminar and turbulent dynamos in chiral
  magnetohydrodynamics-I: Theory},''
  \href{http://dx.doi.org/10.3847/1538-4357/aa886b}{{\em Astrophys. J.}
  {\bfseries 846} no.~2, (2017) 153},
\href{http://arxiv.org/abs/1705.00378}{{\ttfamily arXiv:1705.00378
  [physics.plasm-ph]}}.

\bibitem{Kamada:2018tcs}
K.~Kamada, ``{Return of grand unified theory baryogenesis: Source of helical
  hypermagnetic fields for the baryon asymmetry of the universe},''
  \href{http://dx.doi.org/10.1103/PhysRevD.97.103506}{{\em Phys. Rev.}
  {\bfseries D97} no.~10, (2018) 103506},
\href{http://arxiv.org/abs/1802.03055}{{\ttfamily arXiv:1802.03055 [hep-ph]}}.

\bibitem{Fujita:2016igl}
T.~Fujita and K.~Kamada, ``{Large-scale magnetic fields can explain the baryon
  asymmetry of the Universe},''
  \href{http://dx.doi.org/10.1103/PhysRevD.93.083520}{{\em Phys. Rev.}
  {\bfseries D93} no.~8, (2016) 083520},
\href{http://arxiv.org/abs/1602.02109}{{\ttfamily arXiv:1602.02109 [hep-ph]}}.

\bibitem{Kamada:2016eeb}
K.~Kamada and A.~J. Long, ``{Baryogenesis from decaying magnetic helicity},''
  \href{http://dx.doi.org/10.1103/PhysRevD.94.063501}{{\em Phys. Rev.}
  {\bfseries D94} no.~6, (2016) 063501},
\href{http://arxiv.org/abs/1606.08891}{{\ttfamily arXiv:1606.08891
  [astro-ph.CO]}}.

\bibitem{Kamada:2016cnb}
K.~Kamada and A.~J. Long, ``{Evolution of the Baryon Asymmetry through the
  Electroweak Crossover in the Presence of a Helical Magnetic Field},''
  \href{http://dx.doi.org/10.1103/PhysRevD.94.123509}{{\em Phys. Rev.}
  {\bfseries D94} no.~12, (2016) 123509},
\href{http://arxiv.org/abs/1610.03074}{{\ttfamily arXiv:1610.03074 [hep-ph]}}.

\bibitem{Domcke:2020zez}
V.~Domcke, V.~Guidetti, Y.~Welling, and A.~Westphal, ``{Resonant backreaction
  in axion inflation},''
  \href{http://dx.doi.org/10.1088/1475-7516/2020/09/009}{{\em JCAP} {\bfseries
  09} (2020) 009}, \href{http://arxiv.org/abs/2002.02952}{{\ttfamily
  arXiv:2002.02952 [astro-ph.CO]}}.

\bibitem{Peloso:2022ovc}
M.~Peloso and L.~Sorbo, ``{Instability in axion inflation with strong
  backreaction from gauge modes},''
  \href{http://dx.doi.org/10.1088/1475-7516/2023/01/038}{{\em JCAP} {\bfseries
  01} (2023) 038}, \href{http://arxiv.org/abs/2209.08131}{{\ttfamily
  arXiv:2209.08131 [astro-ph.CO]}}.

\bibitem{Anber:2009ua}
M.~M. Anber and L.~Sorbo, ``{Naturally inflating on steep potentials through
  electromagnetic dissipation},''
  \href{http://dx.doi.org/10.1103/PhysRevD.81.043534}{{\em Phys. Rev.}
  {\bfseries D81} (2010) 043534},
\href{http://arxiv.org/abs/0908.4089}{{\ttfamily arXiv:0908.4089 [hep-th]}}.

\bibitem{Hook:2019vcn}
A.~Hook, J.~Huang, and D.~Racco, ``{Minimal signatures of the Standard Model in
  non-Gaussianities},''
  \href{http://dx.doi.org/10.1103/PhysRevD.101.023519}{{\em Phys. Rev. D}
  {\bfseries 101} no.~2, (2020) 023519},
  \href{http://arxiv.org/abs/1908.00019}{{\ttfamily arXiv:1908.00019
  [hep-ph]}}.

\bibitem{Cuissa:2018oiw}
J.~R.~C. Cuissa and D.~G. Figueroa, ``{Lattice formulation of axion inflation.
  Application to preheating},''
  \href{http://dx.doi.org/10.1088/1475-7516/2019/06/002}{{\em JCAP} {\bfseries
  1906} no.~06, (2019) 002},
\href{http://arxiv.org/abs/1812.03132}{{\ttfamily arXiv:1812.03132
  [astro-ph.CO]}}.

\bibitem{Adshead:2019lbr}
P.~Adshead, J.~T. Giblin, M.~Pieroni, and Z.~J. Weiner, ``{Constraining axion
  inflation with gravitational waves from preheating},''
  \href{http://dx.doi.org/10.1103/PhysRevD.101.083534}{{\em Phys. Rev. D}
  {\bfseries 101} no.~8, (2020) 083534},
  \href{http://arxiv.org/abs/1909.12842}{{\ttfamily arXiv:1909.12842
  [astro-ph.CO]}}.

\bibitem{Adshead:2019igv}
P.~Adshead, J.~T. Giblin, M.~Pieroni, and Z.~J. Weiner, ``{Constraining Axion
  Inflation with Gravitational Waves across 29 Decades in Frequency},''
  \href{http://dx.doi.org/10.1103/PhysRevLett.124.171301}{{\em Phys. Rev.
  Lett.} {\bfseries 124} no.~17, (2020) 171301},
  \href{http://arxiv.org/abs/1909.12843}{{\ttfamily arXiv:1909.12843
  [astro-ph.CO]}}.

\bibitem{Caravano:2021bfn}
A.~Caravano, E.~Komatsu, K.~D. Lozanov, and J.~Weller, ``{Lattice simulations
  of Abelian gauge fields coupled to axions during inflation},''
  \href{http://dx.doi.org/10.1103/PhysRevD.105.123530}{{\em Phys. Rev. D}
  {\bfseries 105} no.~12, (2022) 123530},
  \href{http://arxiv.org/abs/2110.10695}{{\ttfamily arXiv:2110.10695
  [astro-ph.CO]}}.

\bibitem{Caravano:2022epk}
A.~Caravano, E.~Komatsu, K.~D. Lozanov, and J.~Weller, ``{Lattice Simulations
  of Axion-U(1) Inflation},'' \href{http://arxiv.org/abs/2204.12874}{{\ttfamily
  arXiv:2204.12874 [astro-ph.CO]}}.

\bibitem{Sobol:2019xls}
O.~O. Sobol, E.~V. Gorbar, and S.~I. Vilchinskii, ``{Backreaction of
  electromagnetic fields and the Schwinger effect in pseudoscalar inflation
  magnetogenesis},'' \href{http://dx.doi.org/10.1103/PhysRevD.100.063523}{{\em
  Phys. Rev. D} {\bfseries 100} no.~6, (2019) 063523},
  \href{http://arxiv.org/abs/1907.10443}{{\ttfamily arXiv:1907.10443
  [astro-ph.CO]}}.

\bibitem{Cado:2022pxk}
Y.~Cado and M.~Quir\'os, ``{Numerical study of the Schwinger effect in axion
  inflation},'' \href{http://dx.doi.org/10.1103/PhysRevD.106.123527}{{\em Phys.
  Rev. D} {\bfseries 106} no.~12, (2022) 123527},
  \href{http://arxiv.org/abs/2208.10977}{{\ttfamily arXiv:2208.10977
  [hep-ph]}}.

\bibitem{Fujita:2022fwc}
T.~Fujita, J.~Kume, K.~Mukaida, and Y.~Tada, ``{Effective treatment of U(1)
  gauge field and charged particles in axion inflation},''
  \href{http://dx.doi.org/10.1088/1475-7516/2022/09/023}{{\em JCAP} {\bfseries
  09} (2022) 023}, \href{http://arxiv.org/abs/2204.01180}{{\ttfamily
  arXiv:2204.01180 [hep-ph]}}.

\bibitem{Durrer:2013pga}
R.~Durrer and A.~Neronov, ``{Cosmological Magnetic Fields: Their Generation,
  Evolution and Observation},''
  \href{http://dx.doi.org/10.1007/s00159-013-0062-7}{{\em Astron. Astrophys.
  Rev.} {\bfseries 21} (2013) 62},
\href{http://arxiv.org/abs/1303.7121}{{\ttfamily arXiv:1303.7121
  [astro-ph.CO]}}.

\bibitem{Vilenkin:1980fu}
A.~Vilenkin, ``{Equilibrium parity violating current in a magnetic field},''
\href{http://dx.doi.org/10.1103/PhysRevD.22.3080}{{\em Phys. Rev.} {\bfseries
  D22} (1980) 3080--3084}.

\bibitem{Alekseev:1998ds}
A.~{\relax Yu}. Alekseev, V.~V. Cheianov, and J.~Fr{\"o}hlich, ``{Universality
  of transport properties in equilibrium, Goldstone theorem and chiral
  anomaly},'' \href{http://dx.doi.org/10.1103/PhysRevLett.81.3503}{{\em Phys.
  Rev. Lett.} {\bfseries 81} (1998) 3503--3506},
\href{http://arxiv.org/abs/cond-mat/9803346}{{\ttfamily arXiv:cond-mat/9803346
  [cond-mat]}}.

\bibitem{Son:2004tq}
D.~T. Son and A.~R. Zhitnitsky, ``{Quantum anomalies in dense matter},''
  \href{http://dx.doi.org/10.1103/PhysRevD.70.074018}{{\em Phys. Rev.}
  {\bfseries D70} (2004) 074018},
\href{http://arxiv.org/abs/hep-ph/0405216}{{\ttfamily arXiv:hep-ph/0405216
  [hep-ph]}}.

\bibitem{Fukushima:2008xe}
K.~Fukushima, D.~E. Kharzeev, and H.~J. Warringa, ``{The Chiral Magnetic
  Effect},'' \href{http://dx.doi.org/10.1103/PhysRevD.78.074033}{{\em Phys.
  Rev.} {\bfseries D78} (2008) 074033},
\href{http://arxiv.org/abs/0808.3382}{{\ttfamily arXiv:0808.3382 [hep-ph]}}.

\bibitem{Bodeker:2019ajh}
D.~B\"odeker and D.~Schr\"oder, ``{Equilibration of right-handed electrons},''
  \href{http://dx.doi.org/10.1088/1475-7516/2019/05/010}{{\em JCAP} {\bfseries
  05} (2019) 010}, \href{http://arxiv.org/abs/1902.07220}{{\ttfamily
  arXiv:1902.07220 [hep-ph]}}.

\bibitem{Domcke:2020kcp}
V.~Domcke, Y.~Ema, K.~Mukaida, and M.~Yamada, ``{Spontaneous Baryogenesis from
  Axions with Generic Couplings},''
  \href{http://dx.doi.org/10.1007/JHEP08(2020)096}{{\em JHEP} {\bfseries 08}
  (2020) 096}, \href{http://arxiv.org/abs/2006.03148}{{\ttfamily
  arXiv:2006.03148 [hep-ph]}}.

\bibitem{Covi:1996wh}
L.~Covi, E.~Roulet, and F.~Vissani, ``{CP violating decays in leptogenesis
  scenarios},'' \href{http://dx.doi.org/10.1016/0370-2693(96)00817-9}{{\em
  Phys. Lett. B} {\bfseries 384} (1996) 169--174},
  \href{http://arxiv.org/abs/hep-ph/9605319}{{\ttfamily arXiv:hep-ph/9605319}}.

\bibitem{Davidson:2002qv}
S.~Davidson and A.~Ibarra, ``{A Lower bound on the right-handed neutrino mass
  from leptogenesis},''
  \href{http://dx.doi.org/10.1016/S0370-2693(02)01735-5}{{\em Phys. Lett. B}
  {\bfseries 535} (2002) 25--32},
  \href{http://arxiv.org/abs/hep-ph/0202239}{{\ttfamily arXiv:hep-ph/0202239}}.

\bibitem{Buchmuller:2002rq}
W.~Buchmuller, P.~Di~Bari, and M.~Plumacher, ``{Cosmic microwave background,
  matter - antimatter asymmetry and neutrino masses},''
  \href{http://dx.doi.org/10.1016/S0550-3213(02)00737-X}{{\em Nucl. Phys. B}
  {\bfseries 643} (2002) 367--390},
  \href{http://arxiv.org/abs/hep-ph/0205349}{{\ttfamily arXiv:hep-ph/0205349}}.
  [Erratum: Nucl.Phys.B 793, 362 (2008)].

\bibitem{Pilaftsis:1997jf}
A.~Pilaftsis, ``{CP violation and baryogenesis due to heavy Majorana
  neutrinos},'' \href{http://dx.doi.org/10.1103/PhysRevD.56.5431}{{\em Phys.
  Rev. D} {\bfseries 56} (1997) 5431--5451},
  \href{http://arxiv.org/abs/hep-ph/9707235}{{\ttfamily arXiv:hep-ph/9707235}}.

\bibitem{Pilaftsis:2003gt}
A.~Pilaftsis and T.~E.~J. Underwood, ``{Resonant leptogenesis},''
  \href{http://dx.doi.org/10.1016/j.nuclphysb.2004.05.029}{{\em Nucl. Phys. B}
  {\bfseries 692} (2004) 303--345},
  \href{http://arxiv.org/abs/hep-ph/0309342}{{\ttfamily arXiv:hep-ph/0309342}}.

\bibitem{Buchmuller:2004nz}
W.~Buchmuller, P.~Di~Bari, and M.~Plumacher, ``{Leptogenesis for
  pedestrians},'' \href{http://dx.doi.org/10.1016/j.aop.2004.02.003}{{\em
  Annals Phys.} {\bfseries 315} (2005) 305--351},
  \href{http://arxiv.org/abs/hep-ph/0401240}{{\ttfamily arXiv:hep-ph/0401240}}.

\bibitem{Mukaida:2021sgv}
K.~Mukaida, K.~Schmitz, and M.~Yamada, ``{Baryon Asymmetry of the Universe from
  Lepton Flavor Violation},''
  \href{http://dx.doi.org/10.1103/PhysRevLett.129.011803}{{\em Phys. Rev.
  Lett.} {\bfseries 129} no.~1, (2022) 011803},
  \href{http://arxiv.org/abs/2111.03082}{{\ttfamily arXiv:2111.03082
  [hep-ph]}}.

\bibitem{Domcke:2022uue}
V.~Domcke, K.~Kamada, K.~Mukaida, K.~Schmitz, and M.~Yamada, ``{A new
  constraint on primordial lepton flavour asymmetries},''
  \href{http://arxiv.org/abs/2208.03237}{{\ttfamily arXiv:2208.03237
  [hep-ph]}}.

\bibitem{DOnofrio:2014rug}
M.~D'Onofrio, K.~Rummukainen, and A.~Tranberg, ``{Sphaleron Rate in the Minimal
  Standard Model},''
  \href{http://dx.doi.org/10.1103/PhysRevLett.113.141602}{{\em Phys. Rev.
  Lett.} {\bfseries 113} no.~14, (2014) 141602},
\href{http://arxiv.org/abs/1404.3565}{{\ttfamily arXiv:1404.3565 [hep-ph]}}.

\bibitem{DOnofrio:2015gop}
M.~D'Onofrio and K.~Rummukainen, ``{Standard model cross-over on the
  lattice},'' \href{http://dx.doi.org/10.1103/PhysRevD.93.025003}{{\em Phys.
  Rev.} {\bfseries D93} no.~2, (2016) 025003},
\href{http://arxiv.org/abs/1508.07161}{{\ttfamily arXiv:1508.07161 [hep-ph]}}.

\bibitem{Harvey:1990qw}
J.~A. Harvey and M.~S. Turner, ``{Cosmological baryon and lepton number in the
  presence of electroweak fermion number violation},''
  \href{http://dx.doi.org/10.1103/PhysRevD.42.3344}{{\em Phys. Rev. D}
  {\bfseries 42} (1990) 3344--3349}.

\bibitem{Dolgov:2002ab}
A.~D. Dolgov, S.~H. Hansen, S.~Pastor, S.~T. Petcov, G.~G. Raffelt, and D.~V.
  Semikoz, ``{Cosmological bounds on neutrino degeneracy improved by flavor
  oscillations},'' \href{http://dx.doi.org/10.1016/S0550-3213(02)00274-2}{{\em
  Nucl. Phys. B} {\bfseries 632} (2002) 363--382},
  \href{http://arxiv.org/abs/hep-ph/0201287}{{\ttfamily arXiv:hep-ph/0201287}}.

\bibitem{Wong:2002fa}
Y.~Y.~Y. Wong, ``{Analytical treatment of neutrino asymmetry equilibration from
  flavor oscillations in the early universe},''
  \href{http://dx.doi.org/10.1103/PhysRevD.66.025015}{{\em Phys. Rev. D}
  {\bfseries 66} (2002) 025015},
  \href{http://arxiv.org/abs/hep-ph/0203180}{{\ttfamily arXiv:hep-ph/0203180}}.

\bibitem{Laine:1999wv}
M.~Laine and M.~E. Shaposhnikov, ``{A Remark on sphaleron erasure of baryon
  asymmetry},'' \href{http://dx.doi.org/10.1103/PhysRevD.61.117302}{{\em Phys.
  Rev. D} {\bfseries 61} (2000) 117302},
  \href{http://arxiv.org/abs/hep-ph/9911473}{{\ttfamily arXiv:hep-ph/9911473}}.

\bibitem{Kamada:2020bmb}
K.~Kamada, F.~Uchida, and J.~Yokoyama, ``{Baryon isocurvature constraints on
  the primordial hypermagnetic fields},''
  \href{http://dx.doi.org/10.1088/1475-7516/2021/04/034}{{\em JCAP} {\bfseries
  04} (2021) 034}, \href{http://arxiv.org/abs/2012.14435}{{\ttfamily
  arXiv:2012.14435 [astro-ph.CO]}}.

\bibitem{Giovannini:1997gp}
M.~Giovannini and M.~E. Shaposhnikov, ``{Primordial magnetic fields, anomalous
  isocurvature fluctuations and big bang nucleosynthesis},''
  \href{http://dx.doi.org/10.1103/PhysRevLett.80.22}{{\em Phys. Rev. Lett.}
  {\bfseries 80} (1998) 22--25},
  \href{http://arxiv.org/abs/hep-ph/9708303}{{\ttfamily arXiv:hep-ph/9708303}}.

\bibitem{Giovannini:1997eg}
M.~Giovannini and M.~E. Shaposhnikov, ``{Primordial hypermagnetic fields and
  triangle anomaly},'' \href{http://dx.doi.org/10.1103/PhysRevD.57.2186}{{\em
  Phys. Rev.} {\bfseries D57} (1998) 2186--2206},
\href{http://arxiv.org/abs/hep-ph/9710234}{{\ttfamily arXiv:hep-ph/9710234
  [hep-ph]}}.

\bibitem{Kajantie:1996qd}
K.~Kajantie, M.~Laine, K.~Rummukainen, and M.~E. Shaposhnikov, ``{A
  Nonperturbative analysis of the finite T phase transition in SU(2) x U(1)
  electroweak theory},''
  \href{http://dx.doi.org/10.1016/S0550-3213(97)00164-8}{{\em Nucl. Phys.}
  {\bfseries B493} (1997) 413--438},
\href{http://arxiv.org/abs/hep-lat/9612006}{{\ttfamily arXiv:hep-lat/9612006
  [hep-lat]}}.

\bibitem{Brandenburg:1996fc}
A.~Brandenburg, K.~Enqvist, and P.~Olesen, ``{Large scale magnetic fields from
  hydromagnetic turbulence in the very early universe},''
  \href{http://dx.doi.org/10.1103/PhysRevD.54.1291}{{\em Phys. Rev. D}
  {\bfseries 54} (1996) 1291--1300},
  \href{http://arxiv.org/abs/astro-ph/9602031}{{\ttfamily
  arXiv:astro-ph/9602031}}.

\bibitem{Brandenburg:2016odr}
A.~Brandenburg and T.~Kahniashvili, ``{Classes of hydrodynamic and
  magnetohydrodynamic turbulent decay},''
  \href{http://dx.doi.org/10.1103/PhysRevLett.118.055102}{{\em Phys. Rev.
  Lett.} {\bfseries 118} no.~5, (2017) 055102},
  \href{http://arxiv.org/abs/1607.01360}{{\ttfamily arXiv:1607.01360
  [physics.flu-dyn]}}.

\bibitem{Tashiro:2013bxa}
H.~Tashiro and T.~Vachaspati, ``{Cosmological magnetic field correlators from
  blazar induced cascade},''
  \href{http://dx.doi.org/10.1103/PhysRevD.87.123527}{{\em Phys. Rev. D}
  {\bfseries 87} no.~12, (2013) 123527},
  \href{http://arxiv.org/abs/1305.0181}{{\ttfamily arXiv:1305.0181
  [astro-ph.CO]}}.

\bibitem{Tashiro:2013ita}
H.~Tashiro, W.~Chen, F.~Ferrer, and T.~Vachaspati, ``{Search for CP Violating
  Signature of Intergalactic Magnetic Helicity in the Gamma Ray Sky},''
  \href{http://dx.doi.org/10.1093/mnrasl/slu134}{{\em Mon. Not. Roy. Astron.
  Soc.} {\bfseries 445} no.~1, (2014) L41--L45},
  \href{http://arxiv.org/abs/1310.4826}{{\ttfamily arXiv:1310.4826
  [astro-ph.CO]}}.

\bibitem{Tashiro:2014gfa}
H.~Tashiro and T.~Vachaspati, ``{Parity-odd correlators of diffuse gamma rays
  and intergalactic magnetic fields},''
  \href{http://dx.doi.org/10.1093/mnras/stu2736}{{\em Mon. Not. Roy. Astron.
  Soc.} {\bfseries 448} no.~1, (2015) 299--306},
  \href{http://arxiv.org/abs/1409.3627}{{\ttfamily arXiv:1409.3627
  [astro-ph.CO]}}.

\bibitem{Chen:2014qva}
W.~Chen, B.~D. Chowdhury, F.~Ferrer, H.~Tashiro, and T.~Vachaspati,
  ``{Intergalactic magnetic field spectra from diffuse gamma rays},''
  \href{http://dx.doi.org/10.1093/mnras/stv308}{{\em Mon. Not. Roy. Astron.
  Soc.} {\bfseries 450} no.~4, (2015) 3371--3380},
  \href{http://arxiv.org/abs/1412.3171}{{\ttfamily arXiv:1412.3171
  [astro-ph.CO]}}.

\bibitem{MAGIC:2022piy}
{\bfseries MAGIC} Collaboration, V.~A. Acciari {\em et~al.}, ``{A lower bound
  on intergalactic magnetic fields from time variability of 1ES 0229+200 from
  MAGIC and Fermi/LAT observations},''
  \href{http://arxiv.org/abs/2210.03321}{{\ttfamily arXiv:2210.03321
  [astro-ph.HE]}}.

\bibitem{Fujita:2015iga}
T.~Fujita, R.~Namba, Y.~Tada, N.~Takeda, and H.~Tashiro, ``{Consistent
  generation of magnetic fields in axion inflation models},''
  \href{http://dx.doi.org/10.1088/1475-7516/2015/05/054}{{\em JCAP} {\bfseries
  05} (2015) 054}, \href{http://arxiv.org/abs/1503.05802}{{\ttfamily
  arXiv:1503.05802 [astro-ph.CO]}}.

\bibitem{Kamada:2019uxp}
K.~Kamada and C.~S. Shin, ``{Magnetogenesis from a rotating scalar: \`a la
  scalar chiral magnetic effect},''
  \href{http://dx.doi.org/10.1007/JHEP04(2020)185}{{\em JHEP} {\bfseries 04}
  (2020) 185}, \href{http://arxiv.org/abs/1905.06966}{{\ttfamily
  arXiv:1905.06966 [hep-ph]}}.

\bibitem{Co:2019wyp}
R.~T. Co and K.~Harigaya, ``{Axiogenesis},''
  \href{http://dx.doi.org/10.1103/PhysRevLett.124.111602}{{\em Phys. Rev.
  Lett.} {\bfseries 124} no.~11, (2020) 111602},
  \href{http://arxiv.org/abs/1910.02080}{{\ttfamily arXiv:1910.02080
  [hep-ph]}}.

\bibitem{Co:2020xlh}
R.~T. Co, L.~J. Hall, and K.~Harigaya, ``{Predictions for Axion Couplings from
  ALP Cogenesis},'' \href{http://dx.doi.org/10.1007/JHEP01(2021)172}{{\em JHEP}
  {\bfseries 01} (2021) 172}, \href{http://arxiv.org/abs/2006.04809}{{\ttfamily
  arXiv:2006.04809 [hep-ph]}}.

\bibitem{Co:2020jtv}
R.~T. Co, N.~Fernandez, A.~Ghalsasi, L.~J. Hall, and K.~Harigaya,
  ``{Lepto-Axiogenesis},''
  \href{http://dx.doi.org/10.1007/JHEP03(2021)017}{{\em JHEP} {\bfseries 03}
  (2021) 017}, \href{http://arxiv.org/abs/2006.05687}{{\ttfamily
  arXiv:2006.05687 [hep-ph]}}.

\bibitem{Affleck:1984fy}
I.~Affleck and M.~Dine, ``{A New Mechanism for Baryogenesis},''
  \href{http://dx.doi.org/10.1016/0550-3213(85)90021-5}{{\em Nucl. Phys. B}
  {\bfseries 249} (1985) 361--380}.

\bibitem{Dine:1995uk}
M.~Dine, L.~Randall, and S.~D. Thomas, ``{Supersymmetry breaking in the early
  universe},'' \href{http://dx.doi.org/10.1103/PhysRevLett.75.398}{{\em Phys.
  Rev. Lett.} {\bfseries 75} (1995) 398--401},
  \href{http://arxiv.org/abs/hep-ph/9503303}{{\ttfamily arXiv:hep-ph/9503303}}.

\bibitem{Dine:1995kz}
M.~Dine, L.~Randall, and S.~D. Thomas, ``{Baryogenesis from flat directions of
  the supersymmetric standard model},''
  \href{http://dx.doi.org/10.1016/0550-3213(95)00538-2}{{\em Nucl. Phys. B}
  {\bfseries 458} (1996) 291--326},
  \href{http://arxiv.org/abs/hep-ph/9507453}{{\ttfamily arXiv:hep-ph/9507453}}.

\end{thebibliography}\endgroup


\end{document}